\documentclass[twocolumn]{aastex631}

\usepackage[varg]{newtx}
\usepackage{bm}
\usepackage{physics}
\definecolor{PineGreen}{HTML}{019286}

\newcommand{\eps}{\epsilon}

\newcommand{\github}{https://github.com/nuc-astro/WinNet}
\newcommand{\zenodo}{https://zenodo.org/record/8220549}

\usepackage{soul}

\begin{document}

\title{The nuclear reaction network WinNet}

\author[0000-0001-6653-7538]{M. Reichert}
\affiliation{Departament d'Astonomia i Astrof\'{\i}sica, Universitat de Val\`encia, Edifici d'Investigatci\'{o} Jeroni Munyoz, C/Dr. Moliner, 50, E-46100 Burjassot (Val\`encia), Spain}

\author{C. Winteler}
\affiliation{Department of Physics, University of Basel, Klingelbergstrasse 82, CH-4056 Basel, Switzerland}

\author[0000-0003-4156-5342]{O. Korobkin}
\affiliation{Center for Theoretical Astrophysics, Los Alamos National Laboratory, Los Alamos, NM 87545, USA}

\author[0000-0002-6995-3032]{A. Arcones}
\affiliation{Institut für Kernphysik (Theoriezentrum), Technische Universität Darmstadt, Schlossgartenstr. 2, D-64289 Darmstadt, Germany}
\affiliation{GSI Helmholtzzentrum für Schwerionenforschung GmbH, Planckstr. 1, D-64291 Darmstadt, Germany}

\author{J. Bliss}
\affiliation{Institut für Kernphysik (Theoriezentrum), Technische Universität Darmstadt, Schlossgartenstr. 2, D-64289 Darmstadt, Germany}

\author[0000-0002-4445-8908]{M. Eichler}
\affiliation{Institut für Kernphysik (Theoriezentrum), Technische Universität Darmstadt, Schlossgartenstr. 2, D-64289 Darmstadt, Germany}

\author{U. Frischknecht}
\affiliation{Department of Physics, University of Basel, Klingelbergstrasse 82, CH-4056 Basel, Switzerland}

\author[0000-0003-0191-2477]{C. Fröhlich}
\affiliation{Department of Physics, North Carolina State University, Raleigh, NC 27695, USA}

\author[0000-0001-8764-6522]{R. Hirschi}
\affiliation{Astrophysics Group, Lennard-Jones Laboratories, Keele University, Keele ST5 5BG, UK}
\affiliation{Institute for the Physics and Mathematics of the Universe (WPI), University of Tokyo, 5-1-5 Kashiwanoha, Kashiwa 277-8583, Japan}

\author[0000-0001-8168-4579]{M. Jacobi}
\affiliation{Institut für Kernphysik (Theoriezentrum), Technische Universität Darmstadt, Schlossgartenstr. 2, D-64289 Darmstadt, Germany}

\author[0009-0005-5121-7343]{J. Kuske}
\affiliation{Institut für Kernphysik (Theoriezentrum), Technische Universität Darmstadt, Schlossgartenstr. 2, D-64289 Darmstadt, Germany}

\author[0000-0002-3825-0131]{G. Mart\'{\i}nez-Pinedo}
\affiliation{GSI Helmholtzzentrum für Schwerionenforschung GmbH, Planckstr. 1, D-64291 Darmstadt, Germany}
\affiliation{Institut für Kernphysik (Theoriezentrum), Technische Universität Darmstadt, Schlossgartenstr. 2, D-64289 Darmstadt, Germany}

\author[0000-0002-1988-9706]{D. Martin}
\affiliation{Institut für Kernphysik (Theoriezentrum), Technische Universität Darmstadt, Schlossgartenstr. 2, D-64289 Darmstadt, Germany}

\author{D. Mocelj}
\affiliation{Department of Physics, University of Basel, Klingelbergstrasse 82, CH-4056 Basel, Switzerland}

\author[0000-0002-1266-0642]{T. Rauscher}
\affiliation{Department of Physics, University of Basel, Klingelbergstrasse 82, CH-4056 Basel, Switzerland}
\affiliation{Centre for Astrophysics Research, University of Hertfordshire, Hatfield AL10 9AB, United Kingdom}

\author[0000-0002-7256-9330]{F.-K. Thielemann}
\affiliation{Department of Physics, University of Basel, Klingelbergstrasse 82, CH-4056 Basel, Switzerland}
\affiliation{GSI Helmholtzzentrum für Schwerionenforschung GmbH, Planckstr. 1, D-64291 Darmstadt, Germany}



\begin{abstract}
We present the state-of-the-art single-zone nuclear reaction network \textsc{WinNet} that is capable of calculating the nucleosynthetic yields of a large variety of astrophysical environments and conditions. This ranges from the calculation of the primordial nucleosynthesis, where only a few nuclei are considered, to the ejecta of neutron star mergers with several thousands of involved nuclei. Here we describe the underlying physics and implementation details of the reaction network. We additionally present the numerical implementation of two different integration methods, the implicit Euler method and Gears method along with their advantages and disadvantages. We furthermore describe basic example cases of thermodynamic conditions that we provide together with the network and demonstrate the reliability of the code by using simple test cases. With this publication, \textsc{WinNet} is publicly available and open source at \dataset[GitHub]{\github} and \dataset[Zenodo]{\zenodo}.
\end{abstract}



\section{Introduction} \label{sct:intro}

Nuclear reaction networks are crucial to investigate the synthesis of elements and their isotopes in astrophysical events. While the events can vastly differ in their conditions, the procedure to derive their ejecta composition is always similar. The foundation of the understanding of the origin of elements has been outlined already in \citet{abg}, the so-called $\alpha \beta \gamma$-Paper. 

The field of nucleosynthetic calculations encompasses the production of the light elements during the Big Bang \citep[e.g.,][]{bb-peebles1966,bb-wagoner1967,bb-yang1984,bb-boesgaard1985,bb-kawano1988,bb-olive1990,bb-walker1991,bb-smith1993,bb-cyburt2016,bb-coc2017,bb-pitrou2018,Pitrou.ea:2021,Fields.Olive:2022}, the element production during the lifetime of stars \citep[see, e.g.,][]{Kippenhahn.Weigert.Weiss:2013,karakas14review,Karakas.Lugaro:2016,bisterzo17,Kobayashi.Karakas.Lugaro:2020,Busso.ea:2021,Doherty.Gil-Pons:2017,Gil-Pons.Doherty:2018,Leung.Nomoto:2018,Leung.Nomoto.ea:2020,Arnett:1977,Woosley.Weaver:1995,Heger.ea:2003,heger10,Maeder.Meynet:2012,frischknecht:16,thielemann18,Limongi.Chieffi:2018,Arnett.Meakin.ea:2019,Kaiser.ea:2020,Eggenberger.ea:2021}, and more violent explosive events such as classical novae \citep[e.g.,][]{Arnould.Norgaard.ea:1980,Wiescher.ea:1986,Jose.Hernanz.ea:2004,Jose:2016,Vasini.Matteucci.Spitoni:2022}, X-ray bursts \citep[e.g.,][]{Wiescher.ea:1986,Rembges.ea:1997,Schatz.ea:1998,cyburt10,Jose:2016,Meisel.ea:2020}, type Ia supernovae \citep[e.g.,][]{Arnett:1969,Arnett.Truran.Woosley:1971,Iben.Tutukov:1984,Nomoto.Thielemann.Yokoi:1984,Woosley.Taam.Weaver:1986,Mueller.Arnett:1986,Thielemann.Nomoto.ea:1986,Khokhlov.Mueller.Hoeflich:1993,Hoeflich.Wheeler.Thielemann:1998,Roepke.ea:2012,Hillebrandt:2013,Pakmor.ea:2013,Dan.ea:2015,Maeda.Terada:2016,Garcia-Senz.ea:2016, Jiang.ea:2017,Roepke.Sim:2018,Thielemann.Isern.ea:2018,Shen.ea:2018,Leung.Nomoto:2018,Gronow.ea:2021,Lach.ea:2022}, core-collapse supernovae \citep[e.g.,][]{Kotake2012,Burrows:2013,Janka.Melson.Summa:2016,Mueller:2016,Radice.etal:2018,Mueller:2020,Vartanyan.ea:2022} with a focus on nucleosynthesis \citep[e.g.,][]{Woosley.Weaver:1995,Thielemann.Nomoto.ea:1996,woosley-heger2006,heger10,Perego.Hempel.ea:2015,Sukhbold.Ertl.ea:2016,Wanajo.Mueller.ea:2018,Curtis.ea:2019,Witt.etal:2021,Ghosh.ea:2022}, or a focus on $r$-process or neutrino-driven winds in supernovae \citep[e.g.,][]{Qian.Woosley:1996,Cardall1997,hoffman.woosley.qian:1997,Otsuki.Tagoshi.ea:2000,Thompson.Burrows.Meyer:2001,Wanajo2001,froehlich06,Kratz2008,Bliss.etal:2020,Psaltis2022},
magnetorotational supernovae \citep{nishimura06, winteler12b,Nishimura2015,Nishimura2017,Moesta2018,Halevi2018,Reichert.Obergaulinger.ea:2021,Powell2022,Reichert2023}, collapsars \citep{MacFadyen1999,Pruet2003b,Surman2004,McLaughlin2005,Fujimoto2008,Siegel2019,Miller2020,Zenati2020,Barnes2022,Just2022b}, and neutron star mergers \citep[e.g.,][]{freiburghaus99, korobkin12,martin15,Bovard2017,lippuner17b,wu16,Wu2019,Holmbeck2019,Wanajo2021,Rosswog2022,Kullmann2022,Kullmann2022b}. Without nucleosynthesis calculations, a whole layer of information and observables would remain inaccessible. 

Some applications require a complex modeling that takes species diffusion or convective mixing and
nuclear burning simultaneously into account (such as, e.g., the oxygen-burning phase of a star or the rapid accreting white dwarfs;
\citealt{hix99, Denissenkov2019}) and the nuclear reaction network must
therefore be included in a hydrodynamical simulation. This often
has the consequence that only a restricted number of nuclei
are considered in the calculation \citep[from the 13 or 14 alpha nuclei network - 13 or 14$\alpha$ -- developed by Thielemann and
used
e.g. in][]{Mueller1986,Benz.Hills.Thielemann:1989,Livne.Arnett:1995,Garcia-Senz.ea:2013,Garcia-Senz.ea:2016}
over small quasi-equilibrium networks \citep[e.g.][named QE-reduced
or iso7]{Hix.ea:1998,Timmes.Hoffman.Woosley:2000,Hix.ea:2007}, to
slightly enlarged networks beyond 13$\alpha$ -- like net21 -- which
include additional neutron-rich isotopes in the Fe-group in order to
be able to follow $Y_e$ below 0.5 \citep[for a comparison of these
approaches,
see][]{Bravo:2020}. 
Recently such methods have been extended to networks that contain up
to the order of $100$ nuclei
\citep[][]{harris2017,Sandoval2021,Navo2023}. These so-called in situ
networks have the advantage of providing an accurate nuclear energy production as well as more precise nucleon abundances that imply more realistic neutrino opacities for the feedback to the simulation
\citep[e.g.,][]{Mueller1986,Nakamura2014,harris2017,Navo2023}. On the other hand, simplifying assumptions within the nuclear reaction network equations, artificial
numerical diffusion \citep[e.g.,][]{Fryxell1991,hix99,Plewa1999}, and
the reduced set of nuclei in such energy generation networks can make
the predicted ejecta composition, even with extended post-processing
networks, more uncertain \citep[this is nicely shown
in][]{Bravo:2020}.

For astrophysical scenarios with a much larger diffusion timescale compared to the nuclear burning timescale, one can trace the ejecta with passively advected particles whose movements are influenced by the velocity field of the fluid \citep[e.g.,][]{Nagataki1997,Seitenzahl2010,Nishimura2015,harris2017,Bovard2017a,Sieverding2023}. These so-called tracer particles record the thermodynamic conditions as well as the neutrino fluxes in time. In the case that the impact of diffusion on the composition is negligible compared to the burning, each tracer can be calculated individually, and the total ejected matter is the (possibly weighted) average over all tracer particles individually. Reaction networks that are based on individual tracers (or zones) that are unable to interact with each other are called single-zone nuclear reaction networks. The advantage of those codes is that they can include a much more complete set of nuclei and reactions. This enables a calculation of the synthesis of the heaviest known elements, typically with $\sim 7000$ nuclei and $\sim 90000$ reactions involved.

The compilation of a consistent reaction database is especially challenging. Nuclear reactions are often provided in different formats and in different databases that are individually complete. Among others, the largest and publicly available databases are the JINA Reaclib database \citep{cyburt10}, Bruslib \citep{Aikawa2005}, the Starlib database \citep{Sallaska2013}, NACRE \citep{Xu2013}, and KADONIS \citep{Dillmann2006}. However, none of the aforementioned libraries provides a complete set of electron/positron-captures as well as $\beta^+$/$\beta^-$ decays at stellar conditions \citep{Fuller1982,Fuller1985,Oda1994,Langanke2001,Pruet2003,Suzuki2016}, neutrino reactions \citep[e.g.,][]{Bruenn1986,Langanke2002,froehlich06,Sieverding2018,Sieverding.ea:2019}, or fission reactions and fragment distributions \citep{panov05,Goriely.Hilaire.ea:2009,panov10a,petermann2012,eichler15,Vassh.Vogt.ea:2019}. For an almost complete survey of all of these resources, see the JINAWEB collected list.\footnote{\url{https://www.jinaweb.org/science-research/scientific-resources/data}} Therefore, nuclear reaction networks always have to perform a certain amount of merging of the reaction rates if one wants to use a complete as possible set of reaction rates. Doing this rigorously can be a major task, as the consistency depends not only on not adding reactions twice or leaving them out, but also on adding reactions with the same underlying nuclear input, such as the same nuclear masses, which, far from stable nuclei, are theoretically calculated.

From a numerical point of view, reaction networks can be challenging as well. The huge differences in timescales of the reaction rates (e.g., weak decays versus strong reactions) introduce a stiffness into the differential equations. As a consequence, explicit integration methods become unstable and implicit methods have to be applied. A full implicit implementation was first achieved by \citet{Truran1966}, \citet{Truran1967}, \citet{Arnett1969}, \citet{Woosley1973}, \citet{Arnould1976}, and \citet{Thielemann1979}. While nowadays usually the first-order implicit Euler scheme is used within large nuclear reaction networks, tests with higher-order implicit schemes such as the Gear scheme have been performed as well \citep[e.g.,][]{timmes99,longland14}.

There exist a variety of single-zone reaction networks with fully implicit schemes in the literature, e.g., the SantaCruz-code by the Woosley group, going back to \citet{Woosley1973}, which followed \citet{Arnett:1969} and \citet{Truran.Cameron.Gilbert:1966} introducing a complete Newton-Raphson scheme, \textsc{BasNet} (\citealt{Thielemann1979}; for an early comparison of the two codes and the implemented reaction rate libraries, see \citealt{Hoffman.Woosley.ea:1999}), \textsc{Xnet} \citep{hix99}, \textsc{rNET} \citep{Wanajo2001}, \textsc{CFNet} \citep{froehlich2006b}, \textsc{NucNet} \citep{Meyer2007}, \textsc{rjava} \citep{Kostka2014}, \textsc{Torch} \citep{Paxton2015}, \textsc{GSInet} \citep{Mendoza-Temis2015}, \textsc{SkyNet} \citep{lippuner17a}, \textsc{Prism} \citep{Mumpower2018,Sprouse2021}, \textsc{pynucastro} \citep{Smith2023}, and other unnamed reaction networks \citep[e.g.,][]{timmes99,Iliadis2002,Otsuki2003,Koike2004,goriely11a}. \footnote{see also \url{https://cococubed.com/code_pages/burn.shtml}} However, only a small subset of them is publicly available, among them \textsc{pynucastro}\footnote{\url{https://github.com/pynucastro/pynucastro/}}, \textsc{Torch}\footnote{\url{https://cococubed.com/code_pages/net_torch.shtml}}, \textsc{rjava}\footnote{\url{https://quarknova.ca/rJava/index.html}}, \textsc{NucNet}\footnote{\url{https://sourceforge.net/u/mbradle/blog/}}, \textsc{Xnet}\footnote{\url{https://github.com/starkiller-astro/XNet}}, and \textsc{SkyNet}\footnote{\url{https://bitbucket.org/jlippuner/skynet/}}. 

Here we present the single-zone nuclear reaction network code \textsc{WinNet}, an updated version of the reaction network that has been first used in the context of Big Bang nucleosynthesis in \citet{Vonlanthen2009} and later for calculating the synthesis of heavy elements in \citet{winteler12b}. \textsc{WinNet} has a common origin to many other previously mentioned reaction networks such as \textsc{XNet}, \textsc{CFNet}, and \textsc{GSINet} as all of them were influenced by \textsc{BasNet}, which served as an initial template.

\textsc{WinNet} had been already used for different astrophysics problems; however, it was not publicly available. The code has been entirely written in Fortran 90 and has a user-friendly interface. \textsc{WinNet} is able to merge reaction rates from multiple sources and is designed for high-performance computations. It includes two fully implicit schemes, the first-order implicit Euler-backward scheme and the higher-order Gear scheme. This paper presents the basics of nuclear reaction networks as well as provides insight of the implementations within \textsc{WinNet}.

In Section~\ref{sct:Nuclearreactionnetworkfundamentals} we present the
fundamental physic concepts for nuclear reaction networks. This
includes the derivation of the ordinary differential equation (ODE) that is
solved in \textsc{WinNet} (Section~\ref{sct:Nuclearreactionnetwork}), the
principle of detailed balance (Section~\ref{sct:det_balance}),
the concept of nuclear statistical equilibrium
(NSE; Section~\ref{sssct:NSE}), a method to account for nuclear energy
  generation within the temperature evolution (Section~\ref{sct:heating}), and the treatment of
Coulomb corrections (Section~\ref{sct:coulomb_corrections}). The code
structure and included numerical solvers are presented in
Section~\ref{sct:methods}. The different supported reaction rate formats
are introduced in Section~\ref{sct:inputs}. Applications and test cases
are presented in Section~\ref{sct:network_applications}. We close with a summary in
Section~\ref{sct:summary}.

\section{Nuclear reaction network fundamentals}
\label{sct:Nuclearreactionnetworkfundamentals}
\subsection{Nuclear reaction networks}\label{sct:Nuclearreactionnetwork}
The fundamental theory behind nuclear reaction networks reaches back far into the past \citep[see, e.g.,][]{Truran.Cameron.Gilbert:1966,clayton68,Arnett1969,Woosley1973,hix99,Hix2006,winteler12b,lippuner17a}. Here we repeat briefly how to derive the differential equations, but refer the reader to previous publications for more details. 

The cross section of a reaction
\begin{equation}
    \label{eq:cross-section}
\sigma = \frac{\mathrm{number~of~reactions~per~target~and~second}}{\mathrm{flux~of~incoming~particles}},
\end{equation}
is related to the probability of a nucleus $i$ to react with nucleus
$j$.  If (e.g. in laboratory conditions like accelerator experiments) the relative velocity between target $i$ and projectile $j$ is a
constant value $v$, it is given by
\begin{equation}
\sigma=\frac{r/n_i}{n_j v}.    
\end{equation}
Here, $r$ is the number of reactions per volume
  and time, $n_i$ and $n_j$ are the number densities of the target
and projectile, respectively.  In an astrophysical plasma, both
  target and projectile follow specific velocity distributions depending on the environmental conditions like temperature and density \citep[and the reaction cross section is that of a target with thermally populated excited states; e.g.][]{Fowler:1974,Holmes.Woosley.ea:1976,Rauscher.Thielemann:2000, Rauscher:2022}. For
an arbitrary velocity distribution, $r$ can be
expressed as:
\begin{equation} \label{eq:reactionrate}
r_{i,j}=\int \sigma(|\vec{v}_i - \vec{v}_j|) \cdot |\vec{v}_i - \vec{v}_j|\mathrm{d}n_i\mathrm{d}n_j.
\end{equation}
In thermal equilibrium, the velocity (or momentum or energy) distribution depends on the type
of particle, i.e., photons obey a Planck distribution and nuclei obey, in
most cases, a Maxwell-Boltzmann distribution. Therefore, for photons,
$\mathrm{d}n_\gamma$ is given by
\begin{align}
    dn_\gamma=\frac{1}{\pi ^2 (c\hbar)^3}\frac{E_{\mathrm{\gamma}}^2}{\mathrm{exp}\left[ E_{\mathrm{\gamma}}/(k_{\mathrm{B}}T)\right]-1} \mathrm{d}E_{\mathrm{\gamma}}
\end{align}
and for nuclei $\mathrm{d}n_i$ is expressed by
\begin{equation} \label{eq:maxwellboltzmann}
\mathrm{d}n_i = n_i \left( \frac{m_i}{2\pi k_{\mathrm{B}} T}
\right)^{3/2} \mathrm{exp} \left( -\frac{m_i v^2_i}{2k_{\mathrm{B}}
    T}\right)\mathrm{d}^3\vec{v}_i \equiv n_i \phi (v_i) \, \mathrm{d}^3\vec{v}_i,
\end{equation} 
where $m_i$ is the nuclear mass. For reactions between a nucleus and a photon, $r$ is therefore given by
\begin{equation}
r_{i,\gamma} = \frac{n_i}{\pi^2 c^2 \hbar ^3} \int _0 ^{\infty} \frac{\sigma (E_{\mathrm{\gamma}})E_{\mathrm{\gamma}}^2}{\mathrm{exp}\left[ E_{\mathrm{\gamma}} / (k_{\mathrm{B}} T) \right]-1}\mathrm{d}E_{\mathrm{\gamma}}\equiv n_i\, \lambda _{i,\mathrm{\gamma}}(T).
\end{equation}
Reactions of this type are called photodisintegrations. For the case of two nuclei, the $r$ is given by
\begin{align} \label{eq:targetprojnuclei}
r_{i,j}&=n_i n_j \int \sigma(|v_i - v_j|) \times |v_i - v_j| \phi (v_i)\phi (v_j)\mathrm{d}^3\vec{v}_i\mathrm{d}^3\vec{v}_j \nonumber \\
 &= n_i n_j \langle \sigma v \rangle_{i,j},
\end{align}
where $\langle \sigma v \rangle_{i,j}$ stands for the product $\sigma v$ integrated over thermal distributions. Furthermore, an additional factor has to be introduced to avoid double counting of identical project and target nuclei. Equation~(\ref{eq:targetprojnuclei}) becomes 
\begin{equation}
\label{eq:reactivity}
r_{i,j} =  \frac{1}{1+\delta _{ij}} n_i n_j \langle \sigma v \rangle_{i,j}
\end{equation}
with the delta in the usual sense, i.e., $\delta_{i,j}=1$ for $i=j$;
otherwise, $\delta_{i,j}=0$. We get
\begin{align}
r_{j,k,l}&=\frac{1}{1+\delta _{jk}+\delta _{kl}+\delta _{jl}+2 \delta _{jkl}} n_j n_k n_l \langle ijk \rangle \nonumber \\ 
&\equiv \frac{1}{1+\Delta _{jkl}}n_j n_k n_l \langle ijk \rangle.
\end{align}
in case of three participating nuclei, where $\langle ijk \rangle$ stands for three-body reactions \citep[in most cases, a sequence of two two-body reactions and an intermediate reaction product with an extremely short half-life; see, e.g.][]{Nomoto.Thielemann.Miyaji:1985,Goerres.Wiescher.Thielemann:1995}. For example, the triple $\alpha$ reaction, which describes the probability of three helium nuclei to form $^{12}$C has a pre-factor of \mbox{$1/(1+\Delta _{jkl}) = 1/6$}.

Within a fluid that moves with velocity $\vec{v}$, $n_i$ does not only change by nuclear reactions but
also by the net-flow into the volume. We have
\begin{align} \label{eq:cont_nuc}
\frac{\partial n_i}{\partial t} =& -\vec{\nabla}\cdot (n_i \vec{v}) + \sum _j N_j ^i r_j 
+ \sum _{j,k} N^i_{j,k}r_{j,k} \nonumber \\
+&\sum _{j,k,l} N^i_{j,k,l}r_{j,k,l} \nonumber \\
  \coloneqq& -\vec{\nabla}\cdot (n_i \vec{v}) + C_\mathrm{nuc},
\end{align}
where we introduced the factors $N^i_{j}, N^i_{j,k}, N^i_{j,k,l}$
  that account for the number of particles $i$ that gets destroyed
(negative) or created (positive) in the reaction. The
first term in the equation, $-\vec{\nabla}\cdot (n_i \vec{v})$, accounts
for changes due to the fluid flow with velocity $\vec{v}$ and the
last term $C_\mathrm{nuc}$ accounts for changes due to nuclear
reactions. We can reformulate the previous equation by using the
Lagrangian time derivative that is related to the Eulerian time
derivative via
\begin{align}
  \frac{D}{Dt} = \frac{\partial}{\partial t} +\vec{v}\cdot\vec{\nabla}.
\end{align}
We can therefore obtain
\begin{equation}
    \frac{\partial n_i}{\partial t} = \frac{D n_i}{Dt}-\vec{v}\cdot \vec{\nabla} n_i,
\end{equation}
and, as a consequence, Eq.~\eqref{eq:cont_nuc} becomes
\begin{equation}\label{eq:cont_nuc_lagr}
    \frac{D n_i}{Dt} = -n_i\vec{\nabla}\cdot\vec{v} + C_\mathrm{nuc}.
\end{equation}
Using the continuity equation and the Lagrangian
time derivative
\begin{equation}
    \frac{\partial \rho}{\partial t} = - \vec{\nabla}\cdot (\rho
    \vec{v}) = -\vec{v}\cdot\vec{\nabla} \rho - \rho
    \vec{\nabla}\cdot\vec{v} \Rightarrow \frac{D\rho}{Dt}=-\rho \vec{\nabla}\cdot\vec{v},   
\end{equation}
and therefore
\begin{equation}
  \vec{\nabla}\cdot\vec{v} = - \frac{1}{\rho} \frac{D\rho}{Dt}.
\end{equation}
Thus, we can insert into Eq.~\eqref{eq:cont_nuc_lagr} and get
\begin{equation}\label{eq:lagr_deriv_nuc}
    \frac{D n_i}{Dt} - \frac{n_i}{\rho}\frac{D\rho}{Dt} = \rho \frac{D(n_i/\rho)}{Dt} = C_\mathrm{nuc}.
\end{equation}
This derivation has been done previously \citep[see,
e.g.,][]{Mihalas1999} in the context of atomic processes related
  to radiation transport, but as shown here it is also valid in the context of nuclear reactions \citep[see also][]{lippuner17a}.

In order to obtain a density-independent expression instead of utilizing number densities $n_i$, we introduce the  
density (or mass) fraction of nucleus $i$, $X_i$, which can be expressed via 
\begin{equation}
X_i=\frac{\rho_i}{\rho}=\frac{n_im_i}{\rho}=\frac{n_i\mathcal{A}_im_u}{\rho}\approx\frac{n_iA_im_u}{\rho}.
\label{eq:massfrac_def2}
\end{equation}
This includes the mass of nuclei $\mathcal{A}_i m_u$ (where $\mathcal{A}_i$ is the relative atomic mass, which can be with
a permille error approximated by $A_i$, the
mass number of nucleus $i$, and $m_u= m(^{12}\mathrm{C})/12$ is the
atomic mass unit).
Alternatively, one can introduce an abundance, without the inclusion of the weight or mass of a nucleus,
  as the fraction of the number density of nucleus $i$ in comparison to the total number density of nucleons, approximated by $n = \rho/ m_u$, that
is conserved by nuclear reactions
\begin{equation} \label{eq:abundance_def}
   Y_i=\frac{n_i}{n} = \frac{n_i}{\rho/m_u}= \frac{X_i} {A_i}=\frac{n_i}{\rho N_\mathrm{A}} N_\mathrm{A} m_u.
\end{equation}
This definition seems to differ from the traditionally utilized one in nucleosynthesis literature, introduced by
\cite{Fowler.Caughlan.Zimmerman:1967}
\begin{equation}\label{eq:fowler_def}
    Y_i = \frac{n_i}{\rho N_\mathrm{A}},
\end{equation} 
as it includes the Avogadro constant $N_\mathrm{A}$ rather than the nuclear mass unit $m_u$. Eqs.~(\ref{eq:fowler_def}) and (\ref{eq:abundance_def}) differ by the product $M_u=N_\mathrm{A} m_u$, the molar gas constant, which had until 2019 the value $10^{-3}$~kg/mole or 1~g/mole in cgs units, leading in Eq.~(\ref{eq:fowler_def}) to an abundances measure in mole/g and in Eq.~(\ref{eq:abundance_def}) to a dimensionless number.

When utilizing the present values of the natural constants \citep[see Table
XXXI in][]{TiesingaCodata:2021} with
$N_\mathrm{A}=6.02214076 \times 10^{23}$~mole$^{-1}$ (exact) and
$m_u=m(^{12}\text{C})/12=1.6605390660(50)\times10^{-24}$~g with a
relative uncertainty of $3\times 10^{-10}$, one obtains for the
molar mass constant
$M_u = N_\mathrm{A} m_u= 0.99999999965(30)$~g~mol$^{-1}$, i.e. equal to 1
with an uncertainty of $3\times 10^{-10}$. Thus, both expressions are
numerically identical with an extremely high accuracy in cgs units. However, the different
dimensions of Eqs.~(\ref{eq:abundance_def}) and (\ref{eq:fowler_def})
can introduce some confusion \citep[see also][]{Rauscher2020}. In this paper, we continue to utilize the
traditional definition for abundances (Eq.~(\ref{eq:fowler_def}), but
in agreement with Eq.~(\ref{eq:massfrac_def2}) and $Y_i=X_i/A_i$ we will treat mass
fractions $X_i$ as well as abundances $Y_i$ as dimensionless numbers.
When expressing the number densities $n_i$ in terms of abundances
$Y_i$, Eq.~(\ref{eq:lagr_deriv_nuc}) leads to the form
\begin{align}\label{eq:nuclear_reaction_network}
\dot{Y}_i = \frac{D Y_i}{D t} &= \sum _j N_j ^i \lambda _j Y_j && \text{(1-body)} \nonumber \\
&+\sum _{j,k} \frac{N^i _{j,k}}{1+\delta _{jk}}\rho N_{\mathrm{A}} \langle \sigma v \rangle _{j,k}Y_jY_k  && \text{(2-body)} \nonumber \\
&+\sum _{j,k,l}\frac{N^i_{j,k,l}}{1+\Delta_{jkl}}\rho ^2N_{\mathrm{A}}^2 \langle ijk \rangle Y_j Y_k Y_l. && \text{(3-body)},
\end{align}
where the individual terms can be identified with specific
reactions, neglecting
reactions involving four or more participants. The first term,
standing for one-body reactions, usually includes decays,
photodisintegrations, electron- or positron-captures, and
  neutrino absorption. The equation is often called the "nuclear reaction
network equation". It is the fundamental differential equation that is
solved within \textsc{WinNet}. Note that all $\rho N_{\mathrm{A}}$
terms would be replaced by $\rho/m_u$, when utilizing the alternative
definition of abundances $Y_i$, which would replace $N_\mathrm{A}$ by
$m_u^{-1}$. 

\subsection{Detailed balance}
\label{sct:det_balance}
Reverse or backward reactions have a direct relation to the
  forward reaction by the so-called detailed balance theorem. We
denote as forward reaction those with a positive Q-value defined as the
difference between initial and final ground-state masses. The relation
between both can be expressed as
\citep[e.g.,][]{Fowler.Caughlan.Zimmerman:1967}
\begin{align}
\label{eq:det_balance}
    \langle \sigma v \rangle_\mathrm{backward} &= \frac{\Delta_\mathrm{reactants}}{\Delta_\mathrm{products}}\left( \frac{\prod_{i=1} ^{N_\mathrm{reactants}}G_i(T)}{\prod_{j=1} ^{N_\mathrm{products}}G_j(T)}  \right) \\ \nonumber
    &\times \left( \frac{\prod_{i=1} ^{N_\mathrm{reactants}}g_i}{\prod_{j=1} ^{N_\mathrm{products}}g_j}  \right) \left( \frac{\prod_{i=1} ^{N_\mathrm{reactants}} A_i}{\prod_{j=1} ^{N_\mathrm{products}} A_j} \right)^{3n/2} \\ \nonumber
    &\times \left(\frac{m_u k_\mathrm{B} T}{2\pi
      \hbar^2}\right)^{3n/2} \exp\left[ -Q/(k_\mathrm{B}T)\right]
      \langle \sigma v \rangle_\mathrm{forward},
\end{align}
where $\Delta$ is the double counting factor for reactants/products as in Eq.~\eqref{eq:nuclear_reaction_network}, $G$ are the partition functions, $g$ is the spin factor defined as $g=2J+1$, where $J$ is the spin of the ground state, $A_i$ is the mass of nucleus $i$, $Q$ is the Q-value of the reaction, $\langle \sigma v \rangle_\mathrm{forward}$ is the cross section of the forward reaction, and $n$ is the difference between the number of reactants and the number of reaction products. 

Equation~\eqref{eq:det_balance} needs to be modified for
photodisintegration reactions. There, $\langle \sigma\nu
\rangle_\mathrm{backward}$ should be replaced by $\lambda
_\mathrm{backward}$. In this case, $n \ne 0$ and we therefore get the
additional factors that are introduced with $n$ in the exponential in
Eq.~\eqref{eq:det_balance}. This is consistent with literature
\citep[e.g.,][]{Fowler.Caughlan.Zimmerman:1967} and the Reaclib
reverse rates. Therefore, in practice, we can use the above equation
for both cases, capture reactions and
photodisintegrations. Eq.~\eqref{eq:det_balance} is also valid
  for three-body reactions replacing $\langle \sigma v\rangle$ by $\langle ijk\rangle$. It should be mentioned here that the relations in this section include that nuclei in a thermal environment exist with thermally populated excited states.

Within the Jina Reaclib framework, the Q-Values are given for each reaction. Additionally, the mass excesses of all nuclei can be found in a separate file (called "winvn"). Ideally, the mass excess is consistent with the Q-value in the Reaclib; however, as pointed out already in \citet[][]{lippuner17a}, currently there are inconsistencies between these values. Because the reverse rates in Reaclib use the detailed balance principle with the Q-value from the Reaclib, there can be an inconsistency at the transition of NSE to the network equations caused by the inconsistent Q-values (see Fig.~\ref{fig:nse_screen}). Therefore, \textsc{WinNet} is able to calculate detailed balance with the Q-values obtained from the mass excess. We note that there is no optimal solution for this inconsistency. Using the Q-value from the mass excess will make the calculation consistent with NSE, but introduces an inconsistency with the forward rate, as this was calculated on the basis of a different Q-value. Often, it is however more important to be consistent with the equilibrium values. As already mentioned in \citet[][]{lippuner17a}, this inconsistency in the Reaclib database may be resolved in the future. However, one philosophy of the JINA Reaclib database is to have up-to-date nuclear masses. Recalculating all reaction rates whenever a new mass is available may not be feasible. To a certain degree, this inconsistency may therefore always be present \citep{Schatz2022privcom}. In any case, the advantage of an on-the-fly calculation of reverse rates is also given when using tabulated rates. For these rates, a tabulation for forward and reverse rates may break the detailed balance principle, and it can be more consistent to calculate the reverse rate based on the tabulation of the forward rate.

\subsection{Nuclear statistical equilibrium}\label{sssct:NSE}
For high temperatures in explosive environments, typically in excess
of about $T \gtrsim 6$~GK, reactions mediated by the strong and
electromagnetic interaction are in equilibrium. For these conditions,
one can simplify the treatment, replacing the reaction network
equations by utilizing an equilibrium approach, which
can be expressed in terms of the chemical potentials of the nuclei
\begin{equation}
\mu (Z,N) = N \mu _n + Z \mu _p,
\label{chempoteq}
\end{equation}
where $\mu (Z,N)$ is the chemical potential for a nucleus with mass number $A=Z+N$, $\mu _n$ is the chemical potential of neutrons, and $\mu _p$ is the chemical potential of protons. For low enough densities, nucleons (fermions) are nondegenerate and therefore described well by the Maxwell-Boltzmann statistics. Introducing this for the related chemical potentials in Eq.~(\ref{chempoteq}) leads to the so-called Saha equations (for a detailed derivation of NSE, see, e.g. \citealt{hix99,Iliadis2015,lippuner17a}, or for an approach using detailed balance, see, e.g., \citealt{clayton68}):
\begin{multline}
\label{eq:saha}
Y(Z,A) = g_{Z,A} G_{Z,A}(\rho N_A)^{A-1}\frac{A^{3/2}}{2^A}\left(\frac{2\pi \hbar}{m_u k_{\rm B}T}\right)^{\frac{3}{2}(A-1)}\\
e^{B_{Z,A}/k_B T}Y^{A-Z}_n Y^Z_p,
\end{multline}
with the spin factor $g = 2J+1$, where $J$ is the spin of the ground state, the partition function $G_{Z,A}$, and binding energy of a nucleus
$B_{Z,A}$. Furthermore, additional constraints of mass conservation
and charge neutrality hold:
\begin{align}
\sum \limits _i Y_i A_i &=1    && \text{(mass conservation)} \label{eq:massconservation}\\
\sum \limits _i Y_i Z_i &= Y_e && \text{(charge neutrality)} \label{eq:chargeconservation}.
\end{align}
This set of equations has two unknowns, namely the abundances of
protons $Y_p$ and neutrons $Y_n$, because temperature,
density, and electron fraction are assumed to be known quantities (e.g., from a hydrodynamical simulation). The
composition is a function of $Y(\rho ,T ,Y_e)$ only. In particular, no
information of the past behavior is necessary to determine the
composition. 

Within \textsc{WinNet} we solve the system of Equations~\eqref{eq:massconservation} and~\eqref{eq:chargeconservation} either with a Newton-Raphson or identical to \citet{Smith2023} the hybrid Powell method from the MINPACK-I package \citep{More1980} that was translated to Fortran 90 by J. Burkardt\footnote{Accessed from \url{https://people.sc.fsu.edu/~jburkardt/f_src/fsolve/fsolve.html}}. The convergence of the schemes hereby depend often on the initial guess.
In \textsc{WinNet} this guess is obtained by starting to calculate NSE
at a high temperature and descending to lower temperatures, taking the
results of the higher temperatures as initial value for the lower ones. The
initial composition at the starting temperature is assumed to consist of
nucleons only with $Y_n=1-Y_e$ and $Y_p=Y_e$.

Weak reactions are evolved with a simplified
  reaction network that includes only these reactions in
  Eq.~(\ref{eq:nuclear_reaction_network}). After a time step a new
electron fraction is determined using
Eq.~\eqref{eq:chargeconservation} and the composition is recomputed assuming for NSE. This assumes that strong and electromagnetic reactions occur instantaneously following a weak reaction consistently with the NSE assumption. 

The implementation of screening corrections in NSE is discussed in Section~\ref{sct:coulomb_corrections}.

\subsection{Nuclear heating}\label{sct:heating}
A proper consideration of the impact of the energy produced by nuclear processes in the hydrodynamical evolution requires the use of an in situ network as discussed in the introduction. However, in post-processing network calculations, it is commonly assumed that the nuclear energy generation mainly affects the evolution of temperature~\citep[see,
e.g.][]{freiburghaus99,Mueller1986,lippuner17a}. In the following, we describe the general description of energy generation and its treatment in \textsc{WinNet}.

The evolution of a fluid element under exchange of heat with the
surroundings in a local inertial frame comoving with the fluid is
given by the first law of thermodynamics

\begin{equation}
  \label{eq:1stlaw}
   d\varepsilon + p d \left(\frac{1}{n}\right)
  = \delta q\,,
\end{equation}
where $\varepsilon$ is the total energy (including rest-mass energy)
per nucleon, and $\delta q$ is the net heat gained per nucleon. This
includes heat produced by shocks or viscous heating or loss by
neutrinos when weak processes are considered. Alternatively, if the
fluid element is in equilibrium at all times, we have

\begin{align}
  \label{eq:1steq}
       &k_B T d s + \sum_i \mu_i  dY_i + \mu_e dY_e  = \delta q \\
   =\, &k_B T d s + \sum_i \left(\mu_i  +  Z_i \mu_e \right) dY_i\nonumber\,,
\end{align}
where $s$ is the entropy per nucleon in units of $k_B$ and the sum runs
over all nuclear species. The term $\mu_e dY_e$ or  $Z_i \mu_e dY_i$ accounts for the
contribution of electrons and positrons. Typically, the densities we
are interested in are such that matter is transparent to neutrinos. To ensure this, within \textsc{WinNet} we include a user-defined parameter to specify the density below which nuclear heating will be taken into account. The
energy carried by neutrinos per unit of time can be expressed as:

\begin{equation}
  \label{eq:neutrinos}
  \dot{q}_\mathrm{loss} = -\sum_i \langle \varepsilon_\nu\rangle^i \lambda_i Y_i\,,
\end{equation}
where $\langle \varepsilon_\nu\rangle^i$ is the average energy of the neutrinos produced by electron-capture or beta-decay of the nucleus $i$ with rate $\lambda_i$ and abundance $Y_i$. These quantities are provided in tabulations of weak interaction rates at finite temperature and density~\citep[see, e.g.,][]{Langanke2001} and in global calculations of beta-decays for $r$-process nuclei~\citep{Marketin.Huther.ea:2016}. For measured decays, the average neutrino energies are given by the ENSDF database \citep{Brown2018}\footnote{Accessed via the API of \url{https://www-nds.iaea.org/relnsd/vcharthtml/api_v0_guide.html}}. If we consider only beta-decays, we can express the average energy $\langle \varepsilon_\nu\rangle^i$ of the neutrinos as a fraction of the beta-decay Q-value $Q_{\beta,i}$:
\begin{equation}
  \label{eq:nuqbeta}
  \dot{q}_\mathrm{loss} = - \sum_i f_{\nu,i} Q_{\beta,i} \lambda_i Y_i\,.
\end{equation}
Assuming that a constant fraction of the energy is carried by
neutrinos, we have
\begin{equation}
\label{eq:avnuqbeta}
  \dot{q}_\mathrm{loss} = - f_{\nu} \sum_i Q_{\beta,i} \lambda_i Y_i\,.
\end{equation}
A typical value of $f_{\nu}$ for neutron-rich $r$-process nuclei
is \mbox{$f_{\nu}=0.4$}~\citep{Marketin.Huther.ea:2016}, as beta-decays
populate mainly excited states in the daughter nuclei that later
decay by either $\gamma$ or neutron emission. In practice, within \textsc{WinNet} the average energy of neutrinos produced in the reaction can be taken from all aforementioned sources, and in case of an unknown average neutrino energy, a user-defined $f _\nu$ is assumed. Optionally, we also account for escaping thermally produced neutrinos, by, e.g., bremsstrahlung or electron recombination with the analytic fitting formulas of \citet{Itoh1996}\footnote{See also \url{https://cococubed.com/code_pages/nuloss.shtml}}.
Energy cannot only leave the system by neutrinos, but also enter it. When assuming that only neutrino reactions add additional energy to the system, we obtain: 
\begin{equation}
\label{eq:nuadd}
    \dot{q}_\mathrm{gain} = \sum _i \langle \varepsilon_\nu \rangle^i \langle \sigma \rangle_i F_\nu Y_i,
\end{equation}
where $\langle \varepsilon_\nu \rangle$ is the average energy of the absorbed
neutrino, $\langle \sigma \rangle$ is the neutrino average cross
section, and $F_\nu$ the neutrino number flux (see
Section~\ref{ssct:nureac} for more details about the implementation of
neutrino reactions). For the moment, we include $\varepsilon_\nu$ for charged-current reactions on nucleons only. When combining Eq.~\eqref{eq:1steq},
\eqref{eq:avnuqbeta}, and \eqref{eq:nuadd}, we obtain:
\begin{equation}
  \label{eq:sevol}
  \begin{aligned}
      \dot{s} & = -\frac{1}{k_B T} \left(\sum_i \left(\mu_i  + Z_i \mu_e \right) \dot{Y_i} - \dot{q}
      \right) \\
       & = -\frac{1}{k_B T} \left(\sum_i \left(\mu_i  + Z_i \mu_e \right) \dot{Y_i} - (\dot{q}_\mathrm{loss} + \dot{q}_\mathrm{gain})
      \right),
  \end{aligned}
\end{equation}
where we obtain the electron chemical potential from the EOS \citep{Timmes1999b}, and the chemical potentials of nuclei is given by:
\begin{equation}
  \label{eq:chemical}
  \begin{aligned}
    \mu_i & = m_i c^2 + k_B T \eta_i\\
    \eta_i & = -\ln \left[ \frac{g_i G_i(T) m_u}{\rho Y_i}
      \left(\frac{A_i m_u c^2 k_B T}{2\pi\hbar^2}\right)^{3/2}  \right].
  \end{aligned}
\end{equation}
Here $m_i$ is the nuclear mass that we get from the atomic mass excess $\Delta(Z,A)$ by
\begin{equation}
    m(Z,A)c^2 = \Delta(Z,A) + A m_u c^2 - Z  m_e c^2.
\end{equation}
The mass excess from the latest atomic mass evaluation is tabulated in the Jina Reaclib database \citep{cyburt10}.

Under NSE conditions, Eq.~\eqref{eq:sevol} can be expressed as
\begin{equation}
  \label{eq:snseevol}
  \dot{s} = -\frac{1}{k_B T} \left[(\mu_p+\mu_e-\mu_n) \dot{Y}_e -
    \dot{q} \right]\,,
\end{equation}
showing that only reactions that are not in equilibrium, i.e. weak
processes that change $Y_e$ as well as external heating, are responsible for the change in
entropy. This result can be also generalized to $r$-process conditions
for which \mbox{$(n,\gamma)\rightleftarrows(\gamma,n)$} equilibrium is
valid. Hence, reactions in equilibrium do not introduce a
change in entropy. At high densities, neutrinos are characterized by a
chemical potential $\mu_\nu$. In this case, one obtains
$\dot{s} = -\left[(\mu_p+\mu_e-\mu_n-\mu_\nu) \dot{Y}_e - \dot{q}\right]/(k_B T)$, which shows chemical weak equilibrium $\mu_p+\mu_e=\mu_n+\mu_\nu$
corresponds to a maximum of the entropy~\citep{Arcones2010}.

Within \textsc{WinNet} we solve Eq.~\eqref{eq:sevol} explicitly in a
so-called operator splitting method within the same Newton-Raphson as
the nuclear network equations (Eq.~\ref{eq:nuclear_reaction_network}). The initial value of the entropy is determined using the
Timmes EOS~\citep{Timmes1999b}. Within every Newton-Raphson iteration, the newly obtained entropy is translated into a temperature via the EOS assuming that the density and composition remain constant. For
conditions at which $s\approx 1$--5 $k_\mathrm{B}\, \mathrm{nuc}^{-1}$, the entropy is dominated by the
contribution of nuclei and is very sensitive to the
composition. Under these conditions, it is necessary to account for
changes in the composition when searching for a new value of the
temperature. This is currently not implemented in \textsc{WinNet}.

\subsection{Coulomb corrections} \label{sct:coulomb_corrections}
Coulomb effects can significantly influence fusion processes in a hot stellar plasma. Electrons can be attracted by the positive charge of a nucleus and therefore shield and modify the Coulomb interactions between two nuclei. This modifies the nuclear reactions and makes charged particle reactions more likely. The effect can be approximated by correction factors, the so-called screening corrections, which are an important ingredient in nuclear reaction network calculations \citep[e.g.,][]{Salpeter1954}. The calculation of the correction factors depends on the temperature and density of the environment \citep[e.g.,][]{Salpeter1969,Yakovlev1989,Ichimaru1993,Yakovlev2006}. Usually, three different screening regimes are distinguished: the weak screening, the intermediate screening, and the strong screening regime. The regimes are commonly separated in terms of the ion-coupling parameter \citep[e.g.,][]{Kravchuk2014}
\begin{align}
\label{eq:screening_gamma}
    \Gamma_{12} &= 2 \frac{Z_1 Z_2 }{Z_1 ^{1/3}+Z_2 ^{1/3}}\frac{ e^2  (4\pi n_e)^{1/3}}{3^{1/3}k_\mathrm{B}T}  \nonumber \\
    &\approx 4.5494 \times 10^{-4} \frac{Z_1 Z_2 }{Z_1 ^{1/3}+Z_2 ^{1/3}} (\rho Y_\mathrm{e})^{1/3}T^{-1},
\end{align}
where $T$ is the temperature in GK, the electron number density is defined as $n_e=\rho N_\mathrm{A}Y_e$, Z$_i$ is the charge of element $i$, and the elementary charge is $e$. For lower values of $\Gamma_{12}$ the effect of screening becomes smaller. The weak screening regime applies for $\Gamma_{12} \ll 1$, the intermediate regime around $\Gamma_{12} \approx 1$, and the strong regime for larger values. We do not solve the screening corrections numerically, which would be necessary to obtain the corrections for the strong screening regime. Instead, we have implemented a fitted function that was derived within \citet{Kravchuk2014}. 
They express the so-called screening enhancement factor as \citep[Eq.~62 of][]{Kravchuk2014}
\begin{equation}\label{eq:scr_enhancement}
  f_\mathrm{scr} = \exp \left[\Gamma_{12}\left( b_0 + \frac{5}{8}b_2\zeta^{2} + \frac{63}{128}b_4\zeta^{4} \right) \right],
\end{equation}
with $\zeta$ defined as 
\begin{equation}
    \zeta = 3\frac{\Gamma_{12}}{\tau},
\end{equation}
where
\begin{align}
    \tau &= \left(\frac{27\pi^2 \mu(Z_1Z_2)^2 e^4}{2k_\mathrm{B}T\hbar^2}\right)^{1/3} \nonumber \\
         &\approx 4.2487 \times \left( \frac{A_1 A_2}{A_1+A_2}(Z_1Z_2)^2 \frac{1}{T} \right)^{1/3},
\end{align}
with the nucleon number $A_i$ and the reduced mass $\mu$. The fitting parameter $b_0$ is expressed by the difference in Coulomb free energies, which are defined by another fitted function that \citet{Kravchuk2014} took from \citet{Potekhin2000}:
\begin{align}
    f_C(\Gamma) &= C_1 \left[ \sqrt{\Gamma (C_2 + \Gamma)}-C_2 \ln \left( \sqrt{\frac{\Gamma}{C_2}}+\sqrt{1+\frac{\Gamma}{C_2}} \right)  \right]  \nonumber \\
    &+2C_3\left( \sqrt{\Gamma}- \arctan \sqrt{\Gamma}\right) \nonumber \\
    &+D_1 \left[ \Gamma - D_2 \ln \left( 1+ \frac{\Gamma}{D_2}\right)\right] \nonumber \\
    &+ \frac{D_3}{2} \ln \left( 1 + \frac{\Gamma^2}{D_4} \right).
\end{align}
Here, $C_1 = -0.907$, $C_2 = 0.62954$, $C_3 = 0.2771$, $D_1 = 0.00456$, $D_2 = 211.6$, $D_3 = -0.0001$, and $D_4 = 0.00462$, and $\Gamma$ is the ion-coupling parameter for a one-component plasma
\begin{equation}
    \Gamma = \frac{Z^{5/3}e^2 (4\pi n_e)^{1/3}}{3^{1/3} k_\mathrm{B}T}.
\end{equation}
From this, they obtain 
\begin{equation}
    b_0 = \frac{f_C( \Gamma_1 ) + f_C( \Gamma_2 ) - f_C( \Gamma_C )}{\Gamma_{12}},
\end{equation}
where $\Gamma_1$ and $\Gamma_2$ are the ion-coupling parameters of the reacting nuclei, and $\Gamma_C$ is the ion coupling parameter of the compound nucleus. Furthermore, $b_2$ and $b_4$ in eq.~\eqref{eq:scr_enhancement} are defined as
\begin{align}
b_2 &=  - \frac{1}{16} \frac{\left(1+z^{5/3}\right)^3}{1+z}\\
b_4 &=  - \frac{z}{64} \frac{\left(1+z^{5/3}\right)^5}{\left(1+z\right)^{11/3}}.
\end{align}

\begin{figure}
\begin{center}
\includegraphics[width=1.0\linewidth]{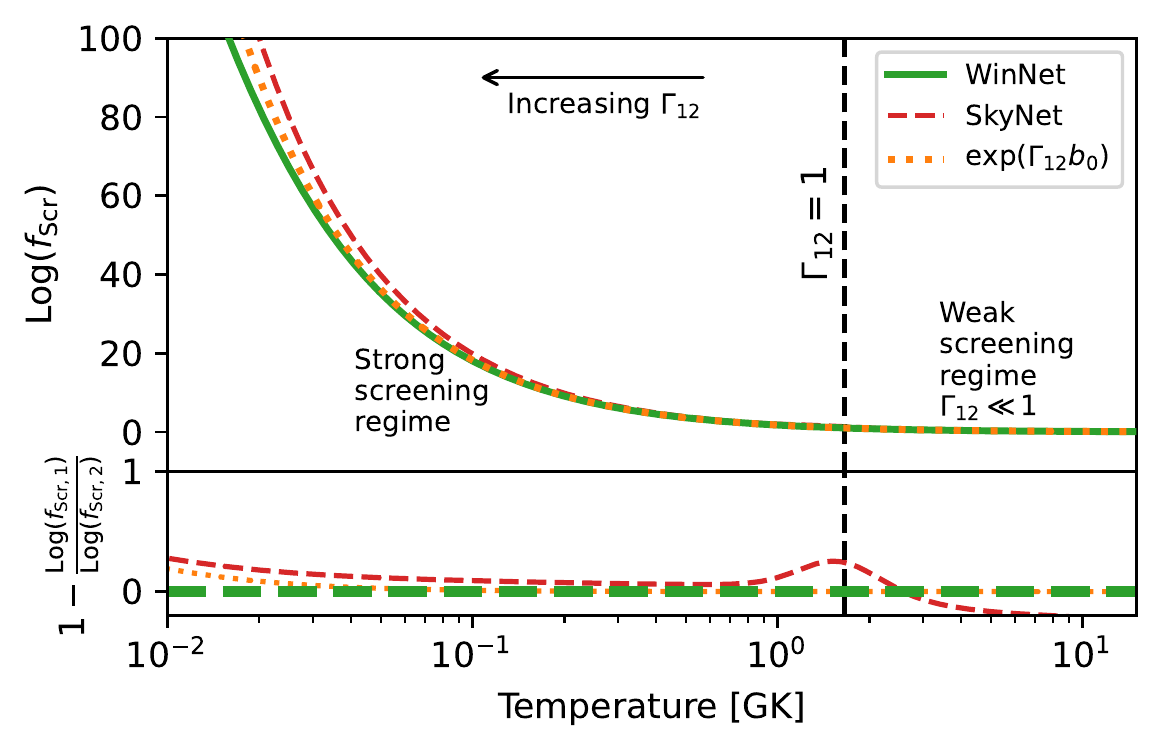}%
\end{center}
\caption{Upper panel: screening correction for the heavy ion reaction $^{12}$C+$^{12}$C for a constant density of $10^8\,\mathrm{g\, cm^{-3}}$ and $Y_\mathrm{e}=0.5$. The screening correction of \citet{Kravchuk2014} that is used in \textsc{WinNet} is shown as the solid green line. The screening correction of \citet{Kravchuk2014} when only using the $b_0$ term that is similar to the original description of \citet{Salpeter1954} is shown as the dotted orange line. The screening correction of \textsc{SkyNet} for a pure carbon composition is shown as the dashed red line. Bottom panel: relative differences of the screening corrections relative to that implemented in \textsc{WinNet}. The vertical dashed line indicates the intermediate screening regime for $\Gamma_{12}=1$.  \label{fig:c12_screen}}
\end{figure}

The differences between the screening correction scheme of \citet[]{Kravchuk2014} that is implemented in \textsc{WinNet} and that of \textsc{SkyNet} \citep{lippuner17a}, which uses a parameterization of \citet{Dewitt1973} are shown in Fig.~\ref{fig:c12_screen}. In the most relevant regime for nucleosynthesis calculations (i.e., $1\le \Gamma_{12} \le 200$) all schemes show good agreement (Fig.~\ref{fig:c12_screen}). For higher values of $\Gamma_{12}>200$, the temperature is usually close to or even below the validity of the reaction rate databases (c.f., $T_\mathrm{min}=10^{-2}\,\mathrm{GK}$ of the Reaclib reaction rate database; \citealt{cyburt10}). 

Screening corrections will modify the reaction rates according to
\begin{equation}
   \langle  \sigma\nu \rangle _\mathrm{scr} = \langle  \sigma\nu \rangle f_\mathrm{scr}.
\end{equation}
The implementation of screening with more than two reactants is realized in several steps. For three reactants, the screening correction of only two reactants is calculated, and, in a next step, the correction of the third reactant with the summed mass and ion number of the first two reactants is calculated. This corresponds to forming a short-lived intermediate nucleus. The total correction is then given by the multiplication of both correction factors $f_\mathrm{scr}$.

\begin{figure}
\begin{center}
\includegraphics[width=1.0\linewidth]{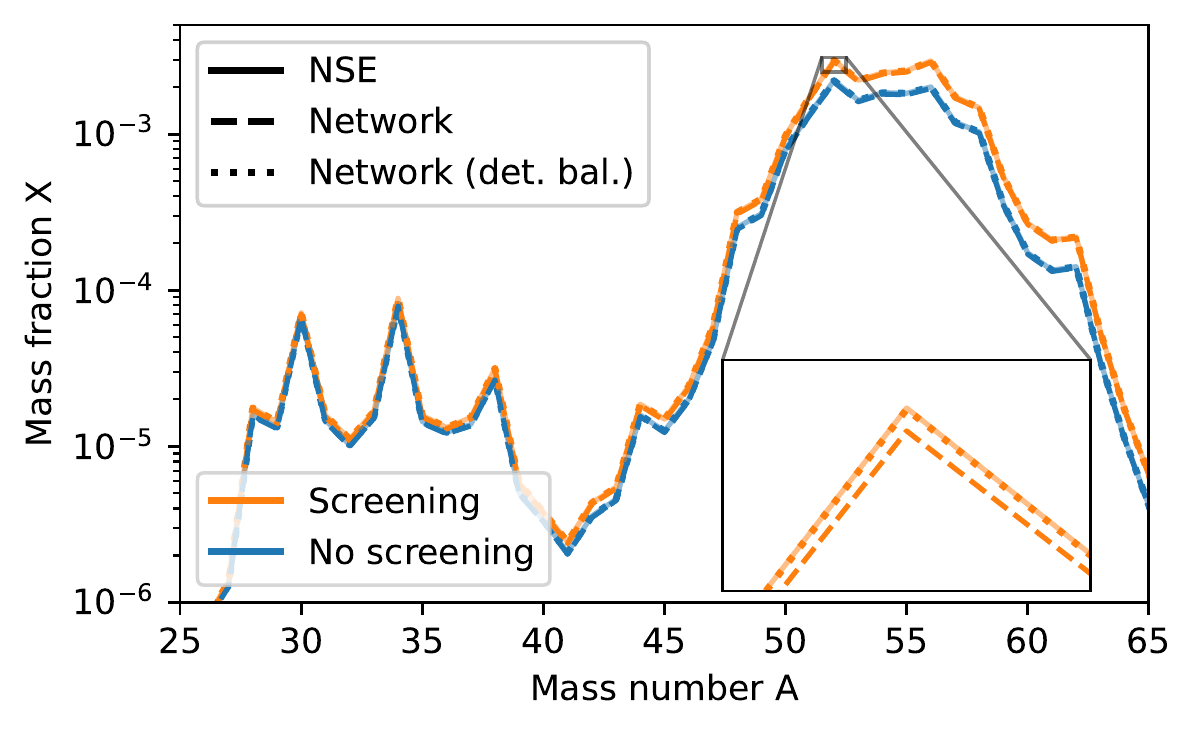}%
\end{center}
\caption{NSE composition with (orange line) and without (blue line) screening corrections in NSE for $T=7 \, \mathrm{GK}$, $\rho = 10^7 \, \mathrm{g\, cm^{-3}}$, and $Y_e = 0.5$ (solid lines). Dashed lines show the result of two hydrostatic network runs with and without screening corrections using strong reaction rates from the Jina Reaclib. Dashed lines show the same, but replacing the reverse reactions of the Jina Reaclib with reverse rates that are calculated with detailed balance using the mass excess of the Jina Reaclib. The hydrostatic calculations start with half neutrons and half protons and are calculated for $10^3\, \mathrm{s}$. This illustrates the consistency of the network at NSE transition with and without screening corrections when using the same nuclear masses for the reactions and NSE.  \label{fig:nse_screen}}
\end{figure}

In the case of NSE (see Section~\ref{sssct:NSE}), screening corrections enter in the form of a change of the binding energy of a charged nucleus, i.e., the difference in the Helmholtz free energy due to the screening. Since all reactions are in equilibrium, we can assume that every nucleus is built by a series of proton captures and neutron captures, where the latter reaction is independent of screening corrections. To obtain a correction for the binding energy of a given nucleus with charge number $Z$ we therefore multiply the screening corrections $f_\mathrm{scr}$ of the necessary amount of ($Z-1$) proton captures\footnote{Note that XNet uses the same approach in NSE; see \url{https://github.com/starkiller-astro/XNet/blob/master/doc/screening/Screening_for_NSE.pdf}}. The impact of screening and the consistency of the network at NSE transition is shown in Fig.~\ref{fig:nse_screen}. Note that there exist other approaches that derive the screening corrections from the detailed balance principle \citep[e.g.,][]{Kushnir2019} or from a global Coulomb correction \citep{Bravo1999,lippuner17a}. All of these approaches are consistent with each other. When taking screening corrections into account, heavier nuclei are synthesized compared to the case without screening. 

\section{Methods and numerical techniques}\label{sct:methods}
\subsection{Code structure and flow diagram}
\begin{figure*}
\begin{center}
\includegraphics[width=0.8\linewidth]{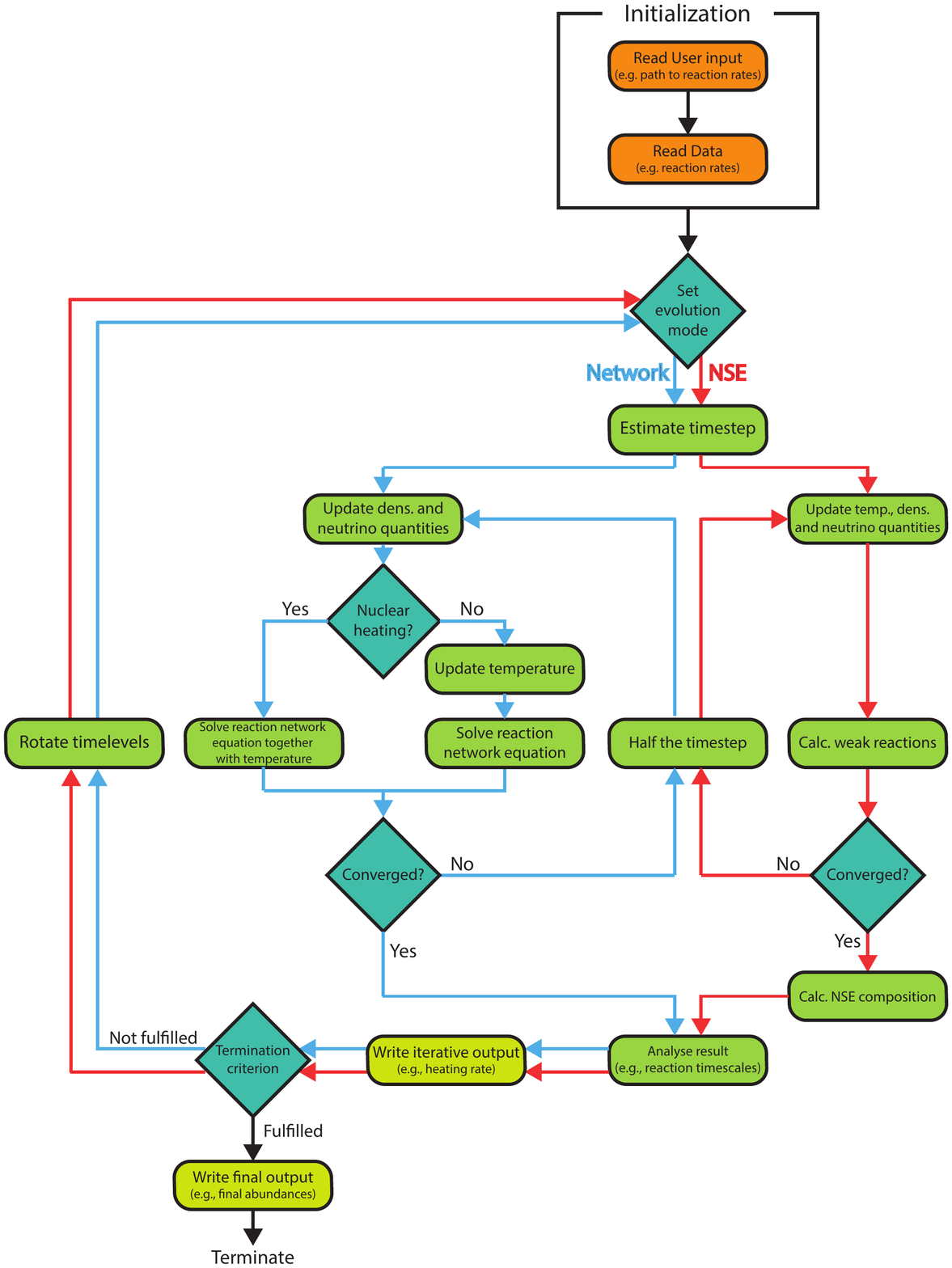}%
\end{center}
\caption{Flow diagram of \textsc{WinNet}. Figure taken from \citet[]{Reichert2021phd}. \label{fig:flowchart}}
\end{figure*}

In the following, we describe the control flow of \textsc{WinNet} (see Fig.~\ref{fig:flowchart}). The code starts by reading a user-defined file in the initialization step. This file contains runtime parameters such as paths to nuclear physics input data and other options. A full list of possible parameters is given in the documentation of the code.

After the initialization, the evolution mode is chosen. This mode is set to either ``Network'' or ``NSE'' and depends on the temperature. The implementation of several modes is necessary as the most efficient approach to determine the composition changes with temperature. Whereas solving the full network equations in a temperature regime where an equilibrium holds can lead to arbitrarily small time steps, solving NSE conditions at too low temperatures can lead to incorrect results. 

For both evolution modes, the temperature, density, and neutrino quantities (i.e., neutrino temperatures or energies and luminosities) are updated using either an interpolation (i.e., linear, cubic, Akima, modified Akima, Pchip) within the thermodynamic data of the Lagrangian tracer particle, analytic equations, or a user-defined extrapolation (i.e., adiabatic, exponential, free). In the network regime, updating the temperature depends on the input settings and includes some special cases. If the user allows feedback of the nuclear energy release on the temperature, a differential equation of the entropy is solved explicitly together with the nuclear reaction network equations (see Section~\ref{sct:heating}). After updating the temperature, density, and neutrino properties, the reaction network equations are solved numerically. For the network regime, the full set of coupled differential equations (including all reactions) is solved. In NSE, Eqs.~(\ref{eq:saha}-\ref{eq:chargeconservation}) are solved for a given temperature, density, and electron fraction. The latter is evolved taking weak reactions into account only.

If no convergence is achieved (the criteria are introduced in Section~\ref{sct:int_schemes}), the step size is halved, and the iteration is repeated. Otherwise, an output is generated, and the time is evolved (indicated by ``rotate timelevels'' in Fig.~\ref{fig:flowchart}). The main loop ends when a user-defined termination criterion is fulfilled. Before the code terminates, final output such as the final abundances and mass fractions are written.

\subsection{Integration schemes}\label{sct:int_schemes}
Due to the stiff behavior of the nuclear reaction network equations (Eq.~\eqref{eq:nuclear_reaction_network}), implicit/backwards methods are necessary to integrate the ODE. The general structure of the network however is independent of the chosen integration method.

Regardless of the chosen integration method, \textsc{WinNet} uses a sparse matrix representation of the Jacobian of the system and the sparse matrix solver PARDISO \citep{Schenk04}, which is OpenMP parallelized. For a detailed description of the sparse format, see e.g., \citet{hix99} or \citet{winteler12a}. This sparse format brings a computational advantage for calculations with more than $N \gtrsim 400$ nuclei. In \textsc{WinNet}, the indices of possible nonvanishing entries are calculated once at the beginning and are updated in a next step when solving the linear system. 
\textsc{WinNet} provides two methods to integrate the system that are outlined subsequently.

\subsubsection{Implicit Euler}
The implicit Euler method (see also, e.g. \citealt{hix99,winteler12a,lippuner17a}) is one of the simplest implicit integration methods. Nevertheless, it is sufficient for most of the calculations, especially when a large number of nuclei is involved in the calculation. For a coupled ODE, we can formulate the problem of integrating the equation by the general form of
\begin{equation}
\frac{\mathrm{D}Y_i}{\mathrm{D}t}=\dot{Y}_i=f_i(t,Y_1,...,Y_N),
\end{equation}
where N is the amount of involved species, and $Y_i$ is the abundance of species $i$. There are two possibilities to discretize this derivative. The simplest approach would be
\begin{equation}
\frac{Y_i(t+h)-Y_i(t)}{h}=f_i(t,Y_1,...,Y_N).
\end{equation}
When choosing a time step $h$, everything except $Y_i(t+h)$ is known, and one can integrate the ODE when knowing an initial value of $Y_i$. However, this approach corresponds to an explicit Euler method, an integration scheme that can be numerically unstable for so-called stiff problems that are present in reaction networks. Therefore, we can discretize the derivative with
\begin{equation}
\frac{Y_i(t+h)-Y_i(t)}{h}=f_i(t+h,Y_1,...,Y_N).
\end{equation}
Note that here $Y_i(t+h)$ as well as $f_i(t+h,Y_1,...,Y_N)$ is unknown. We can derive a iterative formula for the solution of $Y_i(t+h)$. This is given by:
\begin{equation}\label{eq:generalimpliciteuler}
Y_i(t+h)=Y_i(t)+h\, f_i(t+h,Y_1,...,Y_N).
\end{equation}
To get a solution, we have to apply a root-finding algorithm; in \textsc{WinNet} we use the Newton-Raphson method. To apply the Newton-Raphson method, we must reformulate the problem: 
\begin{equation}
    0=Y_i(t)+h\, f_i(t+h,Y_1,...,Y_N)- Y_i(t+h).
\end{equation}
Mathematically, a multidimensional Newton-Raphson can be formulated as:
\begin{equation}
\vec{F}(\vec{x}) = 0,
\end{equation}
which we will later apply and set $\vec{x}$ to $\vec{Y}$. The Taylor series of $\vec{F}$ can be expressed in first order as
\begin{equation}
\label{eq:jac_euler}
F_i(\vec{x}+\delta \vec{x})=F_i(\vec{x}) + \sum \limits ^{N} _{j=1} \frac{\partial F_i}{\partial x_j}\delta x_j + O(\delta \vec{x}^2) =0,
\end{equation}
where $\frac{\partial F_i}{\partial x_j}$ is one entry of the Jacobian matrix containing the partial derivatives of $\vec{F}$, defined as $J_{ij}=\frac{\partial F_i}{\partial x_j}$. Within \textsc{WinNet} this Jacobian is represented in a sparse format. The position of zero entries is reevaluated in every iteration. Furthermore, derivatives of the screening correction factor (Section~\ref{sct:coulomb_corrections}) are considered to be zero, and we calculate the Jacobian analytically as, in this case, it is just the derivative of a polynomial equation (Eq.~\ref{eq:nuclear_reaction_network}).
To find the root of $\vec{F}$ we iterate
\begin{equation}\label{eq:newtonraphson}
\vec{x}^{k+1} = \vec{x}^{k}+\delta \vec{x}^k = \vec{x}^k - J(\vec{x}^k)^{-1} \cdot \vec{F}(\vec{x}^k) 
\end{equation}
until convergence is reached. In a classical Newton-Raphson the convergence criterion is given by $|\vec{x}^{k+1}-\vec{x}^k|<\epsilon_\mathrm{NR}$. In \textsc{WinNet} we implemented a different criterion that is based on mass conservation by using the mass fraction $X_i$ (see Eq.~\ref{eq:massconservation}): 
\begin{equation}
\label{eq:euler_massconservation}
\sum \limits _{i=1} ^{N} Y_i A_i -1 = \sum \limits _{i=1} ^{N} X_i -1 < \epsilon_\mathrm{NR}
\end{equation}
where $\epsilon_\mathrm{NR}<10^{-5}$ is used per default in \textsc{WinNet}. As investigated by \cite{lippuner17a}, this convergence criterion is sufficient for most of the nucleosynthesis calculations. Other convergence criteria such as $|\vec{x}^{k+1}-\vec{x}^k|<\mathrm{\epsilon_\mathrm{NR}}$ are often too strict and slow down the calculation significantly. If the Newton-Raphson does not converge, the calculation is repeated with a halved time step. This is schematically shown by the loop in Fig.~\ref{fig:flowchart}. In the case of nuclear heating being enabled (Section~\ref{sct:heating}), the temperature change relative to the last Newton-Raphson iteration can also be limited in order to assure the convergence of the entropy update. We tested the convergence in more detail in Appendix~\ref{app:convergence_criteria}. By combining Eq.~\eqref{eq:newtonraphson} and Eq.~\eqref{eq:generalimpliciteuler}, we obtain
\begin{equation}
\vec{Y}^{k+1}_{n+1} =\vec{Y}^{k}_{n+1} - \left( \frac{1}{h}\times \bm{1} -\frac{\partial f(\vec{Y}^k_{n+1})}{\partial \vec{Y}^{k}_{n+1}} \right)^{-1} \cdot \left( \frac{\vec{Y}^{k}_{n+1}-\vec{Y}_n}{h} - \vec{f}(\vec{Y}^{k}_{n+1}) \right).
\end{equation}
Compared to other numerical integration methods, within the implicit Euler method no intrinsic error estimation is possible. There exist multistep algorithms that calculate an integration error by comparing the result of the integration after a full time step with the result after two half steps. This however increases the computational cost, and we therefore only estimated the error based on a maximum change of the abundances $\epsilon_\mathrm{Euler}$ which is based on the current derivative. We note that this procedure is similar to the time-step estimate of \textsc{SkyNet} \citep{lippuner17a}. The approximate change within one time step is calculated by:
\begin{align}
\abs{\dot{\vec{Y}}(t)} &= \abs{\frac{\vec{Y}(t+h')-\vec{Y}(t)}{h'}}\\
\epsilon_\mathrm{Euler} &= \max \abs{1-\frac{\vec{Y}(t+h')}{\vec{Y}(t)}}, \label{eq:impl_euler_eps}
\end{align}
therefore, we obtain
\begin{equation}
\abs{\dot{\vec{Y}}(t)} = \abs{\frac{(1-\epsilon_\mathrm{Euler}) \vec{Y}(t)- \vec{Y}(t)}{h'}} \Rightarrow h' = \epsilon_\mathrm{Euler} \max \abs{\frac{\vec{Y}(t)}{\dot{\vec{Y}}(t)}}.
\end{equation}
The default value in \textsc{WinNet} for $\epsilon_\mathrm{Euler}$ is set to a maximum change of 10\%. In order to avoid rapid changes of the time-step, it is limited by the previous step size
\begin{equation}\label{eq:stepsize_impleuler}
h' = \min{ \left(C\, h, \epsilon_\mathrm{Euler}  \max \abs{\frac{\vec{Y}(t)}{\dot{\vec{Y}}(t)}} \right)}
\end{equation}
with the constant $C>1$. Furthermore, only species with abundances higher than a threshold abundance are taken into account in the time step calculation (default in \textsc{WinNet} is $10^{-10}$). Additionally, the step size is restricted to a maximum change of the density within one time step (default value $5\%$) in order to get an adequate resolution for large density gradients. In case that nuclear heating is not enabled, the same restriction is applied to the temperature.

\subsubsection{Gear's Method}
In contrast to Euler's method, Gear's method (\citealt{Gear71}, see also, e.g. \citealt{Byrne75,longland14,martin17}) includes terms of higher orders \citep[see also][for a discussion of the advantages of higher-order solvers for nuclear reaction networks]{timmes99}. In the following, we will denote the highest included order with $q$. It is a so-called predictor-corrector method, where in a first step, a rough solution is guessed, and in a second step, this solution is corrected until a given precision is reached. The first prediction is based on information of the past behavior of the system. Therefore, the so-called Nordsieck vector
\begin{equation}\label{eq:nordsieck}
\vec{z}_n = \left(\vec{Y}_n,h\dot{\vec{Y}}_n,\frac{h^2 \ddot{\vec{Y}}_n}{2!},...,\frac{h^q \vec{Y}_n^{(q)}}{q!} \right)
\end{equation}
is stored, where $\vec{Y}_n$ are the abundances at the current time, $\dot{\vec{Y}}_n,\ddot{\vec{Y}}_n,...,\vec{Y}^{(q)}_n$ are the time derivatives of the abundances, and $h=t_{n+1}-t_n$ is the current step size. In order to obtain the predictor step $\vec{z}_{n+1}^{(0)}$, the Nordsieck vector is multiplied by a $(q+1)\cross (q+1)$ Pascal triangle matrix defined by
\begin{equation}
A^{ij}(q)=\begin{cases}0  & \text{if } i<j\\\begin{pmatrix}i\\j\end{pmatrix} = \frac{i!}{j!(i-j)!} & \text{if } i\ge j\end{cases} \qquad \text{with } i,j \in [0,1,...,q].
\end{equation}
Therefore, the predictor step is given by
\begin{equation}
\vec{z}_{n+1}^{(0)} = \vec{z}_n \cdot A,
\end{equation}
which is the Taylor series of $\vec{Y}_n$ truncated at the order of $q$ in matrix notation. To obtain an accurate solution for $\vec{Y}_{n+1}$ the predictor step is iteratively corrected due to 
\begin{equation}
\vec{z}_{n+1}=\vec{z}_{n+1}^{(0)}+\vec{e}_{n+1}\cdot\vec{\ell},
\end{equation}
with the correction vector $\vec{e}_{n+1}$. $\vec{\ell}$ is a $1\times (q+1)$-vector given by
\begin{equation} \label{eq:gear_ell}
\sum \limits ^{q} _{j=0} \ell _j x^j = \prod \limits _{i=1} ^{q} \left( 1+ \frac{(t-t_{n+1})/h}{(t_{n+1}-t_{n+1-i})/h } \right) =\prod \limits _{i=1} ^{q} \left( 1+ \frac{x}{\xi _i} \right).
\end{equation}
Here, we defined the vector $\vec{\xi}$ storing the information of previous step sizes. The components of \mbox{$\vec{\ell} = [\ell _0 (q),\ell _1 (q),...,\ell _j(q),...,\ell _q (q)]$} are calculated as
\begin{align}
\ell _0 (q) &=1, \nonumber \\
\quad \ell_1(q)&=\sum \limits _{i=1}^{q} \left( \xi _i ^{-1} \right), \nonumber\\
\quad \ell_j (q) &= \ell_j(q-1)+\ell_{j-1}(q-1)/\xi _q, \nonumber\\
\quad \ell _q (q) &= \left( \prod \limits_{i=1}^{q}\xi _i \right) ^{-1}. \nonumber
\end{align}
The correction vector $\vec{e}_n$ is calculated using the same Newton-Raphson scheme as for the solution $\vec{Y}_{n+1}$. To obtain the composition of the next step, 
\begin{align}
\left[ \bm{1}- \frac{h}{\ell _1}J \right] \vec{\Delta} ^{(m)} &= - \left( \vec{Y}_{n+1}^{(m)}-\vec{Y}_{n+1} ^{(0)} \right)+\frac{h}{\ell _1}\left( \dot{\vec{Y}}_{n+1}^{(m)}  -\dot{\vec{Y}}_{n+1}^{(0)} \right),\\
\vec{Y}_{n+1}^{(m+1)} &= \vec{Y}_{n+1}^{(m)}+\vec{\Delta}^{(m)}
\end{align}
is solved. Here, $\vec{Y}_{n+1} ^{(0)}$ and $\dot{\vec{Y}}_{n+1}^{(0)}$ are extracted from the first and second entry of $\vec{z}_{n+1}^{(0)}$. The index $m$ is the number of iterations, $\Delta ^{(m)}$ is an iterative correction, and $J$ is the Jacobian matrix
\begin{equation}
\label{eq:jac_gear}
J_{ij} = \frac{\partial \dot{Y}_{i,n+1}^{(m)}}{\partial {Y}_{j,n+1}^{(m)}}.
\end{equation}
Identically to the implicit Euler integration, the Jacobian is represented by a sparse matrix, for which zero entries are evaluated in every step. Calculating the Jacobian is one of the most expensive steps when solving the ODE. Therefore, some integration packages use so-called 'Jacobian caching' to tackle the problem of recalculating the Jacobian multiple times (e.g., VODE; \citealt{vode}). We investigated a similar technique, Broyden's method \citep{broyden65}, to approximate the Jacobian instead of recalculating it in every iteration. This, however, did not lead to a performance improvement due to rapid changes of the reaction rates and feedback from the nuclear reactions on the temperature. Therefore, more Newton-Raphson iterations were needed to obtain convergence leading to an overall performance loss. After the Newton-Raphson iteration has converged, the correction vector 
\begin{equation}
\vec{e}_{n+1} = \vec{Y}_{n+1}-\vec{Y}_{n+1}^{(0)}
\end{equation}
can be determined. To obtain a sophisticated guess of the time step within a given tolerance, the error can be estimated by the truncation error
\begin{equation}
E_{n+1}(q) = -\frac{1}{\ell _1}\left[ 1+ \prod _{i=2}^{q}\left(\frac{t_{n+1}-t_{n+1-i}}{t_{n}-t_{n+1-i}} \right) \right] ^{-1} \vec{e}_{n+1}.
\end{equation}
The next time step is computed within a certain allowed tolerance $\epsilon_\mathrm{Gear}$ by
\begin{equation}\label{eq:timestep_gear}
h' = h K \left( \frac{\epsilon_\mathrm{Gear}}{\max{\bar{E}_{n+1}(q)}} \right)^{1/q+1},
\end{equation}
where $K$ is a conservative factor usually chosen in the interval $K \in [0.1,0.4]$. As for the calculation of the step size in Eq.~\eqref{eq:stepsize_impleuler}, only abundances above a certain threshold should contribute to the calculation of the new time step. Therefore, the truncation error is rescaled in order to prevent an overweighting of the change of very small abundances, smaller than a threshold $Y_{\mathrm{limit}}$ (default in \textsc{WinNet}: $10^{-10})$,
\begin{equation}
\bar{E}_{i,n+1} = \begin{cases}E_{i,n+1}/Y_i  & \text{if } Y_i > Y_{\mathrm{limit}}\\ E_{i,n+1}/Y_{\mathrm{limit}}  & \text{if } Y_i \le Y_{\mathrm{limit}}\end{cases}.
\end{equation}
In addition to the automatic control of the step size, the order $q$ can be selected automatically as well. For this, we allow only order changes of $q \pm 1$. The error estimates for increasing and decreasing order are calculated by:
\begin{align}
E_{n+1}(q-1) &= - \frac{\prod \limits _{i=1}^{q-1} \xi _i }{\ell _1 (q-1)}  \frac{h^q \vec{Y}_{n+1}^{(q)}}{q!} \\
E_{n+1}(q+1) &= \frac{-\xi_{q+1}(e_{n+1}-Q_{n+1}e_n) }{(q+2)\ell _1 (q+1)\left[1+ \prod \limits_{i=2}^{q} \frac{t_{n+1} - t_{n+1-i}}{t_n-t_{n+1-i}} \right] },
\end{align}
where $Q$ and $C$ are defined as 
\begin{align}
Q_{n+1} &=\frac{C_{n+1}}{C_n}\left( \frac{h_{n+1}}{h_n}\right)^{q+1} \\
C_{n+1} &= \frac{\prod \limits _{i=1}^{q} \xi _{i}}{(q+1)!}\left[1+ \prod \limits _{i=2}^{q} \frac{t_{n+1} - t_{n+1-i}}{t_n -t_{n+1-i}} \right].
\end{align}
To obtain the most efficient way of calculating the solution of the ODE, the step size in Eq.~\eqref{eq:timestep_gear} is calculated for order $q-1$, $q$, and $q+1$, respectively. The order is chosen as the one providing the largest time step, \mbox{$h'=\max (h'(q-1),h'(q),h'(q+1) )$}. Since the Nordsieck vector depends on the step size (see Eq.~\ref{eq:nordsieck}), it must be rescaled whenever the step size is changed:
\begin{equation}
\vec{z}'_{n+1} = \mathrm{diag}(1,\eta,\eta^2,...,\eta^{q})\cdot \vec{z}_{n+1},
\end{equation}
where $\eta = h'/h$. Also, when the order decreases to $q-1$, the Nordsieck vector has to be rescaled. Therefore, we define a correction 
\begin{equation}
\vec{\Delta} ' _i = d_i \vec{z}_{q,n+1},
\end{equation}
where, similar to Eq.~\eqref{eq:gear_ell}, $\vec{d}$ is implicitly defined as
\begin{equation}
\sum \limits _{j=0} ^q d_j x^{j}=x^2 \prod _{i=1} ^{q-2}(x+\xi _i)
\end{equation}
and its components are given by:
\begin{align}
d_0(q) &=d_1 (q)= 0 \nonumber \\
d_2(q)&=\prod \limits _{i=1}^{q-2}\xi _i, \nonumber \\
d_j(q)&=\xi_{q-2} d_j(q-1) +d_{j-1}(q-1), \nonumber \\
d_{q-1}(q) &= \sum \limits _{i=1} ^{q-2}\xi _i, \nonumber \\ 
d_q(q) &= 1. \nonumber
\end{align}
Due to the implementation of higher orders, within Gear's method, one is able to apply larger step sizes compared to the implicit Euler scheme, reducing the amount of iterations drastically without losing accuracy. However, for most of the calculations, more Newton-Raphson iterations are necessary, resulting in similar or even higher computational costs. The difference between the implicit Euler and Gears method is discussed in more detail in Appendix~\ref{app:convergence_criteria}.

\section{Reaction network inputs}\label{sct:inputs}
\subsection{Lagrangian tracer particles}
The nuclear reaction network equations (Eq.~\ref{eq:nuclear_reaction_network}) contain a dependency on the temperature and density of the environment. To get an initial composition from NSE, additionally the electron fraction is necessary. These quantities have therefore to be recorded from a simulation of an astrophysical scenario. This is often done in terms of Lagrangian tracer particles within the hydrodynamic simulation. These particles (also called \emph{trajectories} or \emph{tracer}) are passively advected within the (M)HD simulation, tracing all relevant quantities such as the time, temperature, density, electron fraction, and neutrino properties. \textsc{WinNet} is a so-called single-zone code, i.e., tracer particles cannot interact among one another. This assumption is valid if the nuclear burning timescales are much faster than other timescales changing the abundances (e.g., diffusion). Therefore, for the majority of explosive environments we can use a single-zone reaction network; however, for some cases such as hydrostatic oxygen burning, it has to be taken with care \citep[][]{hix99}. There have been several studies on the uncertainties of a tracer particle method. The necessary amount of tracer particles to achieve convergence has been studied, e.g., in \citet{Seitenzahl2010} and \citet{Nishimura2015}. Also, the initial placement of the tracer particles can have an impact on the convergence of the result \citep[][]{Bovard2017a}. A detailed comparison between setting tracers in contrast to calculating the nucleosynthesis inside the hydrodynamical simulation has been presented in the context of CC-SNe by \citet{harris2017} and by \citet{Navo2023}. Additionally, \citet{Sieverding2023} studied the impact of obtaining tracers in a post-processing step after the calculation of a hydrodynamic model from simulation snapshots.

\subsubsection{Temperature and density regimes}
\begin{figure}
\begin{center}
\includegraphics[width=1.0\linewidth]{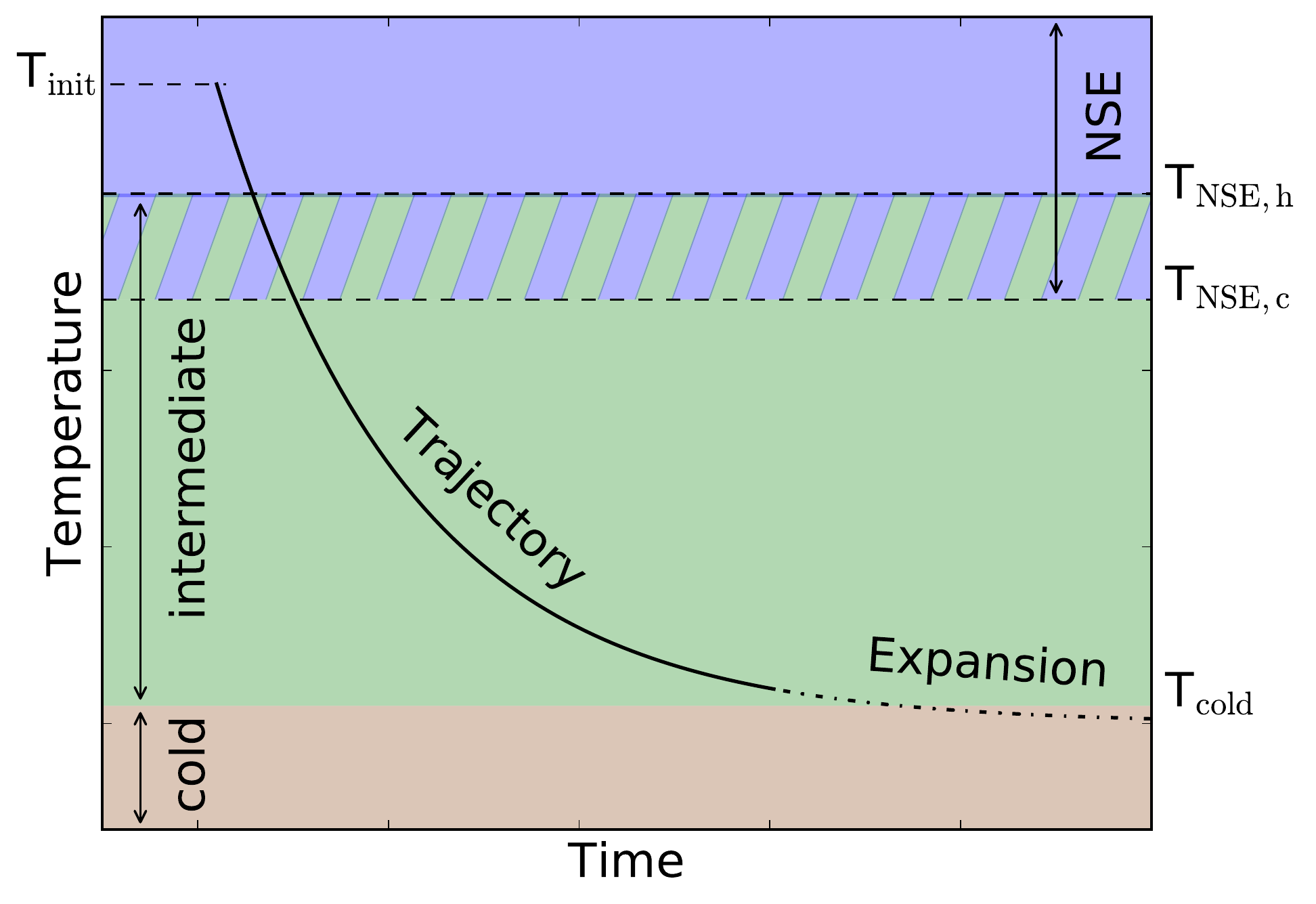}%
\end{center}
\caption{Sketch of different temperature regimes included in WinNet.}
\label{fig:temp_regimes}
\end{figure}

During its evolution, a tracer particle can undergo different temperature regimes, and therefore different approaches are required to obtain the composition within the given time step. In \textsc{WinNet}, there are distinctions between three temperature regimes, the regime of NSE, the intermediate temperature regime, and the cold temperature regime, schematically shown in Fig.~\ref{fig:temp_regimes}. 
In the regime of NSE, the network equations are only solved for weak reactions. Instead of calculating also strong reactions, an equilibrium is assumed (Section~\ref{sssct:NSE}). When the conditions are below a certain temperature threshold $T_\mathrm{NSE}$, the nuclear reaction network is solved for all nuclear reactions. The transition temperature between these regimes can be chosen individually, depending on whether the transition occurs from hot to intermediate temperatures ($T_\mathrm{NSE,c}$) or from intermediate to hot temperatures ($T_\mathrm{NSE,h}$, see Fig.~\ref{fig:temp_regimes}). The exact temperatures of the transitions depend on the environment \citep[e.g.,][]{Khokhlov1991}. The reason for having two transition temperatures is mainly motivated when using a feedback of the nuclear energy on the temperature (Section~\ref{sct:heating}). In this case, a slight inconsistency at the interface between the hot and intermediate regime (see Section~\ref{sct:det_balance}) may cause fluctuations in the temperature that can lead to an infinitesimal time step when using only one transition temperature.

When the temperature drops below $T = 10^{-2}\, \mathrm{GK}$, all reaction rates are frozen to the lower validity limit of the JINA Reaclib reactions \citep[brown region in Fig.~\ref{fig:temp_regimes},][]{cyburt10}. Often, the Lagrangian tracer particle finishes before the nucleosynthesis is completed and an extrapolation of the thermodynamic conditions is required (dotted line in Fig.~\ref{fig:temp_regimes}). The details of these assumptions can have an impact on the final yields \citep[see also][]{harris2017} and should be chosen according to the environment, e.g., a homologous expansion for CC-SNe or a free expansion for the dynamical ejecta of an Neutron star merger (NSM).

\subsection{Reaction rates}\label{sct:reaction_rates}
\begin{figure}
\begin{center}
\includegraphics[width=0.6\linewidth]{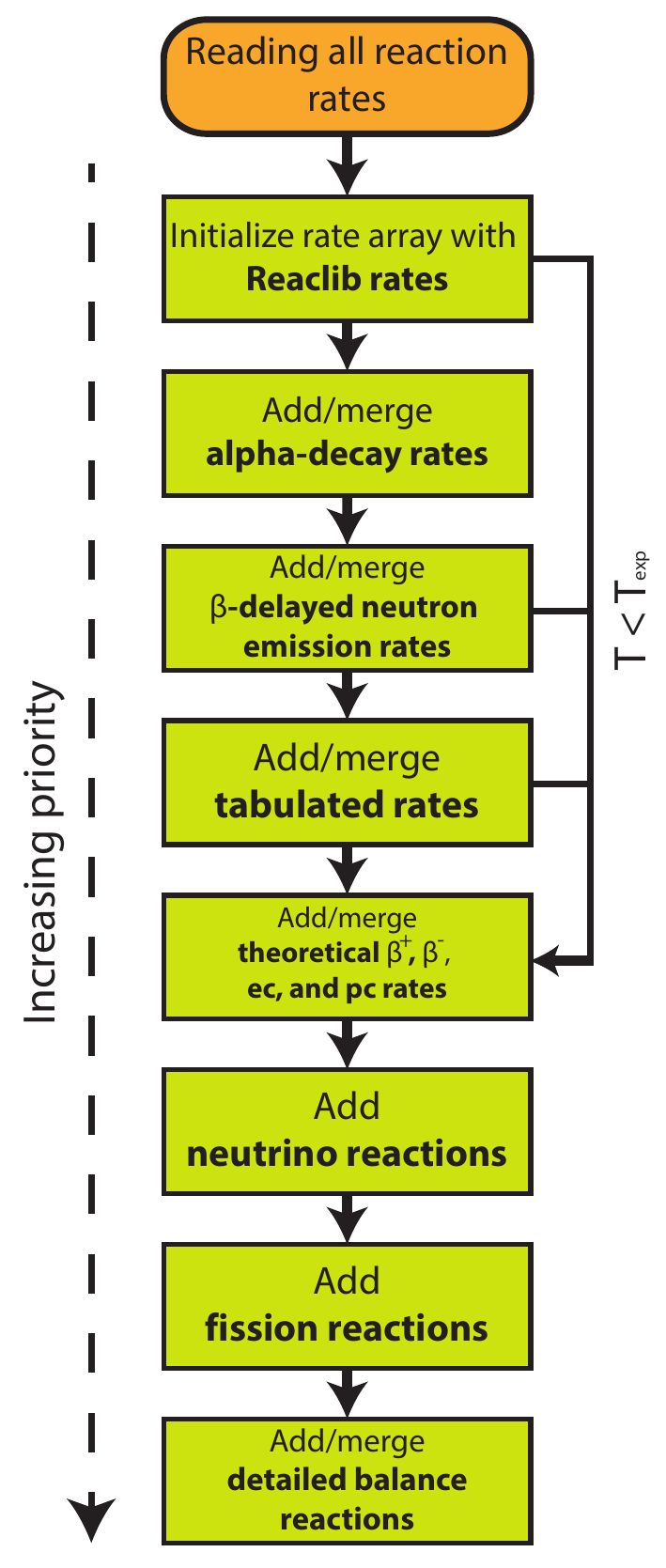}%
\end{center}
\caption{Sketch of the rate replacement procedure. Reaction rates in different formats have different priorities when creating a list with all reactions within \textsc{WinNet}. The priority of the rates increases from the top to the bottom of the plot. At a certain threshold temperature $T_\mathrm{exp}$, theoretical $\beta^+$, $\beta^-$, ec, and pc rates get replaced again as they are only valid above certain temperatures (see text). With the exception of Reaclib rates, all other rates are only optionally used.}
\label{fig:rate_replacement}
\end{figure}

\begin{figure}
\begin{center}
\includegraphics[width=1.0\linewidth]{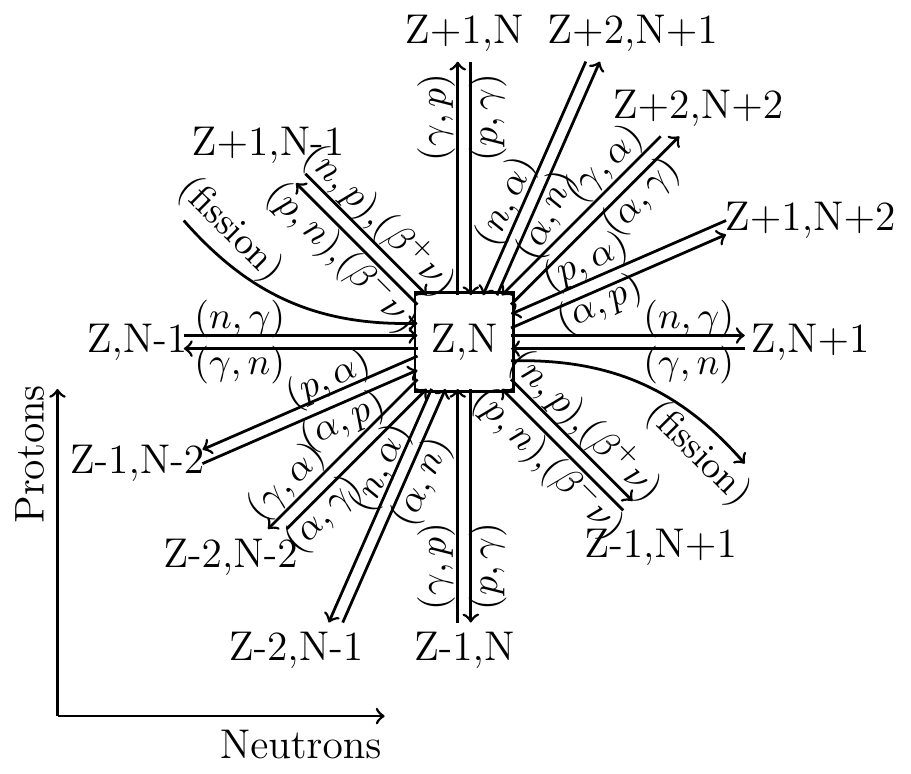}%
\end{center}
\caption{Sketch of the most important nuclear reactions \citep{Reichert2021phd}.}
\label{fig:sketch_reactions}
\end{figure}

Although all nuclei are connected to each other by nuclear reactions, in practice most of the reactions are negligible. The most important reactions for astrophysical environments are given by reactions that involve nucleons or $\alpha$-particles, decays, neutrino reactions, electron- and positron-captures, or fission reactions (Fig.~\ref{fig:sketch_reactions}). There exist many formats of the reaction rates. \textsc{WinNet} is built around the Reaclib reaction rate library, and this library usually contains the majority of reactions \citep[][]{cyburt10}. However, other formats are also supported, e.g., tabulated reaction rates from the TALYS code \citep[][]{Koning2019}. Rates given in different formats are either added or merged into the list of all rates within \textsc{WinNet}. In this case, the different formats have different priorities, starting with the Reaclib reactions with the lowest priority. If, in addition, this rate is also included in the theoretical $\beta^+$, $\beta^-$, ec, and pc rates, it is replaced once again. The priority of the individual rates is shown in Fig.~\ref{fig:rate_replacement}.

We note that \textsc{WinNet} does not perform any evaluation on the reliability of a rate. If a rate is contained multiple times in different formats\footnote{Not to confuse with a rate being contained multiple times in the same format, which can happen due to, e.g., resonances in the rate.}, it is the user's responsibility to choose the desired rate by either fully automatically using the one with the highest priority as in Fig.~\ref{fig:rate_replacement} or by deleting unwanted rates from high-priority formats. The modular structure of \textsc{WinNet} allows for an easy implementation of other popular reaction rate formats. In the following, we give a short overview of the current supported file formats.

\subsubsection{Reaclib file format}
\label{ssct:reaclib}
Most of the nuclear reaction rates are given in form of seven fit parameters, $a_i$, the so-called Reaclib\footnote{See \url{https://reaclib.jinaweb.org/index.php} and \cite{Rauscher.Thielemann:2000,Thielemann1980} for more details about the format and for recent reaction rates} format \citep{cyburt10}. The reaction rate is calculated according to:
\begin{equation}
    R = \mathrm{exp} \left[ a_0 + \sum \limits _{i=1} ^5 a_i T_9 ^{\frac{2i-5}{3}} + a_6 \, \mathrm{ln} T_9 \right].
\end{equation}
Depending on the reaction, $R$ can be either $\lambda$, $N_\mathrm{A}\langle \sigma \rangle_{i,j}$, or $N_\mathrm{A}^2\langle ijk \rangle$ (see Eq.~\ref{eq:nuclear_reaction_network}). Reverse reactions have additionally to be multiplied by the partition functions (see the pre-factor in Eq.~\ref{eq:det_balance}) that are also provided within the Reaclib database in a separate file (with the data originating from \citealt{Rauscher.Thielemann:2000} and \citealt{cyburt10} for more proton-rich nuclei). For higher temperatures ($T>10\,\mathrm{GK}$) partition functions from \citet{Rauscher2003} can be used. The fits of the reaction rates are valid between $10^{-2}\, \mathrm{GK} \le T \le 10^{2}\, \mathrm{GK}$. For lower temperatures, within \textsc{WinNet}, the rates are kept constant. At higher temperatures, usually NSE is assumed that only depends on the binding energies and partition functions. 

Each reaction belongs to a specific chapter in the Reaclib tables as given in Table~\ref{tab:reacl_chapters}. The Reaclib chapters correspond to different one-, two-, and three-body terms in Eq.~\eqref{eq:nuclear_reaction_network}, where each of these terms in the summation can include different numbers of reaction products. Another Reaclib format includes Chapter 8 and 9 together and does not include Chapter 10 and 11. \textsc{WinNet} supports and automatically detects both options. We note that the Reaclib reactions also contain two isomers of $^{26}$Al. \textsc{WinNet} can take these isomers into account when adding their properties into the winvn and in the list of considered nuclei. They will then be treated like all other nuclei.

\begin{table}
\centering
\begin{tabular}{l l l}
\hline
\hline
\textbf{Chapter} & \textbf{No. Reactants} & \textbf{No. Products} \\ 
\hline 
1 &  1 & 1\\ 
2 &  1 & 2\\ 
3 &  1 & 3\\ 
4 &  2 & 1\\ 
5 &  2 & 2\\ 
6 &  2 & 3\\ 
7 &  2 & 4\\ 
8 &  3 & 1\\ 
9 &  3 & 2\\ 
10 &  4 & 2\\ 
11 &  1 & 4\\ 
\hline        
\end{tabular}
\caption{Amount of reactants and products for different Reaclib chapters.}
\label{tab:reacl_chapters}
\end{table}

\subsubsection{Parametric $\alpha$-decays}
\label{sct:alpha_decays}
\begin{figure}
\begin{center}
\includegraphics[width=1.0\linewidth]{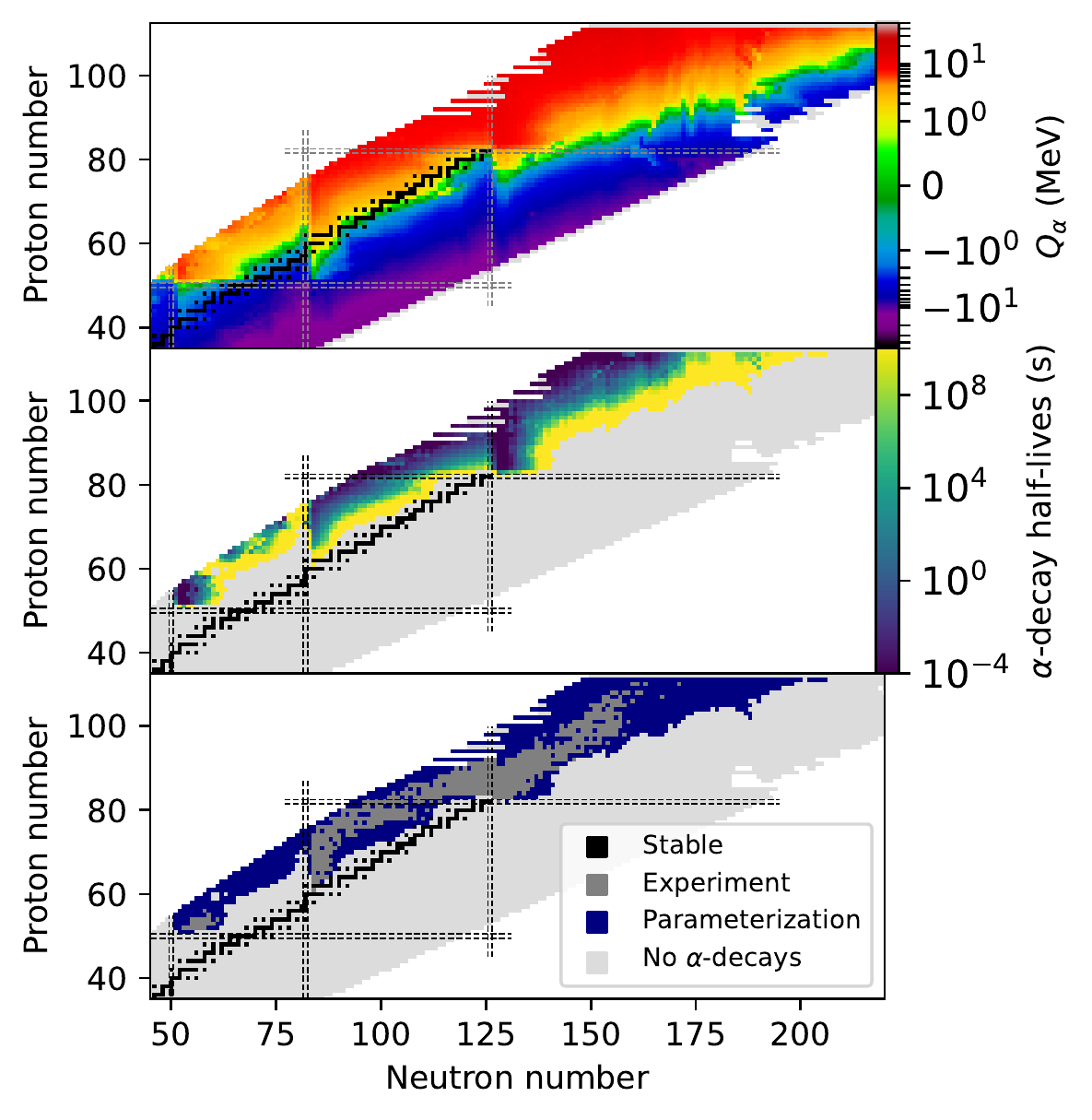}%
\end{center}
\caption{Upper panel: Q-value for $\alpha$-decay using the masses provided with the Jina Reaclib. Second panel: $\alpha$-decay half-lives in seconds. Whenever experimental $\alpha$-decay half-lives are available, we plot these instead of the parameterized ones. Nuclei that have half-lives of \mbox{$T_{1/2}\gtrsim10^{12} \, \mathrm{yr}$} are assumed to not $\alpha$-decay. Bottom panel: distinction between parameterized and experimentally known $\alpha$-decay half-lives included in the Jina Reaclib. Stable nuclei are shown as black squares, experimentally available $\alpha$-decays within the Jina Reaclib are indicated as dark-gray rectangles. Magic numbers of $50$, $82$, and $126$ are shown as dashed lines. All shown rates are publicly available along with \textsc{WinNet}.}
\label{fig:par_alpha_decay}
\end{figure}

The Reaclib reaction rate database contains only experimental $\alpha$-decays. To make the $\alpha$-decays more complete, \textsc{WinNet} is able to calculate additional $\alpha$-decays with the Viola-Seaborg formula \citep[e.g.,][]{viola1966,Sobiczewski1989,Brown1992,Sahu2016}. We provide rate tables of $\alpha$-decays using the parameterization of \citet{Dong2005}. For $Z>84$ and $N>126$, they fitted experimentally determined $\alpha$-decays with 
\begin{equation}
\label{eq:viola_seaborg}
    \log_{10}T_\alpha = (aZ + b)Q_\alpha^{-0.5} + (cZ + d)+h_{log},
\end{equation}
where $Z$ is the proton number of the decaying nucleus, $Q_\alpha$ is the Q-value of the decay, and the parameters $a=1.64062$, $b=-8.54399$, $c=-0.19430$, and $d=-33.9054$ were derived through least-squares fitting. Additionally, the so-called hindrance factor $h_\mathrm{log}$ was fitted:
\begin{equation}
    h_\mathrm{log} = 
    \begin{cases}
      0,      & \text{Z even, N even}\\
      0.8937, & \text{Z even, N odd}\\
      0.5720, & \text{Z odd, N even}\\
      0.9380, & \text{Z odd, N odd}
    \end{cases}.
\end{equation}
An obvious consequence of this parameterization is that $\alpha$-decays happen on shorter timescales if $Q_\alpha$ is large or, in other words, they are more relevant for regions with high $Q_\alpha$ (see, upper and middle panels of Fig.~\ref{fig:par_alpha_decay}). It has been pointed out that this fit is only valid for their fitting regions; other regions need a separate fit. To also obtain a valid fit to the other regions, we use the masses and experimental $\alpha$-decay half-lives of the Reaclib. We therefore use the above parameters only for nuclei with $Z\ge82$ and $N\ge126$, while we use the parameters of Table~\ref{tab:viola_pars} for the other regions.
\begin{figure}
\begin{center}
\includegraphics[width=1.0\linewidth]{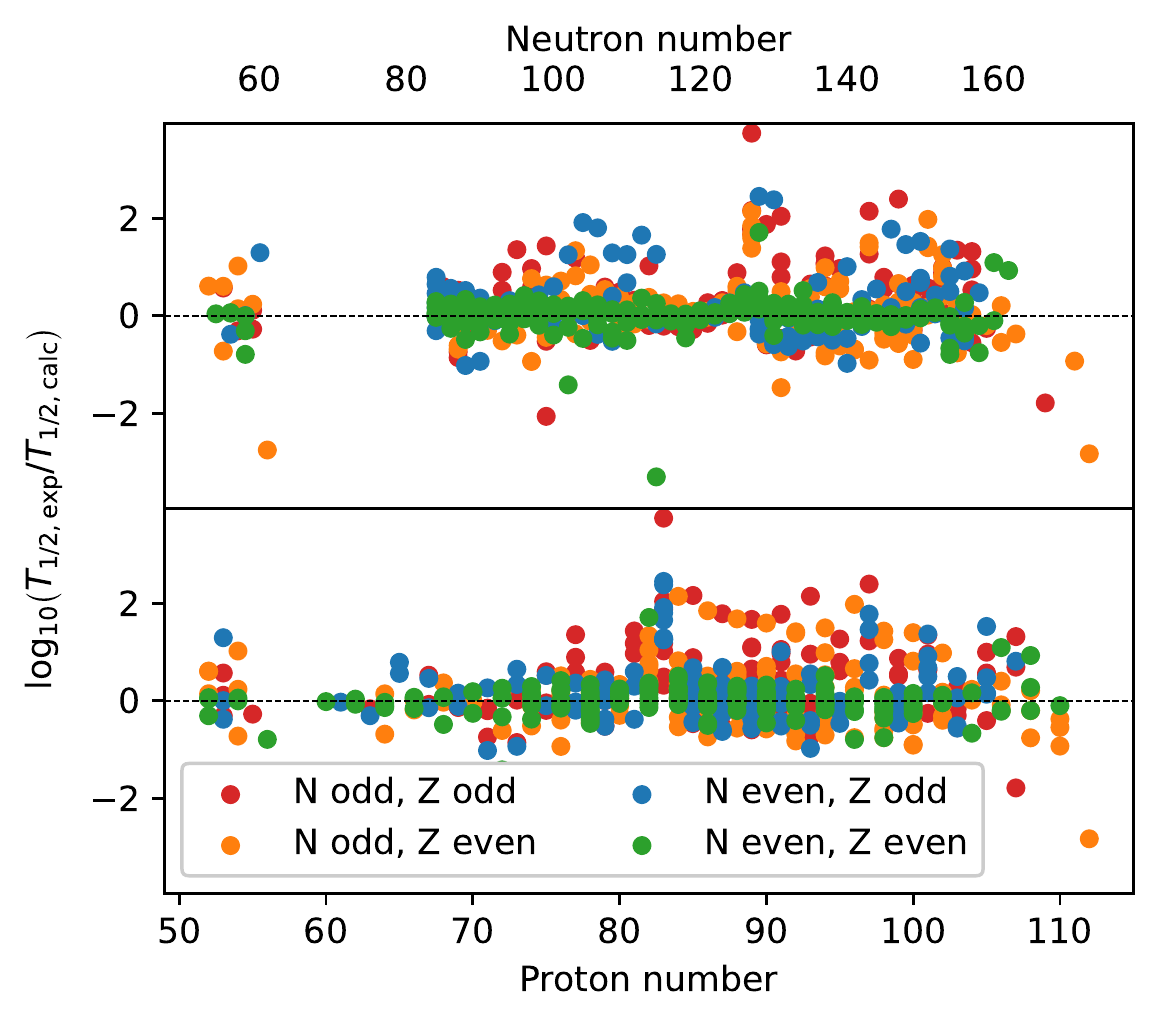}%
\end{center}
\caption{Ratio of calculated and experimental $\alpha$-decay half-lifes. The upper panel shows the ratio versus neutron number, the lower panel versus proton number. The different colors indicate the different types of nuclei as indicated in the legend.}
\label{fig:viola_seaborg}
\end{figure}

\begin{table}
\centering
\begin{tabular}{l l l l l}
\hline
\hline
 & a & b & c & d \\ 
\hline 
\begin{tabular}{@{}l@{}}$Z>82$ \\  $N>126$ \end{tabular} & 1.64062&  -8.54399&  -0.19430& -33.9054\\
\hline
\begin{tabular}{@{}l@{}}$Z>82$ \\  $82<N\le126$  \end{tabular} & 1.71183&  -7.50481&  -0.25315& -30.7028\\
\hline
\begin{tabular}{@{}l@{}}$50<Z\le 82$ \\  $82<N\le126$  \end{tabular} & 1.70875&  -7.52265&  -0.25153& -30.8245\\
\hline
\begin{tabular}{@{}l@{}}$50<Z\le 82$ \\  $50<N\le82$  \end{tabular} & 1.71371&  -7.34226&  -0.24978& -30.6826\\
\hline        
\hline
 & h1 & h2 & h3 & h4 \\ 
 \hline 
\begin{tabular}{@{}l@{}}$Z>82$ \\  $N>126$ \end{tabular}           & 0&  0.8937& 0.5720& 0.9380\\
\hline
\begin{tabular}{@{}l@{}}$Z>82$ \\  $82<N\le126$  \end{tabular}        & 0&  0.0476& 0.1214& 0.3933\\
\hline
\begin{tabular}{@{}l@{}}$50<Z\le 82$ \\  $82<N\le126$  \end{tabular}  & 0&  0.2140& 0.0600& 0.4999\\
\hline
\begin{tabular}{@{}l@{}}$50<Z\le 82$ \\  $50<N\le82$  \end{tabular}   & 0& -0.1242& 1.1799& 0.7166\\
\hline
\end{tabular}
\caption{Fitted parameters for Eq.~\eqref{eq:viola_seaborg}. For $Z>82$, $N>126$ we use the parameterization of \citet{Dong2005}. The lower part of the table shows the hindrance factors and h1 indicates Z even and N even, h2 Z even and N odd, h3 Z odd and N even, h4 Z odd and N odd.}
\label{tab:viola_pars}
\end{table}
This fit over these four individual regions of the nuclear chart that correspond to the regions between magic numbers is in a much better agreement to the experimental half-lives (Fig.~\ref{fig:viola_seaborg}). Still, some deviations of around $1$--$2$ magnitudes are present around the magic numbers. When comparing all available experimental $\alpha$-decays with the calculated ones, we obtain a standard deviation of $\sigma_\mathrm{Z even, N even}=0.38$, $\sigma_\mathrm{Z odd, N even}=1.61$, $\sigma_\mathrm{Z even, N odd}=0.93$, and $\sigma_\mathrm{Z odd, N odd}=0.82$. The large standard deviation of $\sigma_\mathrm{Z odd, N even}$ is driven by the decay of $^{153}\mathrm{Lu}$ whose half-life differs by more than $16$ mag ($3.9\times 10^{16}\, \mathrm{s}$ versus an experimental value of $\sim 1.3\,\mathrm{s}$). Note that $^{153}\mathrm{Lu}$ has a magic neutron number of $82$; nevertheless, the difference between the Viola-Seaborg formula and the experimental value is quite remarkable and indeed possibly a result of an outdated rate in the Jina Reaclib that uses the experimental data last evaluated in 2017. The latest experimental data from 2019 indicates that this nucleus is entirely decaying by an ec/$\beta^+$-decay\footnote{\url{https://www-nds.iaea.org/exfor/servlet/E4sGetIntSection?SectID=14658963&req=2130}} which would agree with the large half-life obtained with our fitted formula. When removing this nucleus from the calculation of the standard deviation, it reduces to $\sigma_\mathrm{Z odd, N even}=0.64$. We therefore have excluded it from our least-squares fit.

On a technical level, within \textsc{WinNet} one can decide if the $\alpha$-decay rates should only supplement the Reaclib rates or also replace them. The latter may become interesting in the future in case other theoretical $\alpha$-decays will be added to the Reaclib. In addition, one can adjust between which proton numbers $\alpha$-decays are added. Within \textsc{WinNet} we provide a file with the $\alpha$-decay rates using the parameterization presented here. For the fit as well as the rates, we used the masses of the Jina Reaclib as an input. 

\subsubsection{Tabulated rates}
\label{ssct:tab_rates}
Another possible format is given in form of a tabulation. This format is common for nuclear reaction codes such as TALYS \citep{Koning2019}. Every rate is tabulated on $30$ temperature grid points from $10^{-4}$ to $10$ GK and, identical to the Reaclib format,  assigned a certain chapter as given in Table~\ref{tab:reacl_chapters}. Reaction rates that are given in tabulated form will replace the respective reaction rates in Reaclib format. Reverse reactions can be given in tabulated form or calculated with the theory of detailed balance within \textsc{WinNet}. These calculations will replace all reverse rates that are given in the reaction rate library.

\subsubsection{Neutrino reactions}
\label{ssct:nureac}
Neutrino reactions are tabulated versus the neutrino temperatures from $2.8$ to $10$ MeV on seven grid points. These reaction rates enter the nuclear reaction network as an additional term in the form of
\begin{equation}
    \frac{\mathrm{D}Y(t)}{\mathrm{D}t} = \langle \sigma \rangle (t) \, F_\nu(t) \, Y(t),
\end{equation}
with the average neutrino cross section integrated over the normalized neutrino spectrum $ \langle \sigma \rangle (t)$ that depends on the neutrino temperature $T_\nu (t)$. Furthermore, \mbox{$F_\nu = L_\nu / \left(4\pi r^2 \langle E_\nu \rangle \right)$} is the neutrino number flux.

\textsc{WinNet} includes a tabulation where the neutrino reactions on nucleons have been calculated as described in, e.g., \citet{Burrows2006} with the weak magnetism and recoil corrections as in \citet{Horowitz:2002}. Within \textsc{WinNet} we provide the rate table as well as a python script to calculate it. In principle the full neutrino energy distribution could be taken from the hydrodynamic simulation and an appropriate neutrino temperature can be calculated based on this. In \textsc{WinNet}, the average neutrino energy $\langle E_\nu \rangle$ is used interchangeably with the neutrino temperature $T_\nu$ by assuming a Fermi-Dirac distribution of the neutrino energies and a zero chemical potential of the neutrinos. For this case,
\begin{equation}
    \langle E_\nu \rangle = \frac{\mathcal{F}_{3}(0)}{\mathcal{F}_{2}(0)}  T_\nu = \frac{7 \pi^4}{180 \, \zeta (3)} T_\nu \approx 3.1513  \, T_\nu
\end{equation}
holds. Here $\zeta$ is Riemann's zeta function, and $\mathcal{F}_{n}$ are the Fermi integrals defined as 
\begin{equation}
   \mathcal{F}_{n}(0) = \frac{1}{\Gamma(n)} \int _0 ^{\infty} \frac{x^n}{\exp \left(x \right)+1 }\mathrm{d}x,
\end{equation}
with the gamma function $\Gamma(n) = (n+1)!$. We note that current CC-SNe simulations hint toward slight deviations of the Fermi-Dirac distribution. Such a deviation can have an impact on the energy integrated neutrino cross sections that we do not take into account with the provided tabulation \citep[e.g.,][]{Tamborra2012,Mirizzi2016,Sieverding.ea:2019}.

\begin{figure}
\begin{center}
\includegraphics[width=1.0\linewidth]{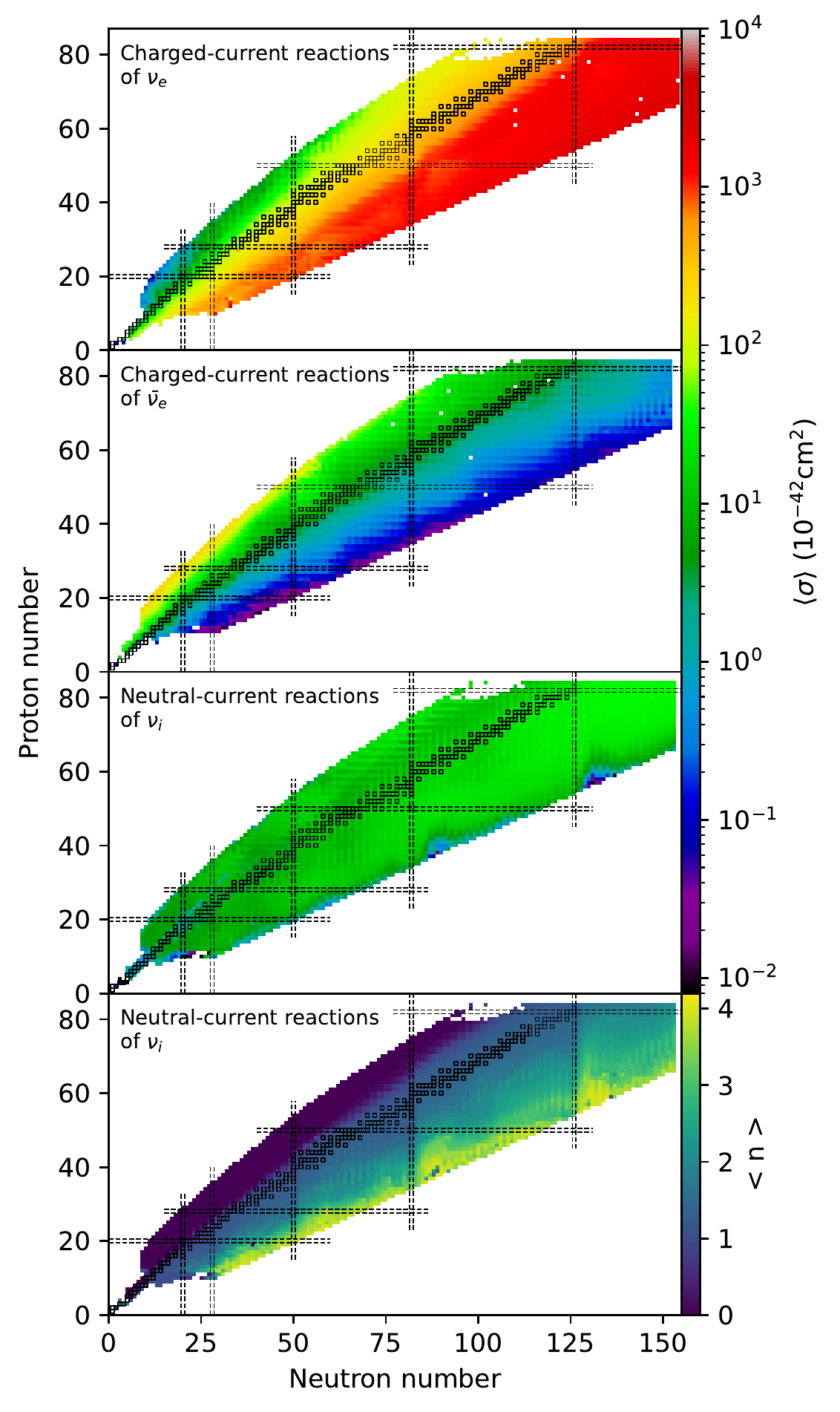}%
\end{center}
\caption{Energy averaged neutrino cross sections from the table of \citet{Sieverding2018} at $T_\nu = 5\,\mathrm{MeV}$. Shown are the summed cross sections of all reaction channels. The individual panels show charged-current reactions of electron neutrinos $\nu_e$, charged-current reactions of electron antineutrinos $\bar{\nu}_e$, neutral-current reactions of any neutrino flavor $\nu_i$, and the average amount of neutrons for neutral-current reactions of any neutrino flavor $\nu_i$. Note that the properties of neutral-current reactions of any antineutrino flavor $\bar{\nu}_i$ are nearly identical to the lower two panels.}
\label{fig:nu_sieverding}
\end{figure}

For neutrino reactions with heavier nuclei, \textsc{WinNet} is able to include neutrino interactions that are provided in a separate file. This file is taken from \citet{Sieverding2018} and includes charged-current as well as neutral-current reactions. All of these reactions contain different reaction channels allowing for the ejection of light particles such as neutrons, protons, and an alpha-particle. An overview of these cross sections is illustrated in Fig.~\ref{fig:nu_sieverding}, where we show the cross sections summed over all reaction channels and the average amount of ejected neutrons for neutral-current reactions at $T_\nu=5\,\mathrm{MeV}$.

Including neutrinos in the calculation requires additional information in form of either a tabulation or a parameterization of these neutrino properties. In the case of charged-current reactions, only the properties of electron neutrinos and antineutrinos have to be provided. Neutral-current reactions need additional properties of muon and tau (anti)neutrinos. Within \textsc{WinNet} it is assumed that the (anti)neutrino energies (or temperatures) for muon and tau neutrinos are identical ($E_{\nu_\mu} = E_{\nu_\tau}$), and they are thus included as species $x$, where $E_{\nu_x}$ has to be provided. Furthermore, the summed luminosities have to be provided ($L_{\nu_x} = L_{\nu_\mu} + L_{\nu_\tau}$) for neutrinos and antineutrinos. Treating muon and tau (anti)neutrinos effectively together as described above may be sufficient, as current CC-SNe simulations do not really distinguish between these neutrino flavors, and little has been done in this direction so far \citep[however, see][]{Bollig2017}. 

\subsubsection{Theoretical weak rates}
\label{sct:twr}
Theoretical models, e.g., shell-model calculations, are used to obtain
weak rates for stellar conditions. These rates are listed on a
temperature and electron density grid
\citep[e.g.,][]{Fuller1985,Oda1994,Langanke2001,Pruet2003,Suzuki2016}.
A direct tabulation of the rates, however, can lead to large
interpolation errors \citep{Fuller1985}. Therefore, the rates are not
tabulated directly, and instead, effective $\log ft_{\text{eff}}$ is
stored. This can be converted to the actual
rate via (see, e.g., \citealt{Langanke2001}):
\begin{equation}
   \lambda = \ln 2 \frac{I}{ft_{\text{eff}}}. 
\end{equation}
Here, $I$ is the phase space integral for ground-state to ground-state transitions
\begin{equation}
    I=\int_{\omega_0=\max(q,1)}^{\infty} \omega^2 (q+\omega )^2
    S(\omega) d\omega,
\end{equation}
with $q=(m_i-m_f)/m_e$ the Q-value in units of the electron mass.
\begin{equation}
    S(\omega) = \frac{1}{\exp \frac{\omega m_e c^2 - \mu_e}{k_\mathrm{B}T}+ 1},
\end{equation}
with the electron chemical potential $\mu_e$. 

We note that these theoretical reaction rates usually neglect atomic electron-capture, which becomes increasingly important for lower temperatures, e.g., for $^{56}$Ni. Therefore, \textsc{WinNet} contains the possibility of replacing all theoretical decays, electron- and positron-captures at low temperatures with the experimental decays provided in the Reaclib. 

\begin{figure}
\begin{center}
\includegraphics[width=1.0\linewidth]{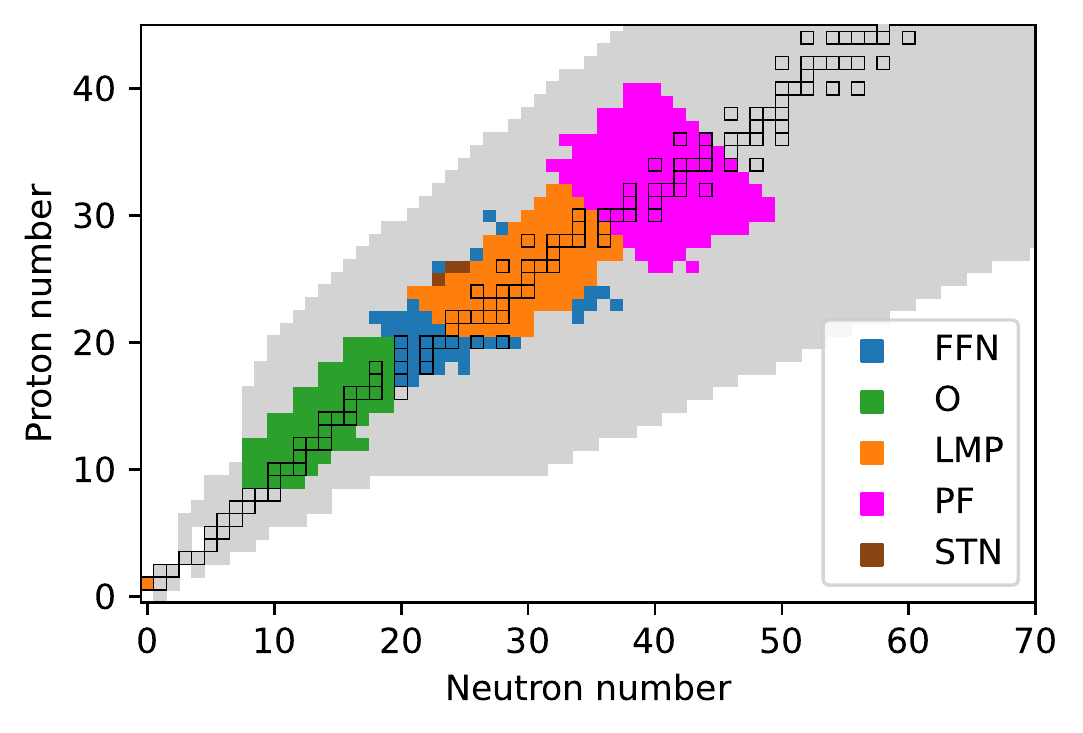}%
\end{center}
\caption{Compiled file of theoretical $\beta^-$-, $\beta^+$-decays, and positron- and electron-capture rates originating from different sources. The sources are FFN \citep{Fuller1985}, O \citep{Oda1994}, LMP \citep{Langanke2001}, PF \citep{Pruet2003}, and STN \citep{Suzuki2016}. Stable nuclei are indicated as black boxes.}
\label{fig:weak_rates}
\end{figure}

\textsc{WinNet} supports an individual grid for each reaction for the tabulation of theoretical $\beta^-$-, $\beta^+$-decays, positron- and electron-capture rates. This is necessary, as different available tabulations were calculated on different temperature and $\log \rho Y_e$ grids. We provide a table that was compiled out of various sources covering different regions of the nuclear chart (Fig.~\ref{fig:weak_rates})\footnote{See \url{https://groups.nscl.msu.edu/charge_exchange/weakrates.html}}. Note that \textsc{WinNet} also uses electron-capture rates on protons as well as positron-captures on neutrons from this table.

Since the reaction rates are tabulated with a dependence on the electron density, in principle the derivative of the reaction rate with respect to the abundances should be nonzero and there should be a term representing this in the Jacobian of the system (Eq.~\ref{eq:jac_euler} and Eq.~\ref{eq:jac_gear}). Within \textsc{WinNet} we ignore this dependence and assume a zero derivative of these reaction rates.

\subsubsection{$\beta$-delayed neutron emission}
The Reaclib file format only allows $\beta$-delayed neutron emissions up to three neutrons (Reaclib chapter 11; see Table~\ref{tab:reacl_chapters}). In practice, decays that emit only up to two neutrons are included. The probability of all other decay channels in the Reaclib format is added to the decay channel with three products. Especially when matter far from the valley of stability on the neutron-rich side is synthesized, $\beta$-delayed neutron emission of more than two neutrons can occur \citep[e.g.,][]{Marketin.Huther.ea:2016,Moeller2019}. Therefore, \textsc{WinNet} supports a file format containing the half-lives of the nuclei and the different channel probabilities up to the $\beta$-delayed emission of $10$ neutrons. Optionally, average emitted neutrino energies can be provided in this file (to account for the energy loss when self-heating is enabled; see Section~\ref{sct:heating}). Duplicates in Reaclib format will be replaced by the reaction rates in this format. Additionally, there exist user-defined parameters to allow for a controlled replacement of rates. With them, one can specify if, e.g., experimentally measured decays should also be replaced.

\subsubsection{Fission reactions and fragments}\label{sct:fission}
There are various fission modes, of which \textsc{WinNet} includes three: spontaneous fission, neutron-induced fission, and beta-delayed fission. In all of these cases, in addition to the probability to undergo fission, the resulting fission fragment distribution is of importance as well.
Investigations for fission barrier heights utilized in astrophysics have been performed from 1980 until today \citep{Howard.Moeller:1980,Myers.Swiatecki:1999,Mamdouh.Pearson.ea:2001,Goriely.Hilaire.ea:2009,Giuliani.Martinez.ea:2018a,Giuliani.Martinez.ea:2018b,Vassh.Vogt.ea:2019,Giuliani.Martinez.ea:2020}. 
Neutron-induced cross section predictions (or also beta-delayed fission) for
astrophysical applications were treated (by, e.g., \citealt{panov05,Martinez-Pinedo.Mocelj.ea:2007,panov10a,Erler.Langanke.ea:2012,Giuliani.Martinez.ea:2018a}. 
Extended compilations have been provided and can be found in several databases\footnote{https://nucastro.org, https://www.jinaweb.org/science-research/scientific-resources/data and https://www-nds.iaea.org, including TALYS results.}.

\begin{figure}
\begin{center}
\includegraphics[width=1.0\linewidth]{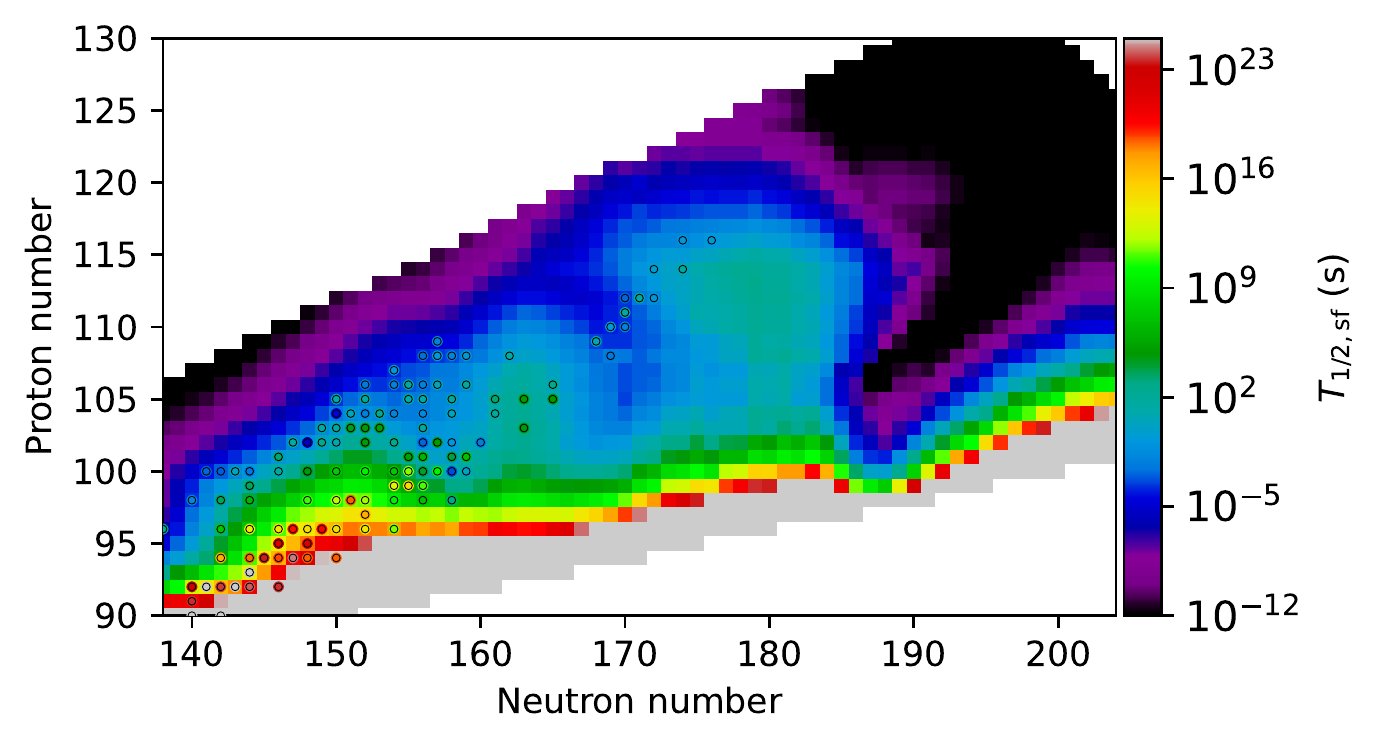}%
\end{center}
\caption{Half-lives of spontaneous fission in the nuclear chart \citep[c.f.,][]{Khuyagbaatar2020}. Colored dots indicate experimentally measured half-lives taken from the ENDFS database.}
\label{fig:sf_halflife}
\end{figure}

In the present paper, we provide only a limited set of fission inputs,
which are available within the \textsc{WinNet} package and
are stored in a separate file. 
The format is similar to the Reaclib file format, but only the name of the parent nucleus is stored. \textsc{WinNet} includes the rates of \citet[]{panov05} for $\beta$-delayed fission, and \citet{panov10a} for neutron-induced fission. Reaction rates for spontaneous fission have been calculated with the semi-empirical formula of \citet{Khuyagbaatar2020}, using the fission barriers provided in \citet{Moeller2015}. These half-lives together with experimentally measured ones are shown in Fig.~\ref{fig:sf_halflife}. While in \citet{Khuyagbaatar2020} spontaneous fission half-lives were fitted to nuclei with even neutron and proton numbers only, we use the same equation for all nuclei. 

The products (or fission fragments) are described by a fission fragment distribution in a probabilistic way. They can either be described by an analytic formula \citep[][]{Kodama1975,Panov2001} or more complicated models \citep[e.g.,][]{Kelic2009,Goriely.Hilaire.ea:2009,Mumpower2020}. Within \textsc{WinNet} we include the fragment distribution of \citet[][]{Kodama1975}, \citet{Panov2001}, and \citet{Mumpower2020}. As pointed out in \citet{Mumpower2020}, the distribution should only be used for $\beta$-delayed and neutron-induced fission. Therefore, \textsc{WinNet} contains these fragment distributions in combination with the ones of \citet[][]{Kodama1975} for spontaneous fission.

\section{Reaction network applications}\label{sct:network_applications}
\subsection{Example cases}
In the following, we discuss several example cases calculated with \textsc{WinNet} that are available together with the code. These examples involve conditions of a variety of scenarios, namely the Big Bang \citep[as described in][]{winteler12a}, the dynamic ejecta of an NSM \citep[from][]{korobkin12,Rosswog2013,piran13a,Bovard2017}, the neutrino-driven wind of an NSM \citep{Perego2014,martin15}, the viscous disc ejecta of an NSM \citep{wu16,lippuner17b}, and the dynamic ejecta of a black hole neutron star merger \citep{korobkin12,Rosswog2013,piran13a}. Additionally, we provide various conditions within MR-SNe \citep{winteler12b,Obergaulinger2017,Aloy2021,Obergaulinger2021,Reichert.Obergaulinger.ea:2021,Reichert2023}, classical novae \citep{Jose1998,jose2022}, the X-ray burst of an accreting neutron star \citep{Schatz.ea:2002}, complete Si burning within a CCSN \citep[with a simple parametric model as described in][]{Nadyozhin2002,Woosley2002} the neutrino-driven wind within a CCSN \citep{Bliss2018}, the detonation phase of a type Ia supernova \citep[with a parametric model as in][]{Meakin2009}, a main $s$-process \citep{Cescutti2018,cescutti_data_2022}, a weak $s$-process \citep{Hirschi2004,Nishimura2017b,pignatari_data_2022}, hydrostatic hydrogen burning, carbon-oxygen burning, and a simple $i$-process model \citep[as described in][]{Dardelet2015}. All of these conditions are examples in \textsc{WinNet} and should guide the user on how to use the code. It is noteworthy that the trajectories represent typical conditions in the scenarios and may be used for sensitivity studies, but they do not necessarily reflect the total yields that can be obtained when calculating often thousands of trajectories of the individual scenarios. Furthermore, a different nuclear physics input is used within the example cases, and we do not aim to exactly reproduce the abundances that have been obtained within the original publications.
All example cases are very diverse in their involved conditions and together they cover a large range of the nuclear chart. In the following sections we present only a subset of the aforementioned examples.

\subsubsection{Big Bang nucleosynthesis}
The synthesis of elements during the first minutes after the origin of our Universe can be calculated with a relatively small network. Following \cite{winteler12a}, we create a trajectory for a flat, isotropic, and homogeneous Universe to describe the conditions during the big bang \citep[see also][]{Vonlanthen2009}. Furthermore, we assume a freeze-out of weak reactions at $T=0.8\, \mathrm{MeV}$. An important quantity is the initial photon-to-baryon ratio, which was measured by the Planck Satellite (\mbox{$5.96 \times 10^{-10} \le \eta \le 6.22 \times 10 ^{-10}$}, \citealt{Planck2016}). 
\begin{figure}
\begin{center}
\includegraphics[width=1.0\linewidth]{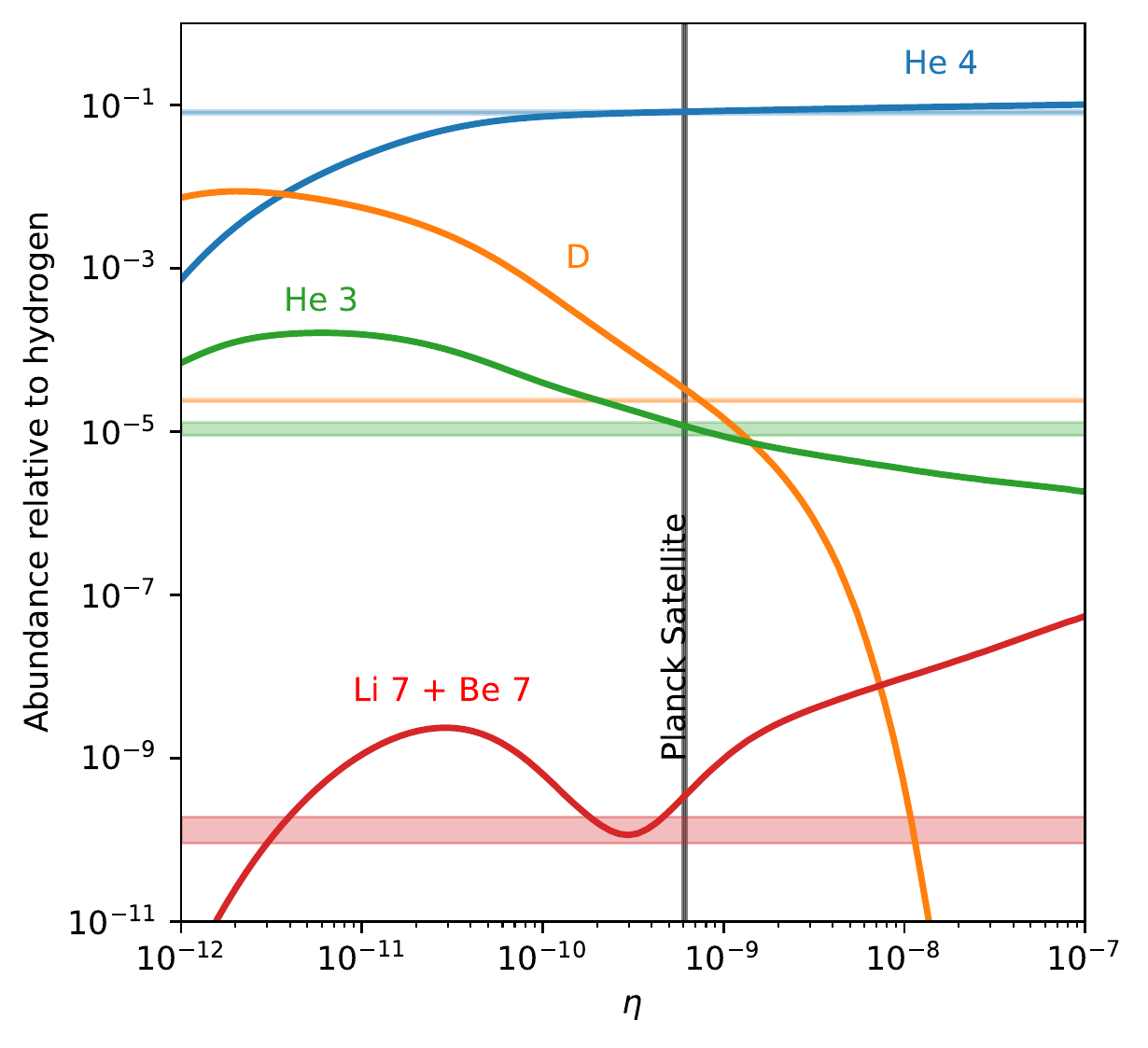}%
\end{center}
\caption{Final abundances relative to hydrogen as a function of the photon-to-baryon ratio $\eta$. Horizontal bands show measurements of the respective isotope.}
\label{fig:reactions_bigbang}
\end{figure}
By creating one trajectory for each baryon to photon ratio we are able to connect the big bang nucleosynthesis with measurements of abundances in stars and therefore probe the conditions of the big bang. For deuterium, the primordial abundance was determined to be $Y(\mathrm{D})/Y(\mathrm{H}) = (2.527\pm 0.03)\times 10^{-5}$ \citep[][orange band in Fig.~\ref{fig:reactions_bigbang}]{Cooke2018}. For deuterium there is a slight discrepancy with respect to the photon-to-baryon ratio determined by the Planck Satellite and observed deuterium abundances. As the deuterium abundance is very sensitive to the d(p,$\gamma$)$^3$He reaction rate, this discrepancy may vanish in the future with new experimentally determined reaction rates \citep{Mossa.ea:2020,Moscoso.ea:2021}. Here, we used the rate of \citet[]{Descouvemont2004} that is included in the JINA Reaclib. Furthermore, observations of $Y(\mathrm{D})/Y(\mathrm{H})$ are also differing \citep[e.g.,][]{Romano2003}. The observed value of $Y(^3\mathrm{He})/Y(\mathrm{H}) = (1.1\pm 0.2)\times 10^{-5}$ \citep[][]{Bania2002} is in perfect agreement with the estimated value. Additionally, the value of $Y(^4\mathrm{He}) = 1/4 \times (0.2561 \pm 0.0108) / Y(\mathrm{H})$ \citep[][]{Aver2010} is in agreement with our calculation (blue band in Fig.\ref{fig:reactions_bigbang}). The observed $^7$Li abundance ($Y(^7\mathrm{Li})/Y(\mathrm{H}) = 1.23\substack{+0.68 \\ -0.32}$, \citealt{Ryan2000}) is in clear discrepancy with the calculated value. This well-known problem is referred to in literature as the lithium problem \citep[see, e.g.,][for  reviews]{Fields2011,Fields.Olive:2022}. 

\subsubsection{Main $s$-process}
\begin{figure}
\begin{center}
\includegraphics[width=1.0\linewidth]{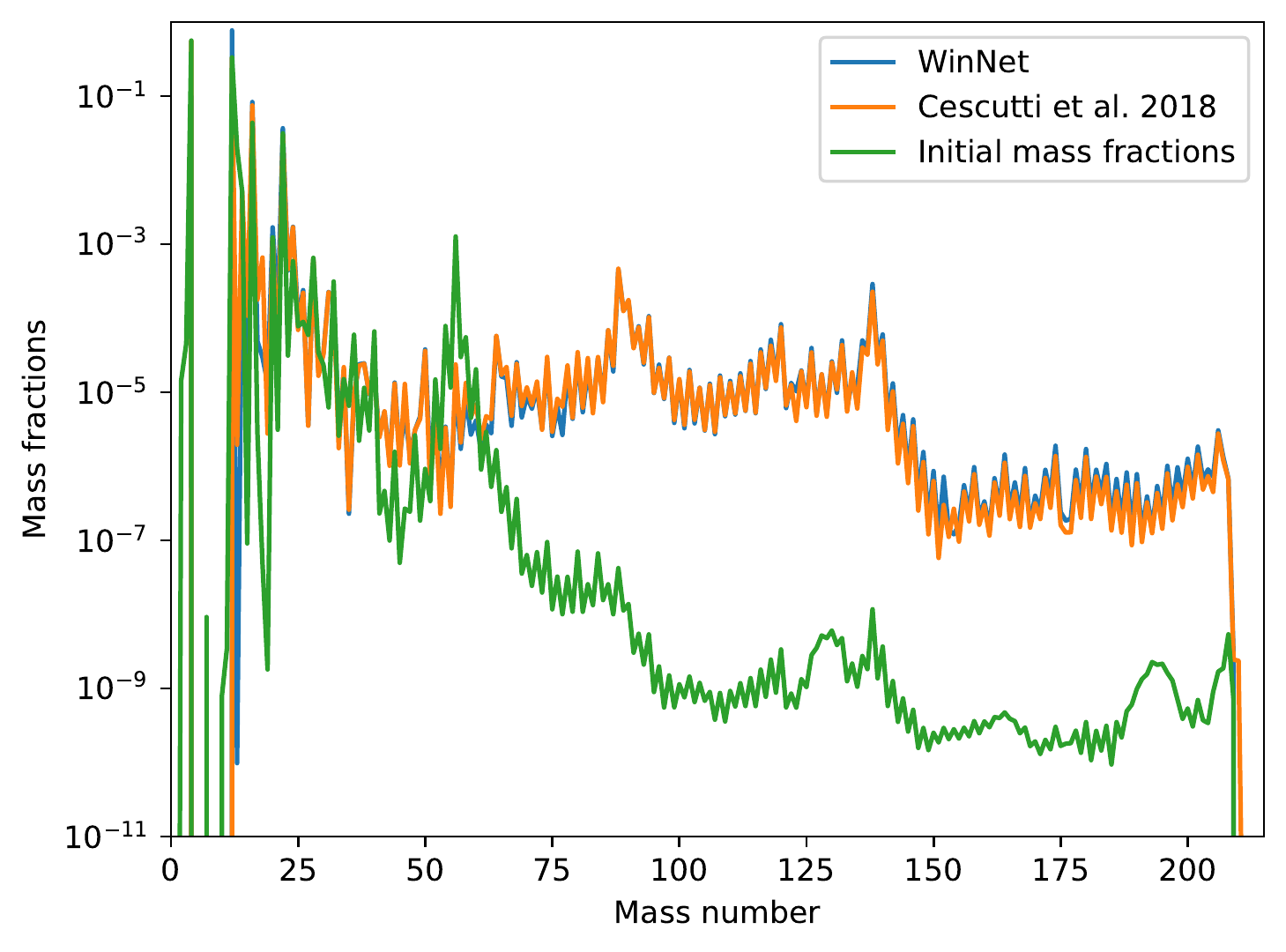}%
\end{center}
\caption{Initial and final mass fractions of a main $s$-process. The trajectory as well as the final mass fractions are taken from \citet{Cescutti2018} accessed via \citet{cescutti_data_2022}. }
\label{fig:main_s}
\end{figure}

We added a trajectory of a main $s$-process to the example cases. This trajectory was used for a Monte Carlo sensitivity study in \citet{Cescutti2018} and can be accessed via \citet{cescutti_data_2022}. The trajectory was extracted from the $^{13}$C pocket after the sixth thermal pulse of a solar metallicity, 3 M$_\odot$ mass AGB star (for more details, see the original publication). The final mass fractions agree well with those of \citet[][see Fig.~\ref{fig:main_s}]{Cescutti2018} given the fact that we do not attempt to use the exact same nuclear input. 

\subsubsection{Complete Silicon burning}
\begin{figure}
\begin{center}
\includegraphics[width=1.0\linewidth]{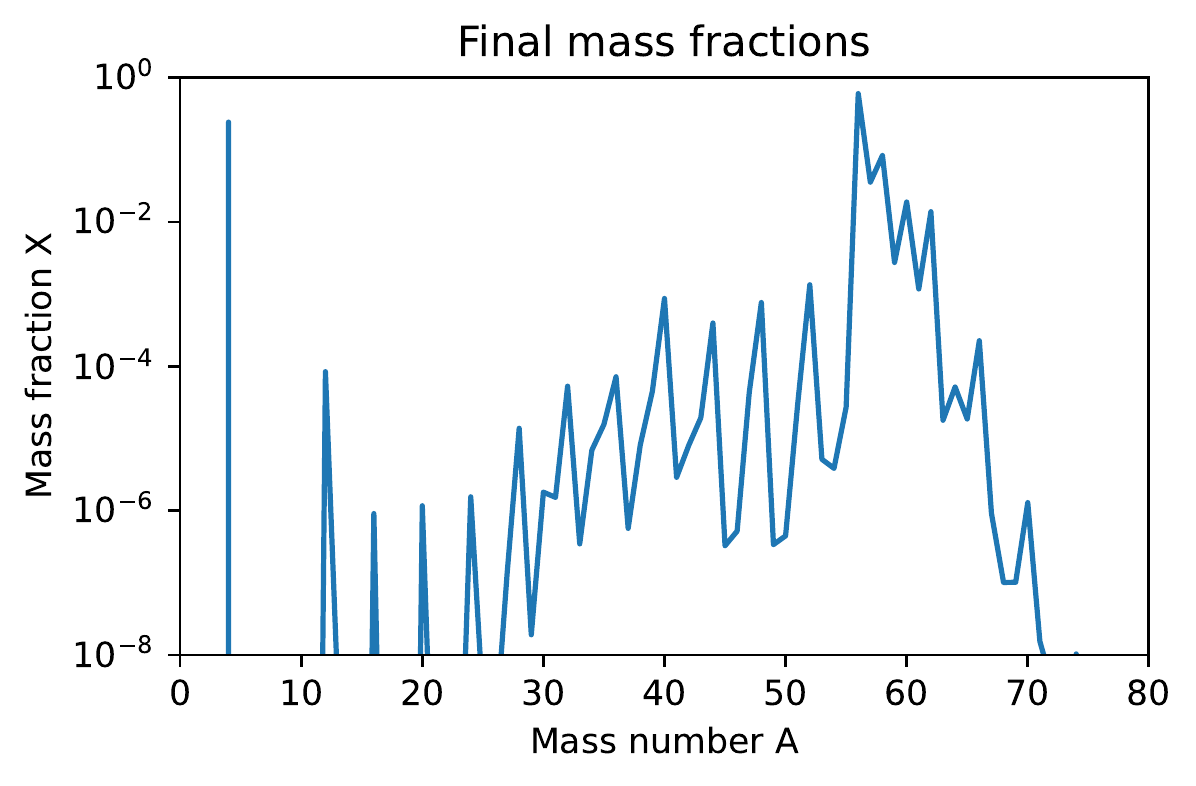}%
\end{center}
\caption{Final mass fractions of complete Si burning obtained with a simple parametric model.}
\label{fig:param_si_burn}
\end{figure}

The complete Si burning can be described by analytical models. For this, we assume that the time scale behaves according to the freefall time scale\citep[e.g.,][]{Arnett1996}:
\begin{align}
    \tau \approx \frac{446}{\sqrt{\rho}}.
\end{align}
The density $\rho$ and the temperature $T$ are assumed to follow:
\begin{align}
    T(t) &= T_S \, e^{-t/(3\tau)} \\
    \rho(t)&= \rho_S \, e^{-t/\tau},
\end{align}
where the shock temperature $T_S$ can be defined as in, e.g., \citet[]{Nadyozhin2002,Woosley2002}
\begin{equation}
    T_S=2.4 \, E_{51}^{1/4} \, R_0^{-3/4}\, \mathrm{GK},
\end{equation}
with the explosion energy $E_{51}$ in $10^{51}\,\mathrm{erg}$, and an initial radius $R_0$ in $10^8\, \mathrm{cm}$. The shock density is given by the jump condition ($\rho_S = 7\rho_0$). For an initial (pre-shock) density of $\rho_0=10^6\,\mathrm{g\,cm^{-3}}$, an initial radius of $R_0=2\times 10^8$, and an explosion energy of $10^{52}\,\mathrm{erg}$ we obtain:
\begin{align}
    T(t) &= 2.4\, (0.2)^{-3/4}\, e^{-t/ (3\tau)}\\
    \rho(t) &= 7\times 10^6  e^{-t/ \tau}.
\end{align}
When we further assume an electron fraction of $Y_e = 0.498$ as typical in the Si shell, we obtain final abundances that are located around $^{56}$Fe (Fig.~\ref{fig:param_si_burn}).

\subsubsection{$\nu$p-process}
\begin{figure}
\begin{center}
\includegraphics[width=1.0\linewidth]{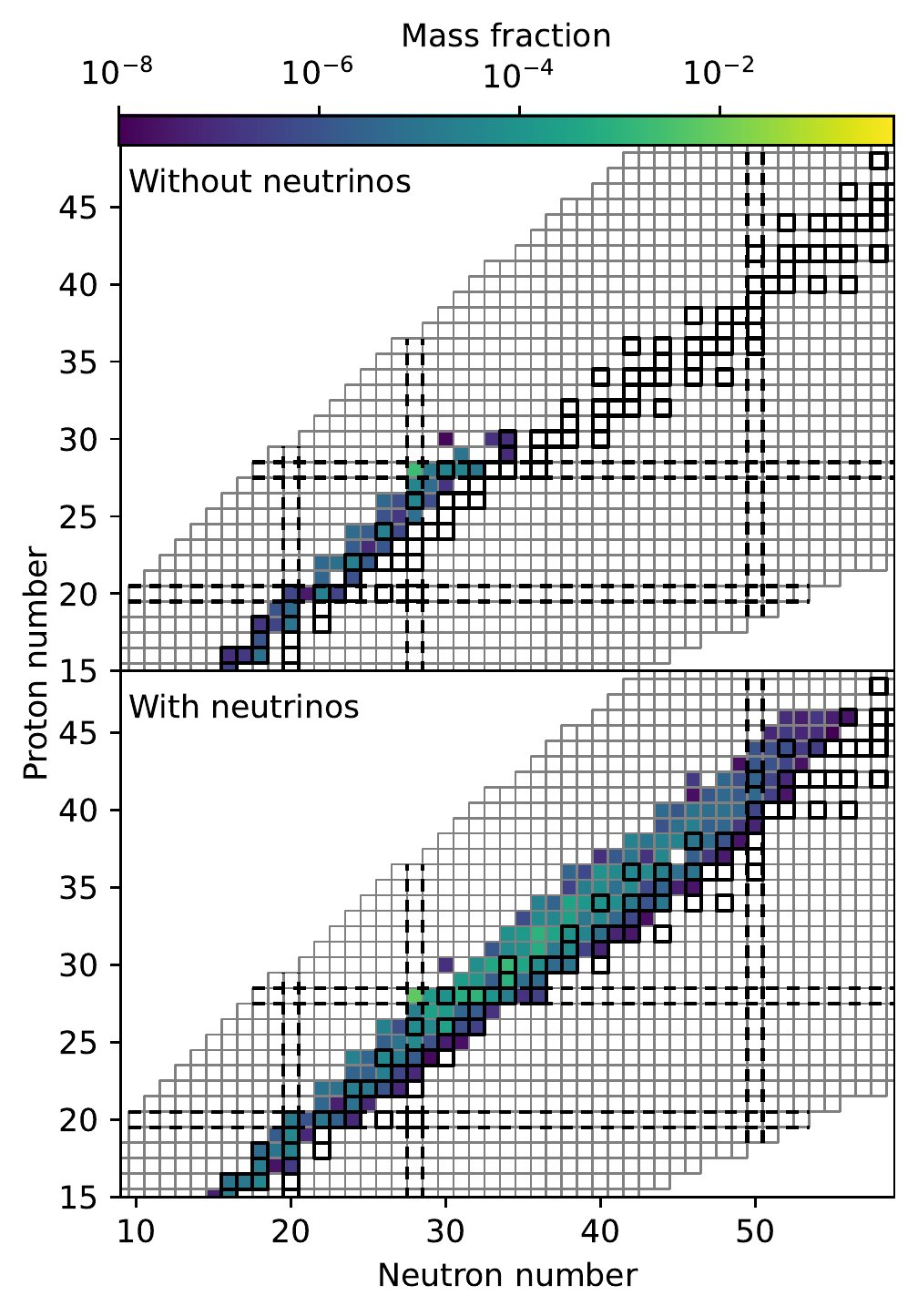}%
\end{center}
\caption{Mass fractions at $t=1.8\times 10^3\,\mathrm{s}$ for one trajectory within the MR-SNe model 35OC-RO of \citet{Obergaulinger2017} and \citet{Reichert.Obergaulinger.ea:2021}. Upper panel: calculation without involving neutrino reactions. Lower panel: calculation using neutrino reactions on nucleons as well as on heavier nuclei \citep{Sieverding2018}.}
\label{fig:nupexample}
\end{figure}

Neutrinos can be crucial to synthesize proton-rich isotopes. If the neutrino flux is strong enough, this can lead to a $\nu$p-process. The conditions for this are, for example fulfilled in the MR-SNe model 35OC-RO of \citet{Obergaulinger2017} and \citet{Reichert.Obergaulinger.ea:2021}. The nucleosynthetic flow with and without neutrinos is shown in Fig.~\ref{fig:nupexample}.

\subsubsection{The weak $r$-process}
\begin{figure}
\begin{center}
\includegraphics[width=1.0\linewidth]{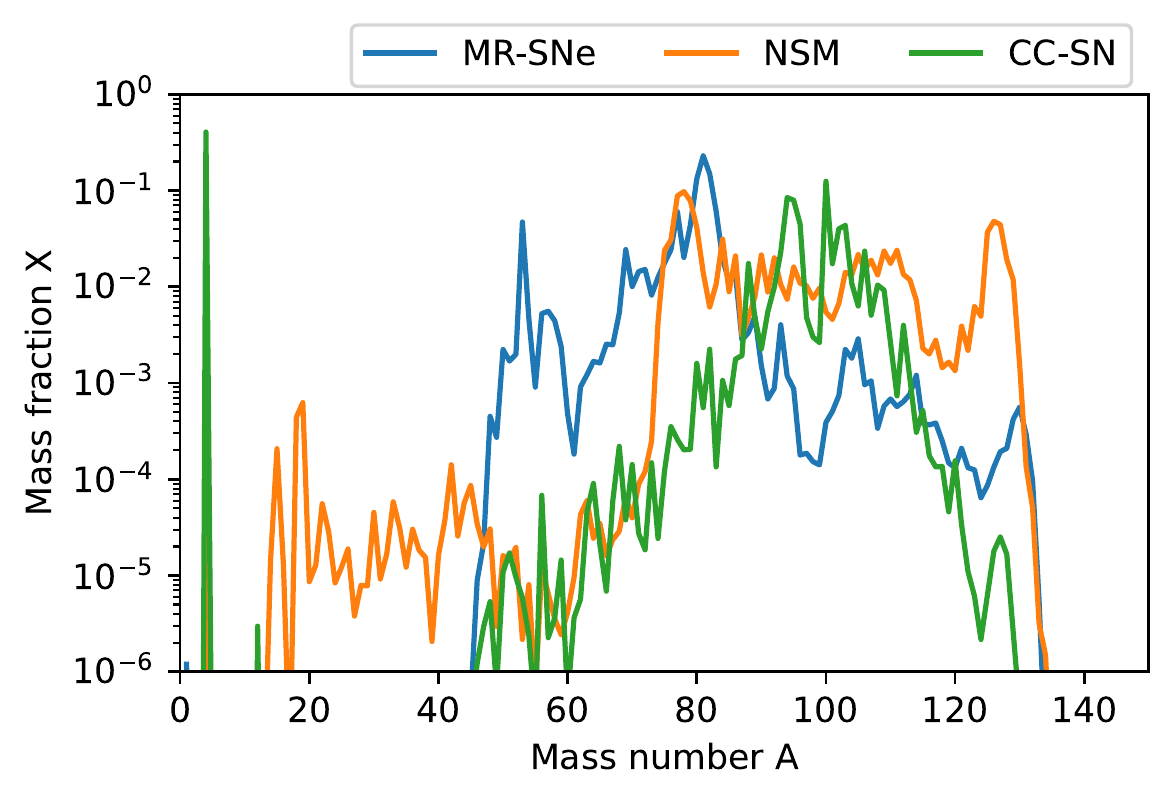}%
\end{center}
\caption{Final mass fractions after $1\,\mathrm{Gyr}$ for a trajectory of an MR-SNe (\citealt{Obergaulinger2017}, \citealt{Reichert.Obergaulinger.ea:2021}), of the neutrino-driven wind of an NSM (\citealt{Perego2014}, \citealt{martin15}), and of the neutrino-driven wind of a CC-SNe \citep{Bliss2018}.}
\label{fig:weak_r}
\end{figure}

The weak $r$-process, i.e., a synthetization of elements up to the second $r$-process peak ($A\sim 130$) can occur in moderately neutron-enriched environments. These conditions can be found in a variety of astrophysical host scenarios. Here, we show an exemplary trajectory from an MR-SNe \citep[][]{Obergaulinger2017,Reichert.Obergaulinger.ea:2021}, the neutrino-driven wind of an NSM \citep[][]{Perego2014, martin15}, and the neutrino-driven wind of a CC-SNe \citep{Bliss2018}. The final mass fractions are shown in Fig.~\ref{fig:weak_r}.

\subsubsection{Strong $r$-process}
\begin{figure}
\begin{center}
\includegraphics[width=1.0\linewidth]{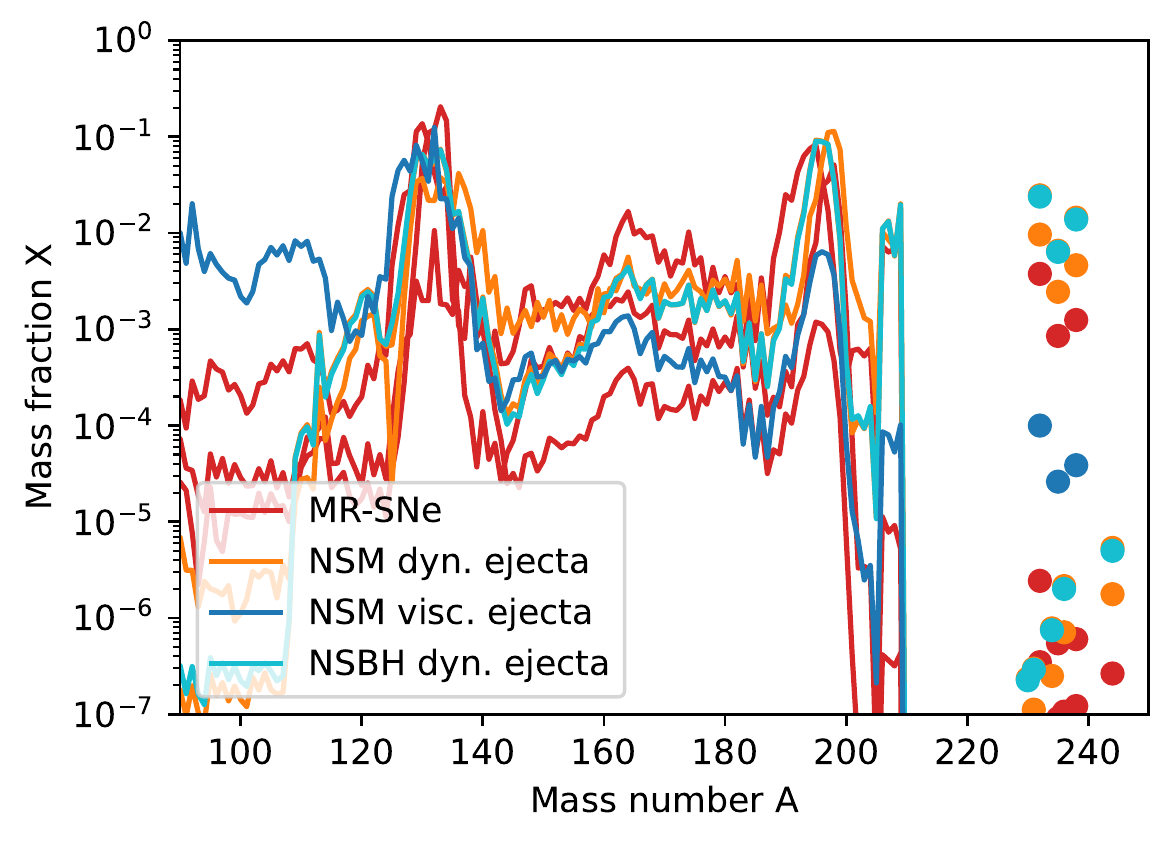}%
\end{center}
\caption{Final mass fractions of various example trajectories. Within MR-SNe, the models used in \citet[][]{winteler12b}, \citet[][]{Obergaulinger2017}, \citet{Reichert.Obergaulinger.ea:2021}, and \citet[][]{Obergaulinger2021}, \citet{Reichert2023} are shown (red lines). For the dynamic ejecta of an NSM (orange lines) we show the simulations of \citet{korobkin12}, \citet{rosswog13a}, \citet[][model ns10ns10]{piran13a}, and \citet[][]{Bovard2017}. Furthermore, we illustrate the viscous ejecta of an NSM from the calculation of \citet[][]{wu16}. The dynamic ejecta of a NSBH merger from  \citet{korobkin12}, \citet{rosswog13a}, \citet[][model BH10]{piran13a} is shown as the cyan line.} 
\label{fig:example_r_process}
\end{figure}

Calculating a full $r$-process is one of the most challenging nuclear reaction network calculations. Here we include $\sim$~$ 6500$ nuclei up to $^{337}$Rg. The astrophysical host event of the $r$-process is not fully understood yet. Very promising candidates are NSMs, NSBH mergers, or MR-SNe. For these scenarios, we show the results of individual trajectories in Fig.~\ref{fig:example_r_process}. These trajectories come from a variety of (M)HD simulations and were presented in
\citet[][]{winteler12b,korobkin12,rosswog13a,piran13a,wu16,Bovard2017,Obergaulinger2017,Reichert.Obergaulinger.ea:2021,Obergaulinger2021,Reichert2023}.

\subsection{Test scenarios}\label{sct:tests}

We have implemented a series of tests in order to monitor the performance and consistency of \textsc{WinNet}. The tests cover a range of numerical and physical scenarios, which we will present in this section. Many of the tests are designed in a way that an analytic calculation of the result is also possible. Furthermore, we implemented technical test cases such as reading the initial composition, the correct reproduction of the input thermodynamic conditions, and correct implementation of the different reaction rate formats.

\subsubsection{$\beta$-decays}
\begin{figure}
\begin{center}
\includegraphics[width=1.0\linewidth]{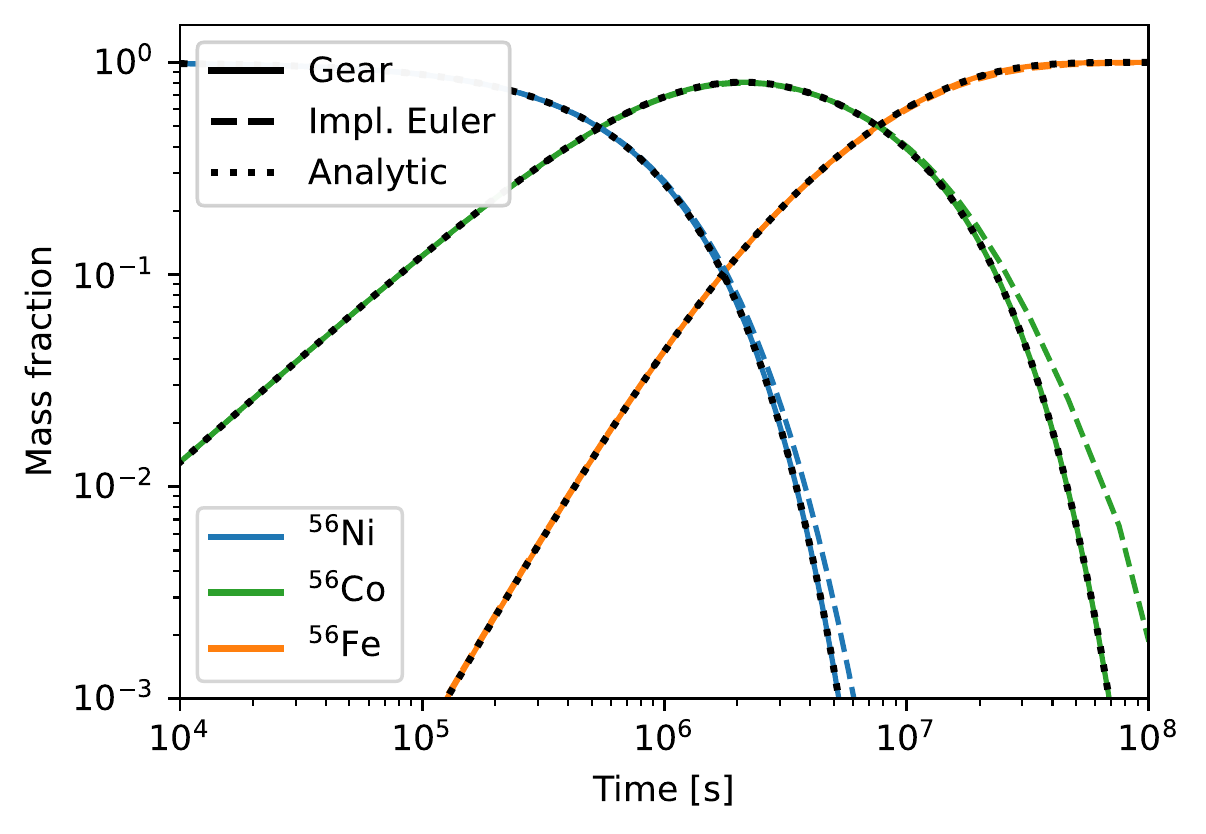}%
\end{center}
\caption{Decay of $^{56}$Ni calculated with the Gear (solid line) and implicit Euler solver (dashed line). The analytic solution is shown with the dotted lines.}
\label{fig:test_ni56}
\end{figure}

A simple nucleosynthesis calculation is given by a $\beta$-decay. We tested the decay of neutrons to protons, as well as the decay chain of $^{56}$Ni. The results give an interesting insight into the accuracy of the integration using an implicit Euler integration scheme. We recall that this scheme does not have any error estimate for the time step and the convergence criterion of the mass conservation (i.e., $\sum_i X_i = 1$; Eq.~\eqref{eq:euler_massconservation}). While this is common practice for calculations involving a large amount of nuclei and reactions \citep[e.g.,][]{lippuner17a,hix99}, within the tested decays it leads to uncertainties. As an example, we show the time evolution of $^{56}$Ni, $^{56}$Co, and $^{56}$Fe in Fig.~\ref{fig:test_ni56}.
The discrepancies between the implicit Euler solution and the analytic solution can be reduced by choosing adapted smaller time steps resulting from smaller $\eps_\mathrm{Euler}$ values (in the example, $\eps_\mathrm{Euler}=10^{-1}$ was used; see Eq.~\eqref{eq:impl_euler_eps}). The example also shows the strength of the adaptive time-step control within the Gear solver, which is able to stay close to the analytic solution.

\begin{figure}
\begin{center}
\includegraphics[width=1.0\linewidth]{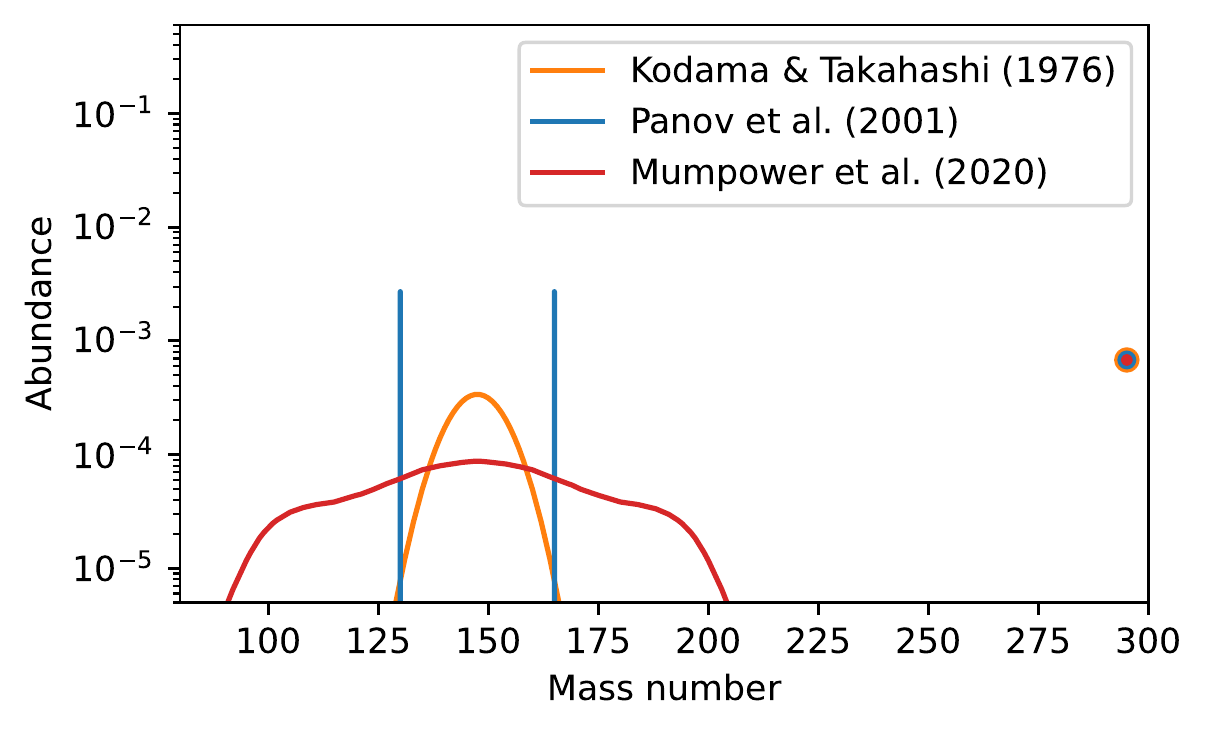}%
\end{center}
\caption{The $\beta$-delayed fission of $^{295}$Am. Shown are the abundances after $10^{-2}\,\mathrm{s}$ for three different fission fragment distributions.}
\label{fig:test_fissfrag}
\end{figure}

Another test is based on the $\beta$-delayed fission of $^{295}$Am. Identical to a normal $\beta$-decay, we can calculate the decay of this nucleus via 
\begin{equation}
Y(t) = Y_0 \, e^{-\alpha t},
\end{equation}
with the decay constant $\alpha$. The products of this decay are determined by the fission fragment distribution that can be calculated analytically, as in \citet[][]{Kodama1975} or \citet[][]{Panov2001}. Additionally, we include the fission fragment distribution of \citet{Mumpower2020} for $\beta$-delayed and neutron-induced fission. This distribution spans a wide range of mass numbers. The different fragments for a simulation time of $t=10^{-2}\,\mathrm{s}$ are shown in Fig.~\ref{fig:test_fissfrag}.
The calculated abundance pattern deviates less than $1\,\%$ from the input fission fragment distributions.

\subsubsection{Equilibrium cases}
Useful scenarios are cases were an equilibrium value is obtained. An equilibrium situation can be challenging for numerical solvers, as constant abundances appear like a reaction time scale that is approaching infinity (e.g., \citealt{hix99,Feger2011phd,lippuner17a}). In the following, we present the case of an (n,$\gamma$)-($\gamma$,n) equilibrium as well as equilibria obtained by electron- and positron-captures and neutrino absorption.

\begin{figure}
\begin{center}
\includegraphics[width=1.0\linewidth]{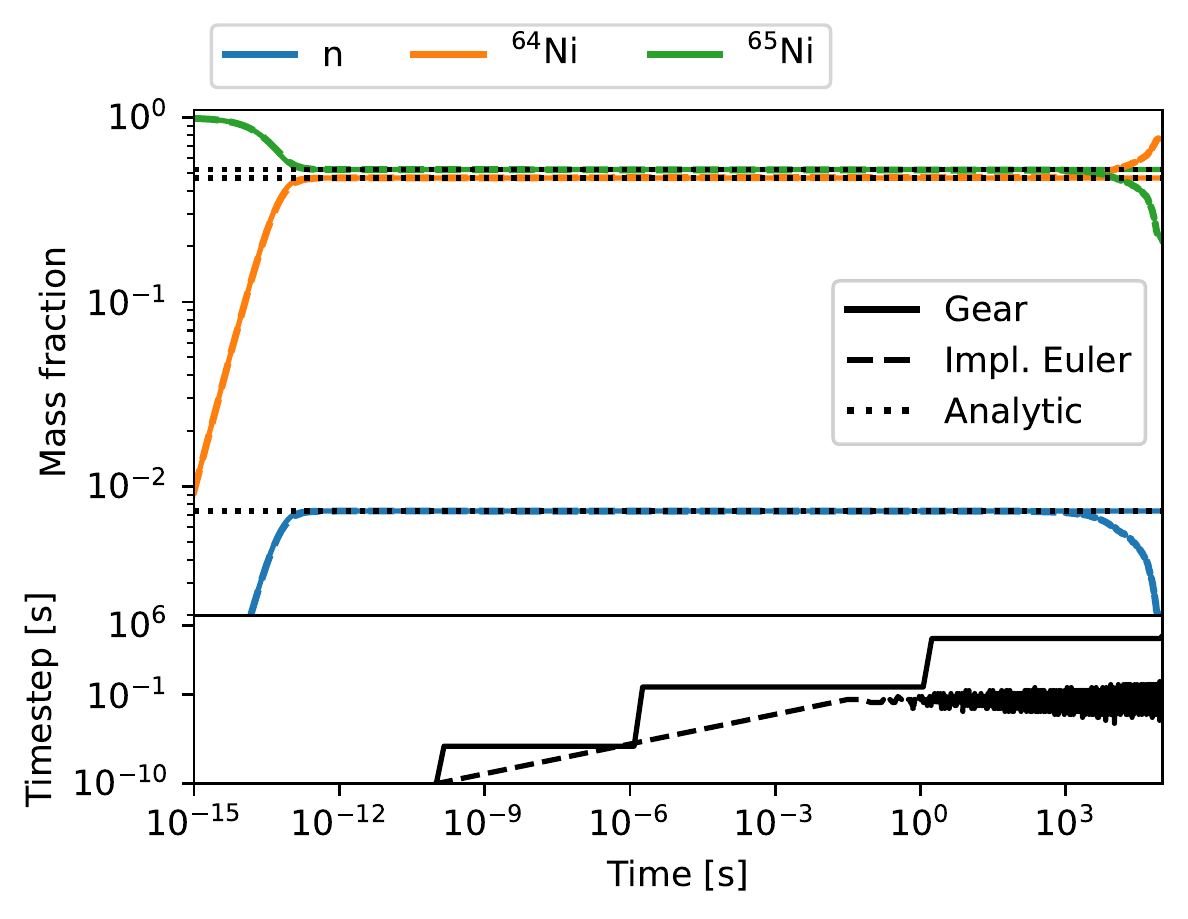}%
\end{center}
\caption{Upper panel: mass fractions of neutrons (blue), $^{64}$Ni (orange), and $^{65}$Ni (green) for the implicit Euler (dashed line) and Gear integration scheme (solid line). The analytic equilibrium solution is shown with the black dotted lines. Lower panel: the time step of the implicit Euler and Gear integration schemes.}
\label{fig:test_ngamma}
\end{figure}

In the case of an (n,$\gamma$)-($\gamma$,n) equilibrium between $^{64}$Ni and $^{65}$Ni, the analytic solution of the equilibrium composition can be derived as:
\begin{align}
Y(\mathrm{n}) &= \frac{\lambda _{\gamma, n} - \sqrt{\lambda _{\gamma ,n}^2 + 4 \rho \, N_\mathrm{A} \, \langle\sigma\nu\rangle _{n, \gamma}  \, \lambda _{\gamma ,n}  /65  }     }{-2  \rho \, N_\mathrm{A} \, \langle\sigma\nu \rangle_{n, \gamma}} \\
Y(^{64}\mathrm{Ni}) &= Y(\mathrm{n}) \\
Y(^{65}\mathrm{Ni}) &= 1/65 - Y(\mathrm{n}),
\end{align}
For $T=8\,\mathrm{GK}$ and $\rho=10^9\,\mathrm{g\,cm^{-3}}$ and matter initially consisting out of pure $Y(^{65}\mathrm{Ni})$ (which introduced the factor $1/65$), we obtain $Y(\mathrm{n})=Y(^{64}\mathrm{Ni}) = 7.35175\times 10^{-3}$ and $Y(^{65}\mathrm{Ni})= 8.03286\times 10^{-3}$. While the integration with the Gear scheme results in an excellent agreement within $0.0015\,\%$, the time step within the implicit Euler becomes very small. This leads to numerical instabilities and a large deviation from the analytic solution after $10^3\,\mathrm{s}$ (see Fig.~\ref{fig:test_ngamma}). This instability is unlikely to be resolved by more restrictive time steps in the implicit Euler scheme as the time step is based of changes in the thermodynamic conditions or abundances. Since both are static, the scheme will always attempt very large (possibly too large) time steps. This continues until large errors have been accumulated and the solution diverges. On the other hand, the Gear solver estimates an integration error, independent on changes in conditions or abundances. As a consequence, the solution is more stable.

Another equilibrium test scenario is given by the equilibrium of electron-, positron-, and neutrino-captures. These equilibria are important to understand the initial electron fraction in $r$-process calculations. In the following, we investigate the situation of the equilibrium electron fraction when only considering electron-/positron-captures on nucleons, neutrino absorption on nucleons, and a combination of both.
\begin{figure}
\begin{center}
\includegraphics[width=1.0\linewidth]{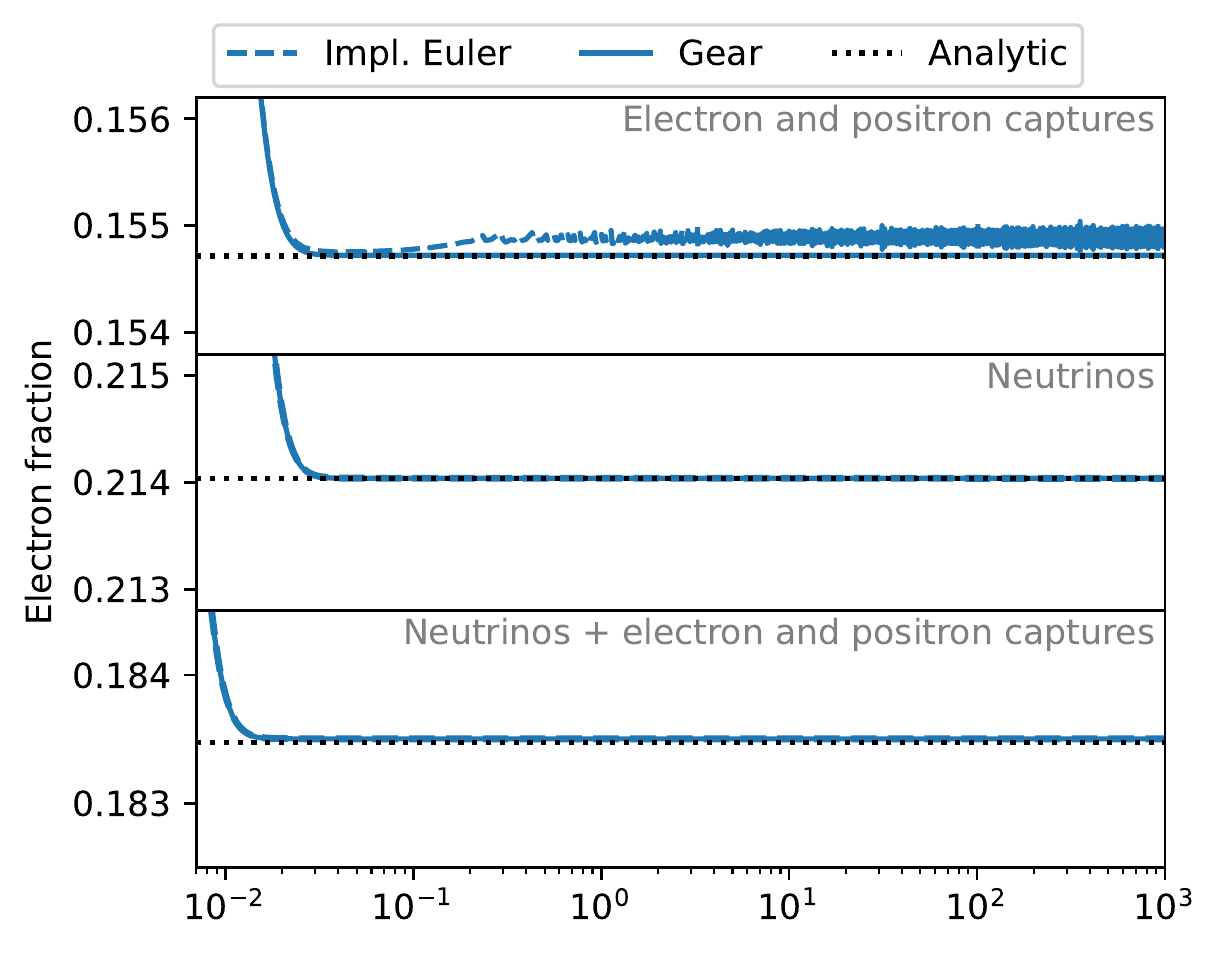}%
\end{center}
\caption{Electron fraction under hydrostatic conditions with $T=30\, \mathrm{GK}$ and $\rho=10^{10}\,\mathrm{g\, cm^{-3}}$. Upper panel: equilibrium case when involving only electron- and positron-captures. Middle panel: equilibrium case when involving only neutrinos with luminosities of $L_\nu = 10^{52}\,\mathrm{erg\,s^{-1}}$,  $L_{\bar{\nu}} = 5\times 10^{52}\,\mathrm{erg\,s^{-1}}$, and neutrino energies of $E_\nu = 25.2 \,\mathrm{MeV}$ as well as $E_{\bar{\nu}} = 31.5 \,\mathrm{MeV}$.}
\label{fig:test_beta_eq}
\end{figure}
Similar to \citet[]{Just2022}, we can calculate the equilibrium electron fractions for all three scenarios. Assuming only electron- and positron-captures, the equilibrium electron fraction for hydrostatic conditions can be obtained by solving:
\begin{equation}
    \lambda_{e^+} Y(\mathrm{n}) - \lambda_{e^-}Y(\mathrm{p}) = 0,
\end{equation}
with the positron and electron-capture rate $\lambda_{e^+}$ and $\lambda_{e^-}$, respectively. For a constant temperature of $30\,\mathrm{GK}$ and density of $10^{10}\,\mathrm{g\,cm^{-3}}$ we obtain $Y_\mathrm{e, em}=0.155$. Both integration schemes obtain a great precision, with the Gear solver agreeing within $0.004\%$, and the implicit Euler agreeing within $0.076\%$. The implicit Euler integration scheme shows again some numerical noise (upper panel in Fig.~\ref{fig:test_beta_eq}) and the result is therefore slightly worse compared to the Gear integration scheme.
For the scenario with only neutrinos irradiating the matter, we can similarly calculate the equilibrium electron fraction:
\begin{equation}
    \lambda_{\nu_e} Y(\mathrm{n}) - \lambda_{\bar{\nu_e}}Y(\mathrm{p}) = 0,
\end{equation}
with the neutrino and antineutrino cross sections $\lambda_{\nu_e}$ and $\lambda_{\bar{\nu_e}}$, respectively. Assuming matter located at a radius of $50\,\mathrm{km}$, irradiated by neutrino luminosities of $L_\nu = 10^{52}\,\mathrm{erg\,s^{-1}}$,  $L_{\bar{\nu}} = 5\times 10^{52}\,\mathrm{erg\,s^{-1}}$, and neutrino energies of $E_\nu = 25.2 \,\mathrm{MeV}$ as well as $E_{\bar{\nu}} = 31.5 \,\mathrm{MeV}$ we obtain $Y_\mathrm{e, abs}=0.214$. The final values of both integration schemes agree within $5\times 10^{-8}\,\%$ (middle panel of Fig.~\ref{fig:test_beta_eq}). 
Combining electron-, positron-, and neutrino-captures, the equilibrium electron fraction can be obtained by solving
\begin{equation}
    (\lambda_{\nu_e}+\lambda_{e^+}) Y(\mathrm{n}) - (\lambda_{\bar{\nu_e}}+\lambda_{e^-})Y(\mathrm{p}) = 0.
\end{equation}
For the conditions assumed here, we obtain $Y_\mathrm{e, \beta}= 0.1835$, which only deviates by $0.014\,\%$ from the equilibrium value (lower panel of Fig.~\ref{fig:test_beta_eq}).

Another equilibrium case is given by NSE (Section~\ref{sssct:NSE}). We tested that the transition from the NSE region to the network region is consistent. Therefore, we calculated the NSE composition for $T=7\,\mathrm{GK}$, $\rho=10^{7}\,\mathrm{g\, cm^{-3}}$, and $Y_e=0.5$ with and without screening. In addition, we calculated the mass fractions after $10^2\,\mathrm{s}$ when starting with neutrons and protons only, using the same hydrostatic conditions and strong reactions only. This system should also approach NSE. Again, we calculate the abundances with and without electron screening corrections (see Fig.~\ref{fig:nse_screen}). As a consequence of the previous outlined tests, we only calculated the test with the Gear integration method. 

\subsubsection{Other tests}
If the nuclear reaction network is sufficiently large, deriving an analytic expression for the solution is often not possible anymore. In these cases, it is beneficial to compare the result with other nuclear reaction networks. 

We calculate a case of hydrostatic carbon-Oxygen burning with $\rho=10^9\,\mathrm{g\, cm^{-3}}$ and a temperature of $3\,\mathrm{GK}$ for $10^{12}\,\mathrm{s}$. The initial composition consisted of $X(^{12}\mathrm{C})= X(^{16}\mathrm{O}) = 0.5$. In total we involve $13$ nuclei in the calculation. We compare the final abundances of \textsc{WinNet} (using Gears integration method) with the results of the nuclear reaction networks \textsc{SkyNet} \citep[][]{lippuner17a}, \textsc{ReNet} \citep{Navo2023}, and \textsc{XNet} \citep[][]{hix99}.
\begin{table}
\centering
\begin{tabular}{l l l l l l}
\hline
\hline
\textbf{A} & \textbf{Z} & \textbf{Y$_\mathrm{WinNet}$} & $\Delta_\mathrm{SkyNet}$ & $\Delta_\mathrm{ReNet}$ & $\Delta_\mathrm{XNet}$\\ 
  &    &  & [$10^{-2}\,\%$] & [$10^{-3}\,\%$]& [$10^{-1}\,\%$]\\ 
\hline 
$4$  &  $2$ & $4.01\times 10^{-09}$  & $0.82$ & $0.13$ & $0.25$\\ 
$12$ &  $6$ & $4.04\times 10^{-18}$ & $0.53$ & $0.08$ & $0.16$\\ 
$16$ &  $8$ & $1.55\times 10^{-16}$ & $8.06$ & $1.27$ & $2.44$\\ 
$20$ & $10$ & $1.89\times 10^{-19}$ & $7.23$ & $1.14$ & $2.19$\\ 
$24$ & $12$ & $1.14\times 10^{-14}$ & $6.61$ & $1.04$ & $2.00$\\ 
$28$ & $14$ & $8.14\times 10^{-09}$ & $5.79$ & $0.91$ & $1.75$\\
$32$ & $16$ & $4.52\times 10^{-08}$ & $4.96$ & $0.78$ & $1.50$\\
$36$ & $18$ & $7.54\times 10^{-09}$ & $4.13$ & $0.65$ & $1.25$\\
$40$ & $20$ & $5.78\times 10^{-07}$ & $3.31$ & $0.52$ & $1.00$\\
$44$ & $22$ & $2.89\times 10^{-09}$ & $2.48$ & $0.39$ & $0.75$\\
$48$ & $24$ & $3.23\times 10^{-07}$ & $1.65$ & $0.26$ & $0.50$\\
$52$ & $26$ & $7.13\times 10^{-05}$ & $0.82$ & $0.13$ & $0.25$\\ 
$56$ & $28$ & $1.78\times 10^{-02}$ & $0.003$& $0.001$ & $0.001$\\ 
\hline        
\end{tabular}
\caption{Final abundances for hydrostatic Carbon-Oxgen test case. Columns 4-6 show the deviation compared to the results of \textsc{SkyNet}, \textsc{ReNet}, and \textsc{XNet}, respectively.}
\label{tab:test_co_hydrostatic}
\end{table}
The final abundances of \textsc{WinNet} deviate by less than $1\%$ to all other reaction networks (Tab.~\ref{tab:test_co_hydrostatic}). The abundant nucleus $^{56}$Ni even agrees with a maximum deviation of $10^{-4}\,\%$ only. We note that we did not tune the specific numerical parameters used in the different codes. More restrictive time steps can therefore lead to an even better agreement.

\begin{figure}
\begin{center}
\includegraphics[width=1.0\linewidth]{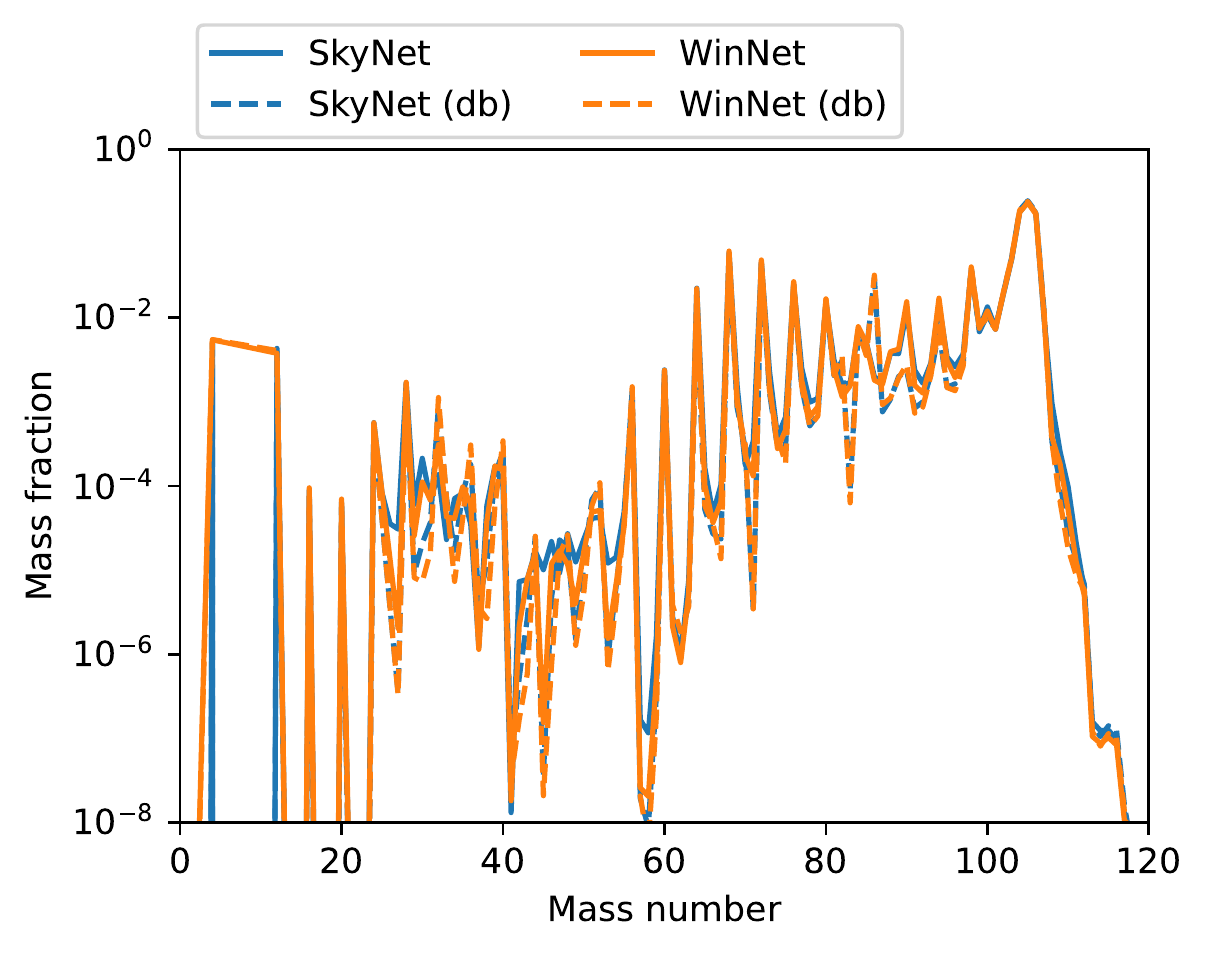}%
\end{center}
\caption{Composition of an X-ray burst \citep[][]{Schatz2001} after $10^3\,\mathrm{s}$. The result is shown for \textsc{WinNet} (orange lines) and \textsc{SkyNet} (blue lines) with (dashed lines) and without (solid lines) calculating reverse reactions via detailed balance.}
\label{fig:xray_burst_skynet}
\end{figure}

To test the implementation of detailed balance (Section~\ref{sct:det_balance}), we repeated the calculation performed in \citet[][]{lippuner17a} with \textsc{SkyNet}. We calculate the nucleosynthesis of a trajectory of an X-ray burst from \citet[][]{Schatz2001}. The result of four calculations is shown in Fig.~\ref{fig:xray_burst_skynet}. There, we use the Reaclib v2.2 and calculate the nucleosynthesis with \textsc{WinNet} using the reverse rates as given by Reaclib (solid orange line). Moreover, we use \textsc{SkyNet} with reverse rates from Reaclib (solid blue line). Additionally, we calculate the same trajectory, but using reverse rates calculated via detailed balance using the Q-value from the mass excess provided within Reaclib (within the winvn file, dashed lines). Both networks agree very well for both cases. The impact of using detailed balance rates with masses from the winvn in contrast to Reaclib reverse rates seems to be larger in \textsc{SkyNet} especially for the smaller mass numbers $\sim 50$.  However, for both networks, there is also a distinct feature visible at $A\sim 85$.

\section{Summary and conclusion}\label{sct:summary}

We have summarized the fundamentals of nuclear reaction networks. The implementation was demonstrated with the single-zone nuclear reaction network code \textsc{WinNet}.

We outlined the differential equations that underlie every nuclear reaction network code. Additionally, we presented two implicit numerical techniques to solve these equations, the implicit Euler and Gear's integration scheme. 

A mandatory ingredient is also the set of reaction rates. The reaction rates can originate from different databases with varying parameterizations. Hereby, one should ensure that the same underlying nuclear physic inputs such as mass models are used. We described the reaction rate formats that are supported by \textsc{WinNet}, namely the Reaclib reaction rate database, a format for $\beta$-delayed neutron emission, tabulated rates, theoretical $\beta^+$, $\beta^-$, electron-capture and positron-capture rates, neutrino reactions, and fission reactions. 

All these different reaction sources get a different priority assigned, and rates that appear in more than one source are replaced by the rate with the highest priority. This priority is chosen arbitrarily without any estimate of the quality of the rate, and there could be still some action required if a user wants to use specific reaction rates.

\textsc{WinNet} is further able to calculate detailed balance reactions on-the-fly, which can be useful especially for tabulated rates, where a tabulation of the reverse reactions could break the detailed balance principle. If included, the detailed balance reactions will replace all reverse reactions in the other reaction rate sources.

This is also especially useful when implementing new reaction rates for which the reverse rates may not always have been published. As an example, \citet{DeBoer.Gorres.Wiescher.ea:2017} published the reaction rate $^{12}\mathrm{C}(\alpha,\gamma)^{16}\mathrm{O}$. While the forward rate can easily be changed as they give Reaclib parameters (Section~\ref{ssct:reaclib}) and tabulated values (Section~\ref{ssct:tab_rates}), the reverse reaction should also be consistently changed (Section~\ref{sct:det_balance}). Instead of calculating this rate by hand, within \textsc{WinNet} one can enable a parameter to calculate the reverse reaction internally.

All charged particle reaction rates can be further altered by electron screening. This correction is implemented with a multiplicative factor to the reaction rates.
 
We presented the energy feedback from nuclear reactions onto the temperature, which is implemented in the form of an operator splitting method.  

Finally, we introduced simple examples and test cases to demonstrate the reliability of the reaction network code \textsc{WinNet}. Using these test cases, we analyzed the advantages and disadvantages of the different implemented numerical integration methods. We conclude that hydrostatic and equilibrium conditions are often more efficient and precise with the Gear integration method. More complex rapidly varying thermodynamic conditions are more efficient with the implicit Euler integration method.

In addition to the reliability, a large focus during the development of \textsc{WinNet} was the usability. The code provides an easy interface to the user by a simple parameter file. Additionally, comments are written entirely in a doxygen\footnote{\url{https://www.doxygen.nl/index.html}} conform format, and the documentation can be accessed along with the code. Due to the modular structure, it is also easy to change the included reactions. Additionally, large effort has been undertaken to supply understandable error messages. To give an example, if an input parameter is misspelled, the error message contains not only that this parameter does not exist, but also points to the most similar existing parameter. 

When deciding for or against favoring the usage of \textsc{WinNet} over other publicly available reaction networks, one should keep in mind the advantages and disadvantages for carrying out the desired task. An obvious point to make here is the code language. Users that want to make changes in the code and that are more familiar with C or C++ may feel more comfortable with using \textsc{NucNet} \citep{Meyer2007} or \textsc{SkyNet} \citep{lippuner17a} rather than the Fortran 90 written codes of \textsc{Torch} \citep{Paxton2015}, \textsc{XNet} \citep{hix99}, or \textsc{WinNet}. For applications that require including the reaction network into a hydrodynamical code, the usage of \textsc{Torch} or \textsc{XNet} may be favored over \textsc{WinNet}. While it is not impossible to include \textsc{WinNet} into a hydrodynamical code, there is more experience with \textsc{Torch} or \textsc{XNet} as both have already been used in hydrodynamical simulations. Furthermore, there has been more effort in parallelizing and optimizing \textsc{XNet} especially when calculating abundances for more than one zone. Since the calculation of the ejecta of astrophysical events often relies on many independent tracer particles, the parallelization of \textsc{WinNet} relies on executing many instances of the reaction network for different tracer particles. No effort has been made in parallelizing the calculation of a single tracer particle, and a single instance of \textsc{WinNet} should always be executed on one CPU only. Similar to \textsc{XNet} and in contrast to \textsc{SkyNet}, \textsc{WinNet} optionally performs an initialization step to bring the included reaction rates into an advantageous shape to minimize the cost of reading them in. Especially when running many tracers, this can reduce the computational cost of the initialization step. In post-processed applications where neutrinos play a major role, \textsc{WinNet} definitely has advantages as it is, to our knowledge, the only public code that is able to use publicly available neutrino reactions on heavier nuclei (Section~\ref{ssct:nureac})\footnote{We note that \textsc{XNet} can also contain neutrino reactions on heavier nuclei; the default rate tables are however not publicly available, and one would have to convert the format of the public rate tables.}. Furthermore, for conditions that span a large range of temperatures and densities, the scheme to use and replace theoretical ec-/pc-/$\beta^+$-/$\beta^-$-reactions with experimental half-lives contained in the Jina REACLIB database (Section~\ref{sct:twr}) is an advantage of \textsc{WinNet}. Regarding calculations of very neutron-rich environments, to our knowledge, \textsc{Torch} was never run in the context of the $r$-process. It is possible to include fission reactions and fragment distributions into \textsc{XNet} (as done in, e.g., \citealt{lippuner17a}). However, including more complex fragment distributions with hundreds of fission fragments might be challenging and not possible without code changes. To date, also \textsc{SkyNet} includes only a relatively simple fragment distribution with mostly only two fragments plus neutron emission. If one wants to carry out a study with more complex fission fragments, \textsc{WinNet} could be a better choice, as an arbitrary amount of fragments can be included (see Section~\ref{sct:fission}). Additionally, we included parameterized $\alpha$-decays into \textsc{WinNet} which is an additional, even though possibly small, effort to include into \textsc{SkyNet} or \textsc{XNet} as well. For testing the impact of newly measured reaction rates in different environments, \textsc{WinNet} is also a good choice, since we consider it relatively easy to exchange reaction rates, but also because \textsc{WinNet} already comes with an extensive set of example cases where the impact of certain rates can be directly tested. Also, for storage critical applications, \textsc{WinNet} includes a very flexible way of turning output on and off. This is not easily possible within \textsc{SkyNet} or \textsc{Torch} without touching the code itself. For example, if one wants to know only the abundances after $1$ day, it is possible within \textsc{WinNet} to exclusively output the abundances at this time. Furthermore, \textsc{WinNet} is able to output either ascii, hdf5, or both files, and we therefore consider the code specifically user and beginner friendly. In contrast, \textsc{SkyNet} necessarily needs to have hdf5 packages installed. Ultimately, for numerical studies where the availability of different numerical solvers is desirable, \textsc{WinNet} or \textsc{XNet} are favored over \textsc{SkyNet}, which only includes a backward Euler integration. On the other hand, \textsc{XNet} and \textsc{SkyNet} include more possibilities for the exploration of matrix inversion packages, as \textsc{WinNet} includes only the sparse PARDISO solver. We also briefly tested the performance of \textsc{WinNet} in comparison to \textsc{SkyNet} for an $r$-process example. In this test, both codes were similarly fast with \textsc{WinNet} being slightly faster. However, to get a firmer and more quantitative comparison of the performance, a more detailed investigation that covers additional numerical parameters and astrophysical conditions would be necessary.

With this work, \textsc{WinNet} will be fully public and available for download at \dataset[GitHub]{\github}\footnote{\url{\github}} and \dataset[Zenodo]{\zenodo}\footnote{\url{\zenodo}} \citep{winnet_zenodo}. This includes not only the code, but also all example and test cases. If you use them, please cite the according publications that can be found in the documentation. 

With future works, we plan to extend \textsc{WinNet} by adding more features. As an example, a sensitivity study that uses \textsc{WinNet} may extend it by the scripts and code extensions to perform this task. To give the individual authors of these parts credit, the corresponding work should be cited when someone makes use of a later added feature. 

\begin{acknowledgments}
We want to thank M. A. Aloy, F. Montes, M. Obergaulinger, T. Psaltis, H. Schatz, A. Sieverding, and M. Ugliano for many beneficial discussions. We further thank all of the people that made example trajectories publicly available and L. Bovard, R. Fernández, M. Obergaulinger, and H. Schatz for giving their consent to include their trajectories along with \textsc{WinNet}. Furthermore, we thank F. Timmes for making many useful tools publicly available and for allowing us to use them. We additionally want to thank the referee for providing useful suggestions that helped to improve the manuscript and certain aspects of \textsc{WinNet} itself. M.R. acknowledges support from the grants FJC2021-046688-I and PID2021-127495NB-I00, funded by MCIN/AEI/10.13039/501100011033 and by the European Union "NextGenerationEU" as well as "ESF Investing in your future". Additionally, he acknowledges support from the Astrophysics and High Energy Physics program of the Generalitat Valenciana ASFAE/2022/026 funded by MCIN and the European Union NextGenerationEU (PRTR-C17.I1). A.A., G.M.P., J.K., and M.J.  acknowledge support by the Deutsche Forschungsgemeinschaft (DFG, German Research Foundation) -- Project-ID 279384907 - SFB 1245 and the State of Hessen within the Research Cluster ELEMENTS (Project ID 500/10.006). A.A., J.K., and M.J. additionally acknowledge support from the European Research Council under grant EUROPIUM-667912. G.M.P. acknowledges support by the ERC under the European Union's Horizon 2020 research and innovation program (ERC Advanced grant KILONOVA No. 885281). O.K. was supported by the US Department of Energy through the Los Alamos National Laboratory (LANL). LANL is operated by Triad National Security, LLC, for the National Nuclear Security Administration of the U.S.DOE (contract No. 89233218CNA000001). This work is authorized for unlimited release under LA-UR-23-25461. C.F. acknowledges support from the by United States Department of Energy, Office of Science, Office of Nuclear Physics (award No. DE-FG02-02ER41216). R.H. acknowledges support from the World Premier International Research Centre Initiative (WPI Initiative), MEXT, Japan; the IReNA AccelNet Network of Networks, supported by the National Science Foundation under grant No. OISE-1927130 and ChETEC-INFRA (grant No. 101008324) supported by the European Union's Horizon 2020 research and innovation program. This publication benefited highly from collaborations and exchange within the European Cost Action CA16117 "Chemical Evolution as Tracers of the Evolution of the Cosmos" (ChETEC) and the "International Research Network for Nuclear Astrophysics" (IReNA).
\end{acknowledgments}

%

\vspace{5mm}


\software{Matplotlib \citep{Matplotlib},  
          Numpy \citep{numpy}, 
          Scipy \citep{Scipy},
          Quadpack \citep{quadpack},
          Timmes EOS \citep{Timmes1999b},
          ReNet \citep{Navo2023},
          XNet \citep{hix99},
          SkyNet \citep{lippuner17a},
          PARDISO \citep{Schenk04}.
          }


\bibliography{sample631,fkt}{}
\bibliographystyle{aasjournal}

\appendix
\section{Code convergence}
\label{app:convergence_criteria}
The accuracy of the nucleosynthesis results not only depends on the nuclear input, but also on numerical parameters. In the following, we investigate in more detail the latter error. For this we use a neutron-rich trajectory from an MR-SNe of the simulations of \citet{winteler12b}. We use reactions from the Jina Reaclib \citep[][]{cyburt10} with additional $\alpha$-decays from the Viola-Seaborg formula, theoretical weak rates from \citet{Langanke2001} that we exchange with experimental reaction rates at $10^{-1}\,\mathrm{GK}$. Fission rates have been used as described in Section~\ref{sct:fission} with the fragment distribution of \citet{Panov2001}.

First we will investigate different values of $\epsilon_\mathrm{NR}$ for the convergence criterion of the root-finding algorithm within the implicit Euler method (Eq.~\eqref{eq:euler_massconservation}). As default in \textsc{WinNet}, we perform at least two root-finding iterations. To avoid a re-adjustment to smaller and smaller time steps due to a not converged root-finding (see Fig.~\ref{fig:flowchart}), we set the maximum amount of allowed root-finding iterations to a large value of $1000$. 

\begin{figure}
\begin{center}
\includegraphics[width=0.5\linewidth]{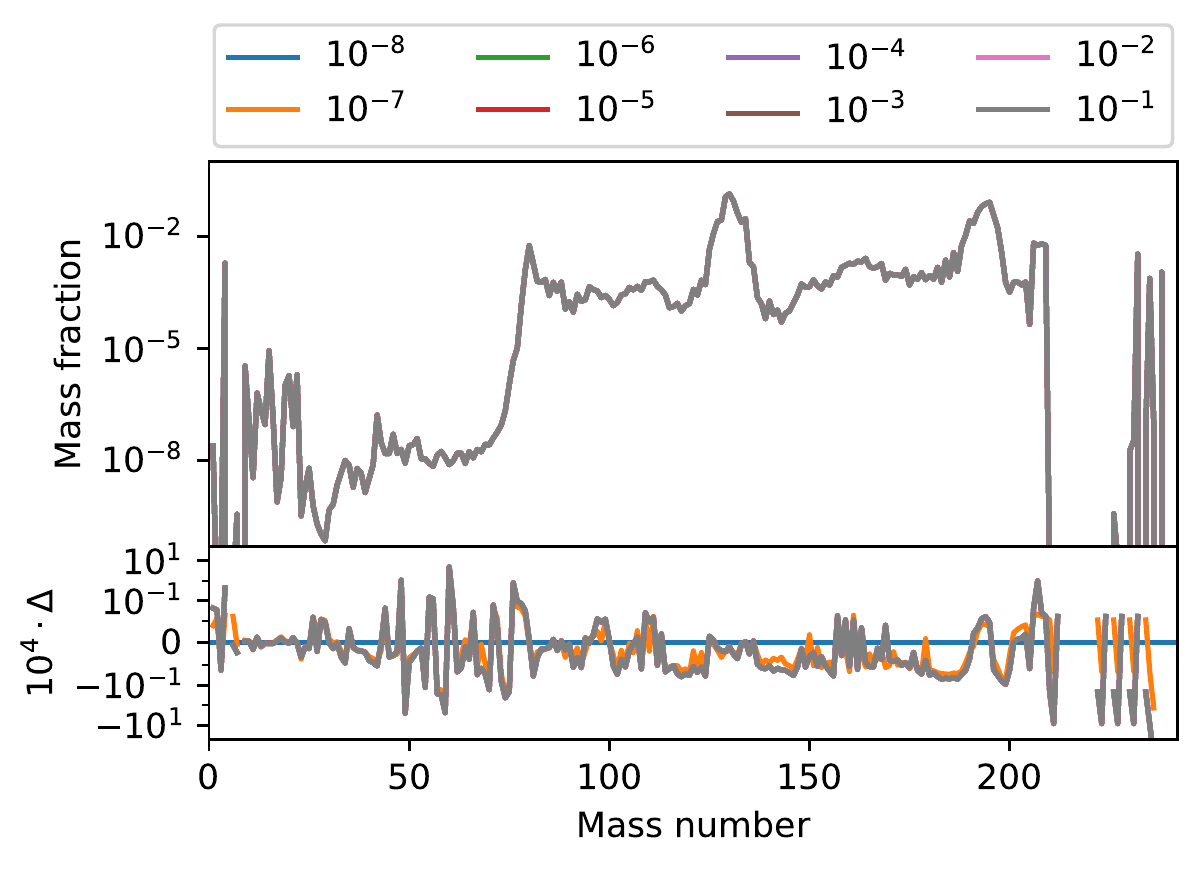}%
\includegraphics[width=0.5\linewidth]{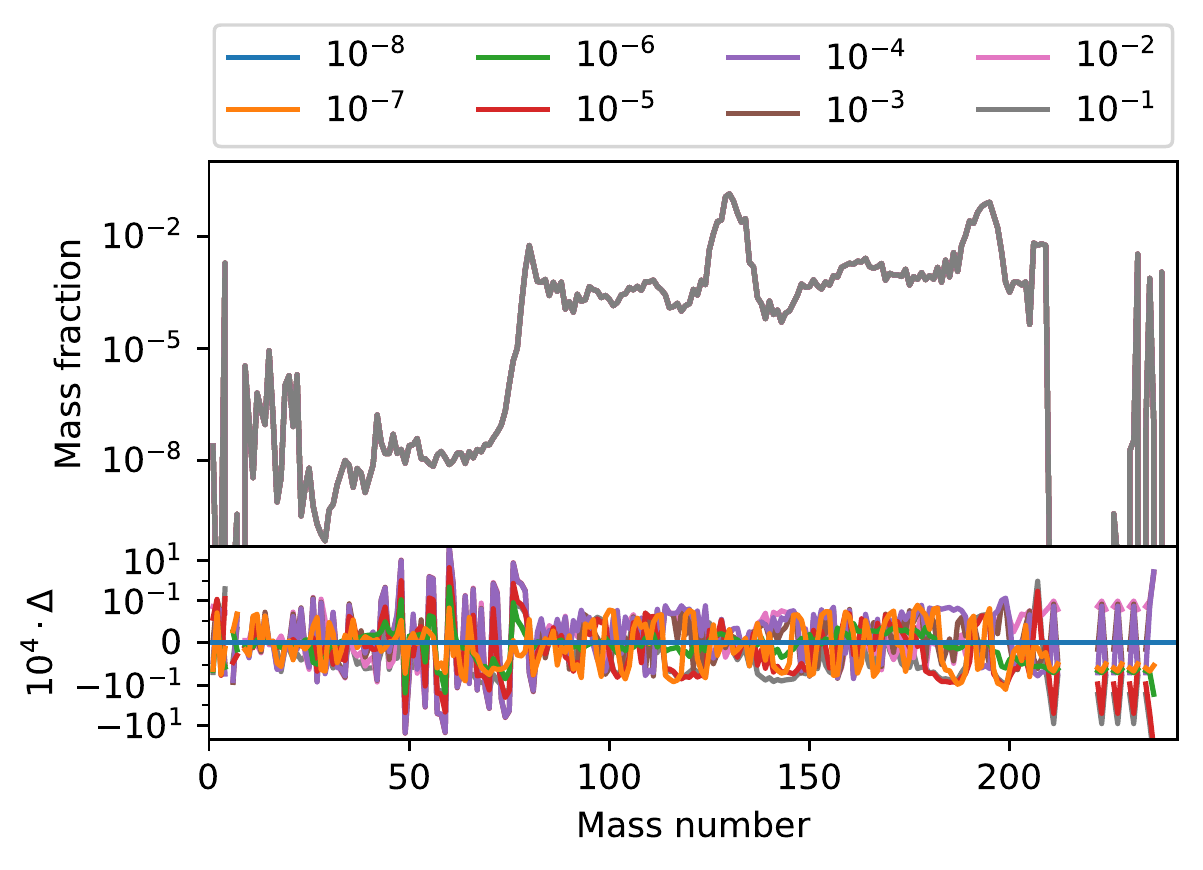}%
\end{center}
\caption{Left plot: calculation using the implicit Euler integration scheme using the Newton-Raphson convergence criteria that is based on baryon conservation (Eq.~\eqref{eq:euler_massconservation}) and different values of $\epsilon_\mathrm{NR}$. Right plot: the same, but using an alternative convergence criteria of the Newton-Raphson $|\max(\vec{x}^{k+1},\vec{x}^k)/\min(\vec{x}^{k+1},\vec{x}^{k+1})-1|<\epsilon_\mathrm{NR}$. The lower panels show the deviation defined as $\Delta = 1- \frac{X_1}{X_2}$, using the most restrictive parameters as reference ($X_1$).}
\label{fig:mass_and_abu_cons_euler}
\end{figure}
The final mass fractions of all runs are shown in the upper-left panel of Fig.~\ref{fig:mass_and_abu_cons_euler}, the difference is defined by $\Delta = 1- \frac{X_1}{X_2}$, where we took $X_1$ as mass fractions from the run with $\epsilon_\mathrm{NR}=10^{-8}$. This is shown in the lower-left panel of Fig.~\ref{fig:mass_and_abu_cons_euler}. The maximum deviation is of the order of $\sim 0.1\%$. Interestingly, it is the iron region that is prone to errors. This part of enhanced errors vanishes completely when not using theoretical weak rates. These rates depend on temperatures as well as densities and can therefore be more challenging to integrate. For values $\epsilon_\mathrm{NR}>10^{-7}$, there is no difference visible. This is due to the fact that the mass for these precisions is already conserved within the minimum of two Newton-Raphson iterations. 

An alternative convergence criteria of the Newton-Raphson that is not based on baryon conservation is $|\max(\vec{x}^{k+1},\vec{x}^k)/\min(\vec{x}^{k+1},\vec{x}^{k+1})-1|<\epsilon_\mathrm{NR}$ for $\vec{x}^{k+1}>10^{-10}$. In other words, every abundance should be converged within a given percentage. The result for this convergence criteria is shown in the right panels of Fig.~\ref{fig:mass_and_abu_cons_euler}. Again, the difference between the most restrictive case and the least restrictive one is of the order of $\sim 0.1\%$, but the parameter has a much more direct impact on the accuracy. The most restrictive scenarios of both convergence criteria agree even within $\sim 0.01\%$ which demonstrates that both criteria can be used interchangeably, and we therefore only include the criterion that is based on baryon conservation, as it has a better performance. 

\begin{figure}
\begin{center}
\includegraphics[width=0.5\linewidth]{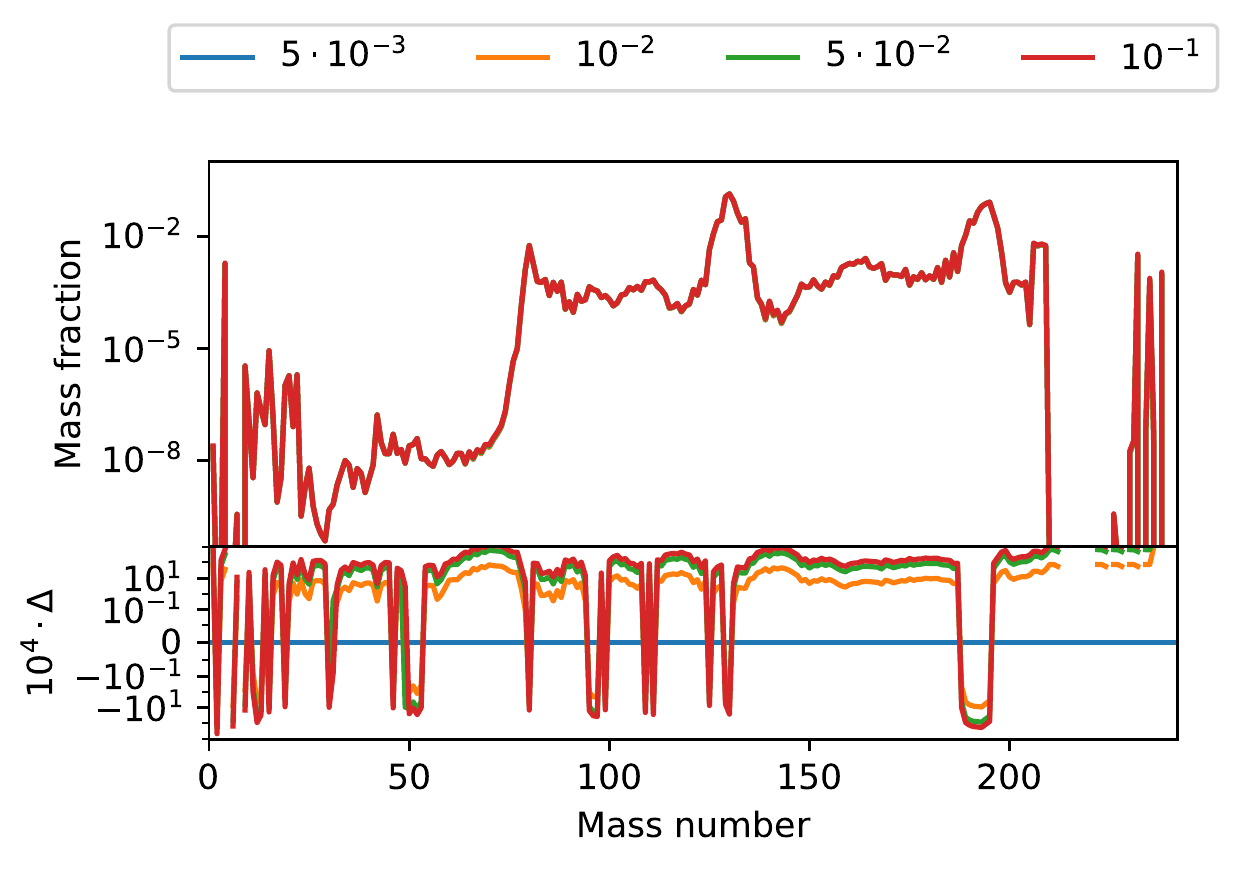}%
\includegraphics[width=0.5\linewidth]{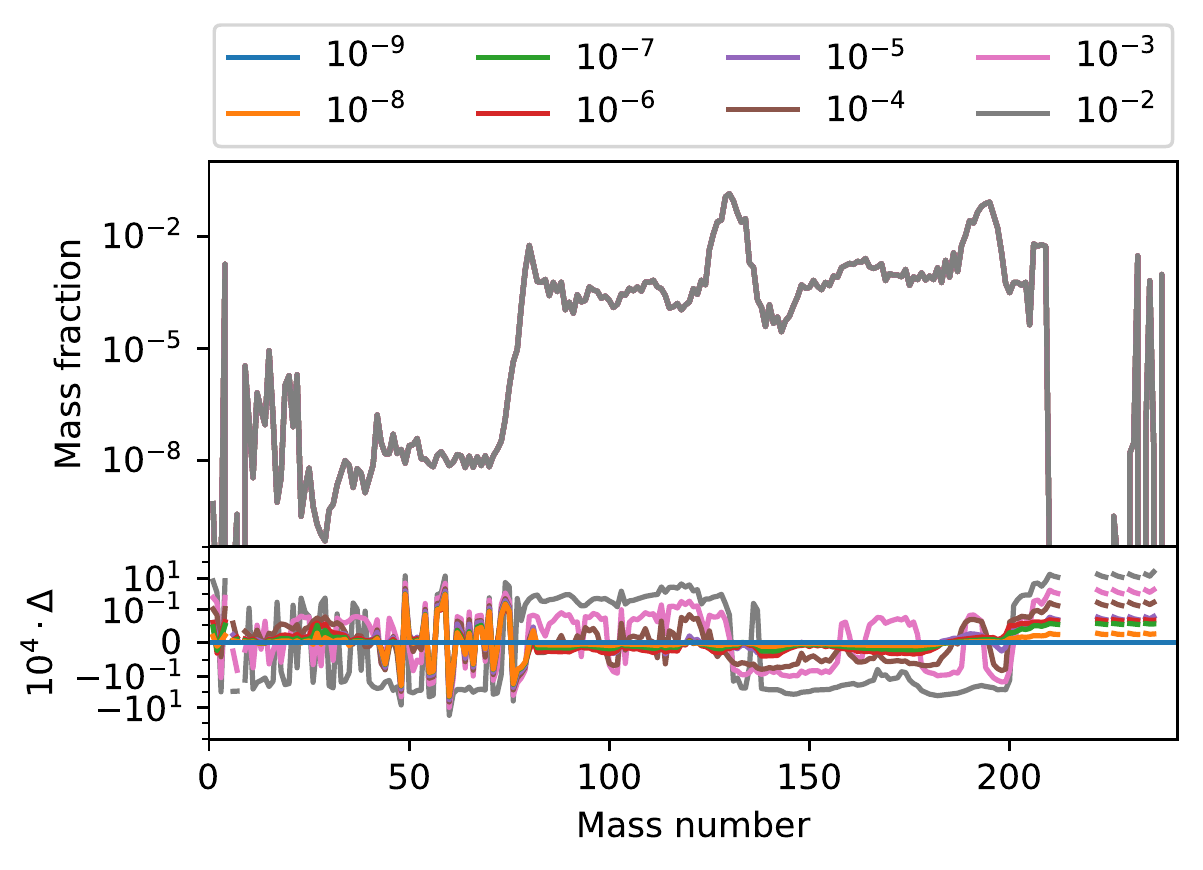}%
\end{center}
\caption{Left plot: calculation using the implicit Euler integration scheme using different time steps by varying $\epsilon_\mathrm{Euler}$ (Eq.~\eqref{eq:impl_euler_eps}). Right plot: the same, but using the Gear solver with different time steps by varying $\epsilon_\mathrm{Gear}$. The lower panels show the deviation defined as $\Delta = 1- \frac{X_1}{X_2}$, using the most restrictive parameters as reference ($X_1$).}
\label{fig:timestep_and_gear_cons}
\end{figure}

All previous calculations were done with the same time step factor of $\epsilon_\mathrm{Euler} = 0.1$ (Eq.~\eqref{eq:impl_euler_eps}). We reduced this factor and tested values of $5\times 10^{-2}$, $1\times 10^{-2}$, and $5\times 10^{-3}$. As shown in the left panels of Fig.~\ref{fig:timestep_and_gear_cons}, the abundances are converged within $\sim 10\%$. In practice, it is not feasible to use a factor of $\epsilon_\mathrm{Euler} = 5\times 10^{-3}$ when calculating many trajectories (c.f., $\sim 3000$ versus $\sim 60000$ time steps for $\epsilon_\mathrm{Euler} = 10^{-1}$ and $\epsilon_\mathrm{Euler} = 5\times 10^{-3}$, respectively).

The Gear integration method, on the other hand, estimates the time step in a more sophisticated way, based on integration errors. This error is controlled by $\epsilon_\mathrm{Gear}$ (Eq.~\eqref{eq:timestep_gear}). When reducing $\epsilon_\mathrm{Gear}$, one directly controls the numerical error (right panels of Fig.~\ref{fig:timestep_and_gear_cons}). The error can be reduced to an almost arbitrary precision, and for all calculated runs it lies within an astonishing precision of $\sim0.1\%$.

\begin{figure}
\begin{center}
\includegraphics[width=0.5\linewidth]{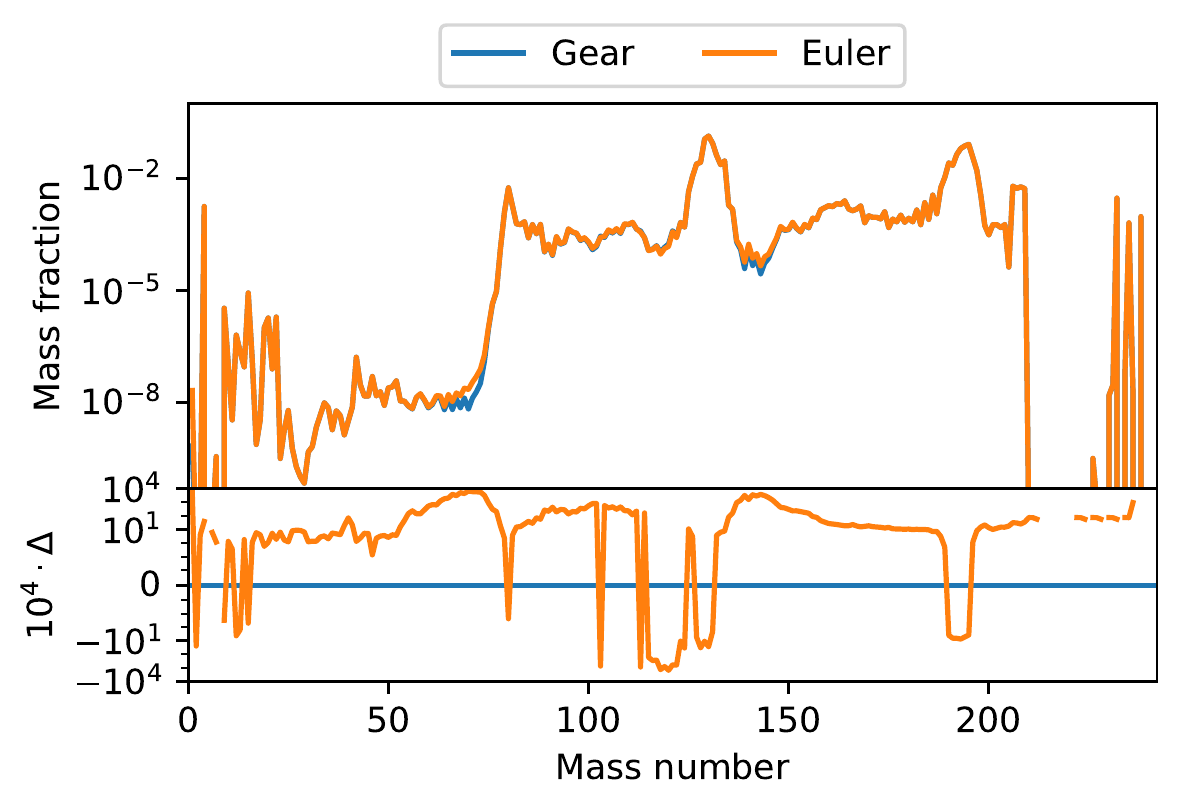}%
\end{center}
\caption{Comparison of a calculation using the implicit Euler method with $\epsilon_\mathrm{Euler}= 5\times10^{-3}$ and a calculation using Gear's method with $\epsilon_\mathrm{Gear}=10^{-9}$. We note that the Gear solver in the lower panel is a horizontal line by definition.}
\label{fig:euler_gear_conv}
\end{figure}
Finally, it is interesting to compare the most precise calculation using the Gear solver ($\epsilon_\mathrm{Gear}=10^{-9}$) with the most precise calculation using the implicit Euler method ($\epsilon_\mathrm{Euler}=5\times 10^{-3}$). This comparison is shown in Fig.~\ref{fig:euler_gear_conv}. The difference between the calculation using the Gear solver and the implicit Euler is for most parts within $~10\%$; however some regions, i.e., around $A\sim 70$ and $A\sim 140$, are differing by a factor of $\sim 2$. However, the largest deviation is visible in the abundance of protons with a factor of $30$ showing that the numerical method can also have a strong impact especially on light nuclei such as neutrons, protons, and alphas.




\end{document}